\documentclass[journal=jctcce,manuscript=article,layout=twocolumn]{achemso}

\usepackage{physics}             % gives access to braket notation, among other things
\usepackage{array}
\usepackage{amsfonts}
\usepackage{amsmath}
\usepackage{mathtools, cuted}

\author{Johann V. Pototschnig}
\affiliation{Department of Chemistry and Pharmaceutical Science, Faculty of Science, Vrije Universiteit Amsterdam, de Boelelaan 1083, 1081 HV Amsterdam, The Netherlands}
\email{j.v.pototschnig@vu.nl}
\author{Anastasios Papadopoulos}
\affiliation{Department of Chemistry and Pharmaceutical Science, Faculty of Science, Vrije Universiteit Amsterdam, de Boelelaan 1083, 1081 HV Amsterdam, The Netherlands}
\email{papadopoulos@kofo.mpg.de}
\author{Dmitry I. Lyakh}
\affiliation{National Center for Computational Sciences, Oak Ridge National Laboratory, Oak Ridge TN, USA}
\author{Michal Repisky}
\affiliation{Hylleraas Centre for Quantum Molecular Sciences, Department of Chemistry, UiT The Arctic University of Norway, N-9037 Troms\o, Norway}
\author{Lo\"ic. Halbert}
\author{Andr\'e Severo Pereira Gomes}
\affiliation{Universit\'e de Lille, CNRS, 
UMR 8523 -- PhLAM -- Physique des Lasers, 
Atomes et Mol\'ecules, F-59000 Lille, France}
\email{andre.gomes@univ-lille.fr}
\author{Hans J{\o}rgen {Aa}. Jensen}
\affiliation{Department of Physics, Chemistry and Pharmacy,
University of Southern Denmark, DK-5230 Odense M, Denmark}
\email{hjj@sdu.dk}
\author{Lucas Visscher}
\affiliation{Department of Chemistry and Pharmaceutical Science, Faculty of Science, Vrije Universiteit Amsterdam, de Boelelaan 1083, 1081 HV Amsterdam, The Netherlands}
\email{l.visscher@vu.nl}

\title{Implementation of relativistic coupled cluster theory for massively parallel GPU-accelerated computing architectures}

\abbreviations{}

\begin{document}

%%%%%%%%%%%%%%%%%%%%%%%%%%%%%%%%%%%%%%%%%%%%%%%%%%%%%%%%%%%%%%%%%%%%%
%% The "tocentry" environment can be used to create an entry for the
%% graphical table of contents. It is given here as some journals
%% require that it is printed as part of the abstract page. It will
%% be automatically moved as appropriate.
%%%%%%%%%%%%%%%%%%%%%%%%%%%%%%%%%%%%%%%%%%%%%%%%%%%%%%%%%%%%%%%%%%%%%
%\begin{tocentry}
%\includegraphics[width=9cm]{pics/TOC.eps}
%\end{tocentry}

%%%%%%%%%%%%%%%%%%%%%%%%%%%%%%%%%%%%%%%%%%%%%%%%%%%%%%%%%%%%%%%%%%%%%
%% The abstract environment will automatically gobble the contents
%% if an abstract is not used by the target journal.
%%%%%%%%%%%%%%%%%%%%%%%%%%%%%%%%%%%%%%%%%%%%%%%%%%%%%%%%%%%%%%%%%%%%%

\begin{abstract}
In this paper, we report a reimplementation of the core algorithms of relativistic coupled cluster theory aimed at modern heterogeneous high-performance computational infrastructures. The code is designed for efficient parallel execution on many compute nodes with optional GPU coprocessing, accomplished via the new ExaTENSOR back end. The resulting ExaCorr module is primarily intended for calculations of molecules with one or more heavy elements, as relativistic effects on electronic structure are included from the outset. In the current work, we thereby focus on exact 2-component methods and demonstrate the accuracy and performance of the software. The module can be used as a stand-alone program requiring a set of molecular orbital coefficients as starting point, but is also interfaced to the DIRAC program that can be used to generate these. We therefore also briefly discuss an improvement of the parallel computing aspects of the relativistic self-consistent field algorithm of the DIRAC program.
\end{abstract}
%%%%%%%%%%%%%%%%%%%%%%%%%%%%%%%%%%%%%%%%%%%%%%%%%%%%%%%%%%%%%%%%%%%%%
%% Start the main part of the manuscript here.
%%%%%%%%%%%%%%%%%%%%%%%%%%%%%%%%%%%%%%%%%%%%%%%%%%%%%%%%%%%%%%%%%%%%%
\section{Introduction}
Computational chemistry is a standard tool in the analysis, design, and synthesis of molecular systems\cite{Thiel2013}. In particular, density functional theory (DFT) is used in a routine fashion in academic and industrial applications. While often sufficiently accurate, DFT does not allow for molecule-specific validations of the accuracy of its predictions. However, this is possible for the wave-function based methods, such as coupled cluster (CC) theory, for which extensions of the single-particle basis combined with an increase of the excitation level in the CC ansatz leads to a systematic improvement of the accuracy. For organic molecules, CC methods can nowadays predict molecular structures to a precision that is better than one picometer in bond lengths and better than one degree in bond angles\cite{Bak2001,Coriani2005}. Furthermore, the efficient equation-of-motion (EOM) treatment of electronically excited states\cite{Bartlett2012, shee2018equation} makes it possible to study photochemical processes and aid the interpretation of spectroscopic data. The standard approaches to compute ground-state energies, molecular properties and electronically excited states have all been generalized to relativistic theory as well, yielding methods that can provide very high accuracy in the electronic structure part of a calculation. This is demonstrated in numerous small molecule applications\cite{Haase2020, Kervazo2019, Denis2019, Zelovich2017, Stralen2002} for which the steep scaling with the system size of the coupled cluster algorithm is not an issue. This rapid increase in computational requirements does, however, in practice prevent application to systems that contain more than about 10 atoms.

Further improvements of the relativistic algorithms are well possible, however, as many reduced-scaling techniques from non-relativistic algorithms can be taken over in a slightly modified form. 
One example is the use of the density fitting (DF)\cite{Whitten1973,Vahtras1993,Foerster2020} or Cholesky decomposition\cite{Beebe1977,Pedersen2009,Peng2017,Helmich-Paris2019} to reduce the size of the two-electron integral tensors. 
Here, relativistic treatments require handling of the density of the small components of the Dirac wave functions or, equivalently, fitting of the relativistic correction terms to a two-electron operator in the two-component formulation\cite{Sikkema2009}.
Another example is the use of the Laplace transform in M\o{}ller-Plesset perturbation theory,\cite{Helmich-Paris2019} where the effects of spin-orbit coupling are visible in the form of (quaternion) imaginary contributions to the density matrices. In both examples one observes a steep increase in the computational cost of the algorithm, but also notes that the formal scaling with the system
size is identical to that of the non-relativistic algorithm. Because numerically small contributions to tensor elements can be neglected by the use of screening techniques, and many additional terms are only significant in the vicinity of heavy atoms, scaling can in principle be further improved. On the other hand, one may observe that inclusion of only a single heavy atom already presents a challenge due to the number of electrons that has to be
correlated in a coupled cluster treatment. This can be illustrated by comparing the CO$_2$ molecule to the uranyl ion, UO$_2^{2+}$. Both are linear triatomic systems with the oxygen atoms contributing a total of 12 valence electrons. In CO$_2$ this yields a total of 16 valence electrons that are to be correlated, while in uranyl one needs to correlate at least 24 electrons\cite{Jong1998} and preferably 34 electrons\cite{Infante2004}. The difference is caused by the so-called semicore 5d, 6s and 6p orbitals. While the effects on chemical bonding due to electron correlation with electrons in the compact 1s shell can be safely neglected in CO$_2$, correlation with electrons in the 6s and 6p orbitals is non-negligible. Additionally, the correlation with the 5d shell may also give a quantitatively important contribution.

Both the increase of the number of electrons to be correlated and the switch from real to complex algebra make relativistic calculations rather demanding. However, they are very  well suited  for  the deployment on supercomputers, because the key algorithms can mostly be formulated as contractions of large tensors, which  can  be  carried  out  with  a  high  computational efficiency. To be able to realize the full potential of both reduced-scaling techniques and parallel computing, it is advantageous to create a modern implementation of the relativistic coupled cluster algorithm.
The legacy CC code of DIRAC, RELCCSD,\cite{visscher1996formulation} allows for parallelization\cite{Pernpointner2003}, but doesn't scale well on a larger number of nodes as it was designed for clusters of the early 2000s. An advantage of this code is the use of spatial symmetry, which reduces the computational cost and is helpful in interpreting molecular spinors and electronic transitions. Both aspects are less relevant when applying the coupled cluster approach to large molecular systems that possess (almost) no symmetry. In our reimplementation, we therefore do not consider symmetry but instead focus on data and compute parallelism. The ExaCorr implementation that we describe here is based on the ExaTENSOR library\cite{lyakh2019domain}, a scalable numerical tensor algebra library for GPU-accelerated HPC platforms developed at the Oak Ridge Leadership Computing Facility (OLCF).

The main body of post-Hartree-Fock quantum-chemical machinery is based on numerical tensor algebra. For the commonly used coupled cluster with singles and doubles (CCSD) model, it is possible to formulate\cite{Papadopoulos2019} all operations as tensor contractions of at most 4-dimensional tensors. This strict adherence to formulation in terms of tensor contractions is the key to the computationally efficient implementation that we present here. As this implementation does not yet introduce approximations such as rank reduction\cite{Parrish.Martinez.2013pxo, Peng2017a} or Laplace transformation\cite{Almlof1991}, it should be regarded both as a platform for future developments and a tool with which to generate reference data to validate approximate methods in which the large number of two-electron integrals and/or excitation amplitudes is reduced.
 
The paper is structured in the following way. In the theory section, we briefly summarize the coupled cluster algorithms that we consider in the current work. This is followed by the implementation section in which the implementation of the algorithms is discussed. The next section is devoted to the details of the computations we used to test the implementation. In the results and discussion section, we present calculations for validation of the correctness of the results by comparing with the reference RELCCSD implementation as well as calculations aimed at showing the computational scaling. The conclusion follows which includes a discussion of follow-up work.

\section{Theory}

\subsection{Relativistic theory}
Prerequisite for a relativistic coupled cluster calculation is a set of two- or four-component molecular spinors obtained by solving the relativistic Dirac-Hartree-Fock equation. In the four-component case, this equation reads
\begin{strip}
\begin{align} 
\hat{f}_D \psi =
\left[ 
\begin{array}{c c}
 \hat{V}_{eN} + \hat{J} - \hat{K} &
  c \left(\boldsymbol{\sigma}\cdot\hat{\mathbf{p}}\right) - \hat{K}\\
 c \left(\boldsymbol{\sigma}\cdot\hat{\mathbf{p}}\right) - \hat{K}& 
\hat{V}_{eN} + \hat{J} - \hat{K} -2 m c^2 \\
\end{array}
 \right]
 \left[ 
\begin{array}{c}
\psi^L \\
\psi^S
\end{array}
 \right]
 =
 E
  \left[ 
\begin{array}{c}
\psi^L \\
\psi^S
\end{array}
 \right].
\label{eq:DIRACFOCK}
\end{align}
\end{strip}
in which $\hat{V}_{eN}$ represents the nuclear-electron interaction, usually defined with a Gaussian model of the nuclear charge distribution\cite{Visscher1997a}, and the local $\hat{J}$ and non-local $\hat{K}$ operators describe the electron-electron interaction in the mean-field approximation,
\begin{strip}
\begin{align} 
\hat{J} = \sum_j^{N_o} \int\psi_j^\dagger(\mathbf{x}_2)
\hat{g}\left(\mathbf{x}_1,\mathbf{x}_2 \right)
\psi_j(\mathbf{x}_2)d\mathbf{x}_2 \\
\hat{K}\psi_i^X(\mathbf{x}_1) = \sum_j^{N_o} \sum_Y^{L,S} \int\psi^{\dagger Y}_j(\mathbf{x}_2)
\hat{g}\left(\mathbf{x}_1,\mathbf{x}_2 \right)
\psi^X_i(\mathbf{x}_2)d\mathbf{x}_2 \psi^Y_j(\mathbf{x}_1)
\label{eq:JK}
\end{align}
\end{strip}
in which $N_o$ denotes the number of occupied spinors. The 2-electron interaction operator $\hat{g}\left(1,2 \right)$ is the Coulomb(-Gaunt) operator
\begin{align} 
\hat{g}\left(1,2 \right) = \frac{e^2}{4 \pi \varepsilon_0} \frac{1}{r_{ij}} 
\left( - \frac{e^2}{4 \pi \varepsilon_0} \frac{\mathbf{\alpha}_i \cdot \mathbf{\alpha}_j}{r_{ij}} \right).
\label{eq:Coulomb-Gaunt}
\end{align}
Before proceeding to the coupled cluster stage, all operators are transformed using the exact 2-component (X2C) method that allows re-expression of the 4-component spinors into a 2-component picture. 
Three main variants of the X2C method are used in the current work. 
The first one, termed as X2C-1e, is based on the simple X2C transformation of the one-electron Dirac Hamiltonian that is combined with the non-relativistic Coulomb operator to describe the electron-electron interactions~\cite{Ilias2007,Konecny2016}. 
Since X2C-1e omits all two-electron relativistic corrections and leaves the relativistic scalar and spin-orbit coupling operators associated with the nuclear potential unscreened, the second variant extends X2C-1e by an explicit addition of the atomic mean-field two-electron potential (done via AMFI code~\cite{prog:amfi}). This approach is termed as X2C-AMFI and is the default X2C Hamiltonian in DIRAC. 
In both X2C approaches, the transformation to the 2-component picture is carried out \textit{before} the Hartree-Fock procedure and therefore the two-electron molecular integrals that involve small component basis functions are never computed. 
In contrast, these types of integrals do enter in the third variant, named X2Cmmf~\cite{Sikkema2009}, as the X2C transformation is carried out \textit{after} solving the Hartree-Fock equations, and therefore the full molecular potential is used to define the X2C transformation. This makes X2Cmmf more accurate than the X2C-AMFI (or X2C-1e) approach.
Moreover, the two-electron spin-orbit contributions of electrons which will not be explicitly taken into account in the correlation treatment (hereafter referred to as core, or frozen electrons) are treated exactly.

Although (obviously) more expensive due to the mean-field part of the calculations, the X2Cmmf procedure has the same favorable computational characteristics as the X2C-AMFI (or X2C-1e) procedure in the post-HF steps, with the advantage that it yields results that are very close to the full 4-component treatment~\cite{Sikkema2009,shee2018equation}. In the current implementation, the X2Cmmf approach functions as the high-level reference method, while in the DIRAC code the use of the X2C-AMFI and a non-relativistic treatment (to compare with other coupled cluster implementations) are also supported. Currently, the X2Cmmf approach also allows for an approximate inclusion of the Gaunt interaction,~\cite{Sikkema2009} and an implementation of the full Dirac-Coulomb-Gaunt operator for the use in very precise benchmark calculations is planned as well.

All the aforementioned methods apply the no-virtual pair approximation such that the Hamiltonian to be used for the coupled cluster treatment is written in the second quantization as

\begin{strip}
\begin{align}
\hat{H} = 
E^{core}
+ \sum_{pq}^{valence} h^{core}_{pq} a^\dagger_p a_q
+ \frac{1}{4} \sum_{pqrs}^{valence} V^{pq}_{rs} a^\dagger_p a^\dagger_r a_s a_q ,
\label{eq:Hamiltonian}
\end{align}
\end{strip}
with $V^{pq}_{rs}$ being the anti-symmetrized two-electron integrals
\begin{align} 
V^{pq}_{rs} = 
\left\langle pr \right| \hat{g} \left| qs \right\rangle - \left\langle pr \right| \hat{g} \left| sq \right\rangle,
\label{eq:two_electron}
\end{align}
and valence in the summation indicating the spinors that are active in the coupled cluster calculation (omitting frozen occupied spinors as well as deleted virtual spinors). The $E^{core}$ constant contains the energy of the core electrons as well as the nuclear repulsion term. The operator $h^{core}$ describes the interaction between the frozen core electrons and the valence electrons and
contains the Dirac kinetic energy and nuclear-electron interaction terms defined above as well. The main difference with the non-relativistic treatments that use an identical expression, is the fact that the tensors
$\mathbf{h}^{core}$ and $\mathbf{V}$ are defined in complex algebra, whereas in the non-relativistic treatments it is usually possible to employ real algebra.

    In non-relativistic quantum chemistry the Hamiltonian is spin-free, which makes it possible to separate the spatial and spin degrees of freedom and solve equations for spatial orbitals. In relativistic computations such a separation is not possible because relativistic spinors cannot be written as a simple product of a spatial and spin function. Though, in the absence of magnetic fields, one may still use time-reversal symmetry, also known as Kramers symmetry, as each spinor can be related to another with the same energy\cite{Kramers1930,Saue1999}. 
    Use of this symmetry implies a Kramers-restricted (KR) algorithm in which the occupation of each of the two spinors that comprise a Kramers pair is kept identical in defining the mean field potential.
    In contrast, a Kramers unrestricted (KU) algorithm treats the spinors independent from each other and allows to obtain a so-called spin polarization effect~\cite{Repisky2020}.
    
    As the use of Kramers symmetry has little advantage in coupled cluster calculations\cite{Visscher1995, visscher1996formulation}, and we intend to keep the implementation modular and independent of the program used to generate the spinors, we will henceforth assume that all spinors are unrelated to each other, demanding only orthogonality between them.

\subsection{Coupled cluster algorithms}

The wave-function in the coupled cluster method is defined as
\begin{align}
\left| \Psi_{\textnormal{CC}} \right\rangle =  e^{\hat{T}}  \left| \Phi_{\textnormal{0}} \right\rangle ,
\label{eq:cc_wf}
\end{align}
where $\left| \Phi_{\textnormal{0}} \right\rangle$ is the single-determinant wave function. The cluster operator $\hat{T}$ is most commonly restricted to the single and double hole-particle excitations
\begin{strip}
\begin{align}
\hat{T}=\hat{T_1}+\hat{T_2} ; \;
\hat{T_1} = \sum_i \sum_a t^{a}_{i} a^\dagger_{a} a_i ; \;
\hat{T_2} = \sum_{ij} \sum_{ab} t^{ab}_{ij} a^\dagger_{a} a^\dagger_{b} a_i a_j,
\label{eq:t_amp}
\end{align}
\end{strip}
defining the coupled cluster singles and doubles method (CCSD). The energy and cluster amplitudes are computed using the equations
\begin{align}
\left\langle \Phi_0 \right| \hat{\bar{H}} \left| \Phi_0 \right\rangle = E ,
\\
\left\langle \Phi_l \right| \hat{\bar{H}} \left| \Phi_0 \right\rangle = 0 ;
 \left| \Phi_l \right\rangle = \hat{\tau_l} \left| \Phi_0 \right\rangle ,
\label{eq:CCSD}
\end{align}
with $\hat{\tau_l}$ denoting a generic excitation operation yielding any (singly, doubly, $\ldots$) excited determinant $\left|\Phi_l\right\rangle$, and where the similarity-transformed Hamiltonian
\begin{align}
\hat{\bar{H}}=e^{-\hat{T}}\hat{H}e^{\hat{T}}
\end{align}
is employed. The working equations for this formalism are well-known and can for instance be found in the paper\cite{visscher1996formulation} describing the RELCCSD program that we used as a reference implementation.
In contrast to this code, for the current implementation we assume that the working memory of the parallel computer is large enough to keep all tensors in memory. Furthermore we formulate all operations as tensor contractions to enable efficient use of the ExaTENSOR library. Some intermediates were therefore also altered, resulting in the working equations listed in Appendix \ref{ap:CC}. To allow for faster calculations the CC2 approximation is implemented according to the working equations of Appendix \ref{ap:cc2}. In order to speed up the convergence of the CC solver, the direct inversion in the iterative subspace (DIIS) algorithm\cite{Pulay1980} was implemented. Triple excitations are necessary to achieve chemical accuracy, but they are computationally expensive. A widely applied compromise is thus to add them perturbatively\cite{Raghavachari1989, Deegan1994, visscher1996formulation}. The relevant working equations can be found in Appendix \ref{ap:triples}.

In order to obtain first-order molecular electronic properties at the coupled cluster level we use the Lagrange formalism,\cite{shee2016analytic} which requires solving the equations for the Lagrange multipliers $\{\mathbf{\lambda}\}$
\begin{strip}
\begin{align}
L^{CC}\left(\mathbf{t},\mathbf{\lambda} \right)= 
\left\langle \Phi_0 \right| \hat{\bar{H}} \left| \Phi_0 \right\rangle 
+\sum_l \lambda_l \left\langle \Phi_l \right| \hat{\bar{H}} \left| \Phi_0 \right\rangle 
\end{align}
\end{strip}

These are obtained from the stationary conditions
\begin{align}
\frac{\partial L^{CC}}{\partial \mathbf{\lambda}} = 0 ,
\label{eq:CCSD_LAMBDA}
\\
\frac{\partial L^{CC}}{\partial \mathbf{t}} = 0 ,
\label{eq:LAMBDA}
\end{align}
where Eq. \ref{eq:CCSD_LAMBDA} represents the CC equations. Note that this definition of the Lagrangian neglects orbital relaxation, which is assumed to be partly covered by the $\hat{T_1}$ operator. Equation \ref{eq:LAMBDA} is solved to obtain the values of the Lagrange multipliers after which the expectation value of any one-body operator $\hat{O}$ can be computed by computing the one-body density matrix $\boldsymbol{\gamma}$\cite{shee2016analytic}
\begin{align}
\gamma_{p}^{q} & = \left\langle \Phi_0 \right| 
\left( 1 + \Lambda \right)e^{-\hat{T}} a^{\dagger}_q a_p e^{\hat{T}} \left| \Phi_0 \right\rangle \\
\left\langle \hat{O} \right\rangle & = \sum_{p,q} \gamma_{p}^{q} o_{pq}
\end{align}
The symmetrized one-body density matrix is transformed to the atomic orbital basis and then contracted with the matrix representation of the appropriate property operator $\hat{O}$. The resulting working equations are listed in appendices \ref{ap:lambda} and \ref{ap:density}.

\section{Implementation}

In this part the details of the implementation are presented. 
In order to run coupled cluster computations, molecular spinors for a reference state are required. This computation of these molecular spinors is described in section~\ref{sec:mol_orb}, as well as changes to improve performance for the larger systems that become feasible with the new implementation. 
The implementation utilizes two separate libraries to perform the compute-intensive operations.  The calculation of two-electron integrals in the atomic basis is  performed by the efficient InteRest library (section~\ref{sec:InterRest}), while the tensor contractions that comprise the majority
of the coupled cluster algorithm are performed with the ExaTENSOR library(section~\ref{sec:ExaTENSOR_TALSH}) 
Input handling and interfacing to SCF programs is discussed in section~\ref{sec:Interface}, while
transformation of two-electron integrals from the atomic to the molecular basis is described in section~\ref{sec:AOtoMO}. 
Finally, details regarding the coupled cluster code are discussed in section~\ref{sec:CC_implem}.

\subsection{Generation of the molecular spinors}
\label{sec:mol_orb}
    
Molecular spinors are required for the ExaCorr coupled cluster module and they thus need to be efficiently generated for large system sizes. Because of the fast evaluation of 2-electron integrals by the InteRest module, the efficient parallel implementation of the AO-to-MO transformation, and the fast solution of the CC equations described below, for DIRAC calculations the Fock matrix diagonalizations required in the self consistent field (SCF) stage became a bottleneck. As this step was not parallelized, it became excruciatingly slow for large AO basis spaces.

% =============== QDIAG implementation ===

Historically, before DIRAC the well-known double point groups as formulated by Wigner\cite{Wigner} were used in the pioneering 4c relativistic molecular codes.
When the SCF optimization was implemented in DIRAC\cite{Saue1996,Saue1999}, we used instead a more general quaternion description, which in fact relies on the simpler (single) point group irreps for quaternion basis function components\cite{Saue1996}.
This implementation has, with small adjustments, been used until work on ExaCorr started, and each SCF-DIIS iteration has thus been based on a direct MPI parallel construction of Fock matrices based on the DALTON implementation,\cite{daltonprogram} followed by a sequential quaternion generalization of the Fock matrix diagonalization (see appendix D in \citenum{Saue1996}). This procedure gave a satisfactory scaling with number of MPI nodes for
calculations of up to approximately 2000 AOs, that were feasible with the RELCCSD module.
However, the new ExaCorr CC module described in this paper allows for larger applications and significantly large AO spaces and it became paramount that one should be able to do SCF calculations with 1000-5000 AO basis functions (and more in the future) in a small fraction of the wall time needed for the AO-to-MO transformation and the CC calculations.
Analysing the SCF performance for such larger systems with large numbers of compute nodes, it turned out that the parallel Fock matrix constructions is acceptably efficient, but it was no surprise that the sequential quaternion matrix diagonalization needed in the MO based DIIS algorithm needed revision.

In this subsection we describe how this diagonalization bottleneck was removed by tuning of the sequential QDIAG code and addition of OpenMP structures in QDIAG. 
In relativistic quantum chemistry there mostly two approaches used for diagonalization. 
On one hand quaternions can be used to get matrix representation in real numbers, which is used here and in a recent publication dealing with large scale quaternion matrix diagonalization.\cite{Shiozaki2017}
On the other hand, complex numbers and routines can be used which is applied in ReSpect. 

The implementation of the QDIAG routines in DIRAC by Saue\cite{Saue1996} in 1995 was based on his clever quaternion generalization of the complex diagonalization routines in EISPACK, 
where the EISPACK routines are direct transcriptions of the original ALGOL versions. However, ALGOL just as C and C++ uses a row-major storage of matrices while FORTRAN uses a column-major. Therefore the EISPACK routines were very inefficient for larger matrices because of many cache misses caused by the large strides in memory. A necessary first step was consequently to rewrite the QDIAG routines by transposing the access to all matrices followed by improvements of the logical structure. This change by itself already caused significant improvement in the sequential performance. The resulting implementation was then suitable for OpenMP parallelization. Initial timings of a large application on the TITAN supercomputer that was performed with 800 cores indicated that OpenMP parallelization with just 8 OpenMP threads was sufficient to reduce the time spent in diagonalization to less than 11 minutes, compared to an overall wallclock time of 66 minutes in one SCF iteration (outputs of this and other benchmark runs are provided in the supporting information). Additional timings on the SUMMIT supercomputer also showed that the wall time spent in diagonalization is much less than needed for other steps like Fock matrix construction, and therefore it was deemed unnecessary to also program additional GPU and/or MPI parallelization. 

\subsection{InteRest Integral Library}
\label{sec:InterRest}

In the ExaTENSOR library (described below) it is possible to call an external library to initialize a particular tensor with the desired values. This mechanism allows for efficient parallel computation of the electron repulsion integrals (ERIs). Prerequisite is, however, that this external library is sufficiently modular, a requirement that could not be met by the legacy HERMIT integral generator used in DIRAC. We therefore interfaced the InteRest library~\cite{Repisky2018} to enable parallel computation of the ERIs arising from relativistic and nonrelativistic theories.

As discussed in Ref.~\citenum{Repisky2020}, all commonly applied basis types in relativistic calculations are of a multi-component spinor nature and can uniformly be formulated in terms of real quaternion functions ($\mathbb{H_{R}}(\mathbb{R}^{3})$) or complex quaternion functions ($\mathbb{H_{C}}(\mathbb{R}^{3})$) over the field of real numbers $\mathbb{R}$. A product of any two quaternion basis functions $X_{a}(\vec{r})$ and $X_b(\vec{r})$ defines the so-called quaternion overlap distribution function $\Omega_{ab}(\vec{r})\equiv X^{\dagger}_{a}(\vec{r})X_{b}(\vec{r})$ in terms of which one can formulate and design an efficient algorithm for evaluation of non-relativistic and relativistic ERIs.~\cite{Repisky2020}
For instance, if $X(\vec{r})=X^{\text{RKB}}(\vec{r})\in\mathbb{H_{R}}(\mathbb{R}^{3})$ refers to a restricted kinetically balanced (RKB) basis,~\cite{Stanton1984} then $\boldsymbol{\Omega}\in\mathbb{H_{R}}(\mathbb{R}^{3})$
comprises of four real quaternion components. Then, a single quadruplet of ERIs, defined similarly to the non-relativistic case as 
\begin{equation}
   \label{eq:ERI}
   \left[\Omega_{ab}|\Omega_{cd}\right]
   \equiv
   \iint
   \frac{\Omega_{ab}(\vec{r}_{1})\Omega_{cd}(\vec{r}_{2})}{r_{12}}~d\vec{r}_{1}d\vec{r}_{2},
\end{equation}
requires the evaluation and processing of 25 times more real scalar integrals than in the non-relativistic case. InteRest utilizes the Obara–Saika integration technique over Cartesian Gaussians~\cite{Obara1986} to compute all these scalar integrals in parallel and it groups them into four integral classes [LL$|$LL], [SS$|$LL], [LL$|$SS], and [SS$|$SS] according to their values, which gradually decrease in powers of $c^{-2}$.~\cite{Repisky2020} At the expense of going from real to complex quaternion functions, the presented uniform formalism for relativistic ERI evaluation can also be applied in solid-state domain.~\cite{Kadek2019} Additional basis requirements needed for magnetic property calculations, such as the restricted magnetic balance~\cite{Komorovsky2008} (RMB) in combination with the gauge-including atomic orbitals (RMB-GIAO),~\cite{Komorovsky2010} can also be handled with the discussed integral scheme. A thorough discussion about this topic has been given in Ref.~\citenum{Repisky2020}.

\subsection{ExaTENSOR and TAL-SH Backends}
\label{sec:ExaTENSOR_TALSH}

The ExaCorr module provides two distinct implementations of coupled cluster methods, one intended for execution on a single shared-memory node with an optional GPU acceleration and another one for execution on many such nodes (distributed parallelism), thus supporting a broad variety of computer platforms, from simple workstations to leadership HPC systems. Both implementations use the ExaTENSOR library\cite{lyakh2019domain} as a massively parallel GPU-accelerated processing backend for numerical tensor algebra operations, although there are some differences in the interface between the single- and multi-node API. For the single-node runs (possibly with use of OpenMP and /or GPU acceleration), only the single-node component of ExaTENSOR, the TAL-SH library\cite{TALSH}, is used. The single-node implementation is more efficient when MPI parallelization is not needed. It also serves as a validation reference for the corresponding multi-node implementation. The ExaTENSOR library is written in a mix of Fortran-2008 and C/C++. It depends on BLAS, LAPACK, OpenMP, CUDA, and MPI (MPI is not necessary for single-node runs while CUDA is only necessary for the GPU-enabled builds).

\begin{figure*}[htb]
\centering
\includegraphics[width=0.7\textwidth]{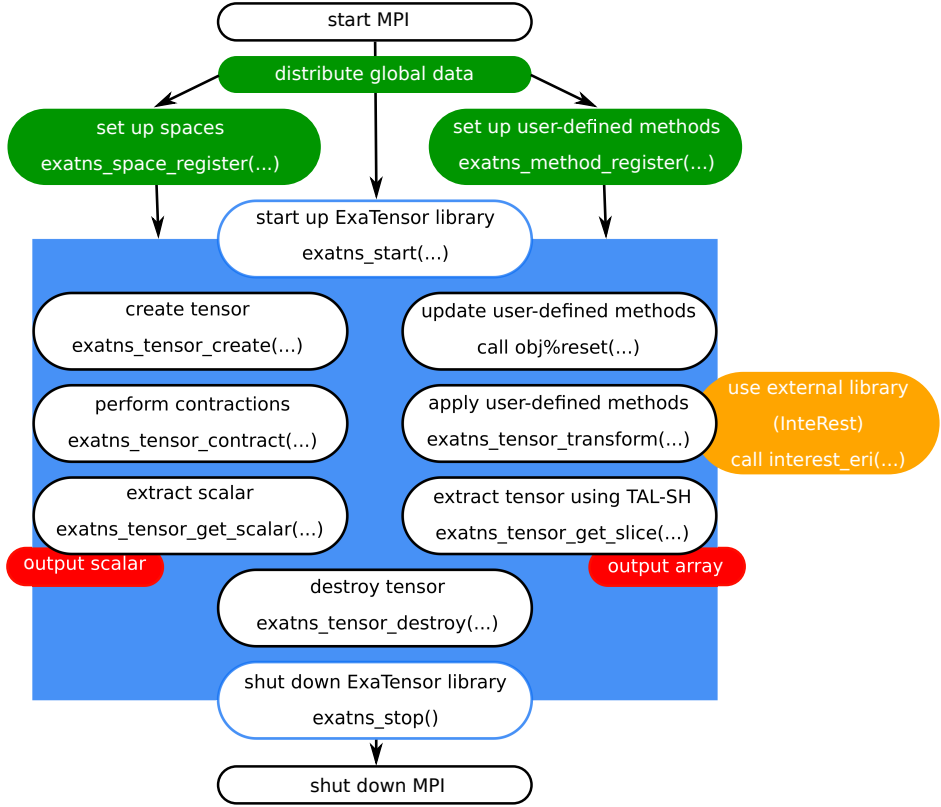}
\caption{ExaCorr computational workflow based on the ExaTENSOR library}
\label{fig:exatensor_outline}
\end{figure*}

Figure \ref{fig:exatensor_outline} outlines the computational workflow where the ExaCorr module offloads all computationally expensive operations (primarily tensor contractions) to the ExaTENSOR library. Essentially, the high-level interface of the ExaTENSOR library allows for creation, destruction, addition and contraction of distributed tensors via a single API call per operation, thus making it possible to directly translate tensor equations into the library calls. Such direct translation of quantum many-body equations into a human-readable code drastically accelerates the implementation of new coupled cluster methods in the ExaCorr module. Additionally, ExaTENSOR provides API for user-defined transformations on distributed tensors, which are often necessary in the coupled-cluster algorithms. As described by Figure \ref{fig:exatensor_outline}, the general computational workflow of a coupled cluster method implemented in ExaCorr starts with a replication of some global data, like molecular spinor coefficients, diagonal elements of the Fock matrix, etc., which normally do not consume much memory. Then all necessary vector spaces for the many-body tensors are defined, such as the space of atomic orbitals, occupied molecular spinors, virtual molecular spinors, etc. After that, all necessary ExaCorr-specific unary tensor transformations are registered. These extensions of the ExaTENSOR library are implemented as extensions of an abstract tensor transformation class provided by the ExaTENSOR interface. Once this is done, the ExaTENSOR parallel runtime (domain-specific virtual processor\cite{lyakh2019domain}) is started within a provided MPI communicator. After initialization, ExaTENSOR will begin accepting commands to perform distributed tensor algebra operations which realize a given coupled cluster algorithm. Importantly, the ability to implement user-defined tensor transformations facilitates the use of external libraries within ExaTENSOR, for example the InteRest library\cite{Repisky2018}, which was straightforwardly integrated with ExaTENSOR to enable parallel computation of the Coulomb integrals. Finally, once the given coupled cluster workload has been executed to completion, a local copy of the resulting scalar (e.g., energy, property) or tensor (e.g., density matrix) of interest can be retrieved. At the very end, the ExaTENSOR parallel runtime is explicitly shut down and control is handed back to the stand-alone ExaCorr or DIRAC program.

In ExaTENSOR, a tensor is formally defined as a vector from a linear space constructed as a direct product of basic vector spaces. Such a multi-indexed vector (tensor) is represented by an array of complex numbers. The number of basic spaces in the direct product space defines the \textit{order} of the tensors living in that direct product space (note that in physics the tensor \textit{order} is usually called the tensor \textit{rank}). Each tensor dimension is thus associated with a specific basic space (or its subspace) from the defining direct product space. One must explicitly register all necessary basic vector spaces by calling \texttt{exatns\_space\_register} API function provided by ExaTENSOR. In order to construct a basic vector space, one simply needs to provide a basis for that space or just specify its dimension. One can also construct a subspace of a registered basic vector space, thus enabling construction of tensor slices. The definition of the basic vector spaces requires their splitting into a user-defined number of subspaces, thus inducing the splitting of tensors into tensor slices. These slices are called elementary tensor blocks. All tensors are stored as collections of such elementary tensor blocks. In the current implementation, the segment size used for splitting a basic vector space into a direct sum of its subspaces can be controlled by a keyword (see supplemental information).

Another prerequisite of coupled cluster algorithms is the necessity of custom tensor transformations (or initializations), like initialization of the Coulomb integral tensor, import of pre-existing many-body tensors (e.g., the Fock matrix), Jacobi preconditioning during amplitude updates, etc. Each such an initialization or transformation can easily be injected into ExaTENSOR by implementing user-defined tensor classes extending the abstract class \texttt{tens\_method\_uni\_t}, followed by their registration with \texttt{exatns\_method\_register}. These user-defined subclasses can further be classified as either static or dynamic. The objects of static subclasses do not change their internal state after the registration with the ExaTENSOR runtime whereas the objects of dynamic subclasses are allowed to change their internal state after the registration, thus enabling further flexibility and dynamic behavior during the execution of a tensor algorithm.

Once all necessary basic vector spaces/subspaces and user-defined tensor methods have been pre-registered, one may proceed to the execution of the actual tensor operations on distributed tensors. A tensor is created via calling \texttt{exatn\_tensor\_create} where a user provides which space/subspace each tensor dimension is associated with. Inside the ExaTENSOR parallel runtime, each tensor is recursively decomposed into smaller slices which are distributed across all nodes. A tensor can then be initialized to either a scalar value or some custom value via a user-defined initialization method (\texttt{exatns\_tensor\_init}). There are three main tensor operations currently provided by ExaTENSOR: User-defined unary tensor transformation (\texttt{exatns\_tensor\_transform}), tensor addition (\texttt{exatns\_tensor\_add}), and tensor contraction (\texttt{exatns\_tensor\_contract}). These are sufficient for implementing the majority of coupled cluster algorithms. Both tensor addition and tensor contraction API take symbolic strings specifying the addition/contraction index pattern, for example
\begin{verbatim}
  S(a,b,i,j) += V(a,b,c,d) * T(c,d,i,j)
\end{verbatim}
for a partial contraction over indices $c$ and $d$, or
\begin{verbatim}
  E() += V+(a,b,i,j) * T(a,b,i,j)
\end{verbatim}
for a full contraction over all indices in which the complex conjugate values of the tensor $V$ are used (indicated by the $+$ symbol). Tensor reordering is usually not necessary but can be achieved with the following specification:
\begin{verbatim}
  A(a,c,b,d) += B(a,b,c,d)
\end{verbatim}
This reordering is employed in the creation of the antisymmetrized ERIs of equation \ref{eq:two_electron} after the AO-to-MO transformation is completed.

In principle, the tensor operations submitted to ExaTENSOR are processed asynchronously but the processes can be explicitly synchronized by calling \texttt{exatns\_sync} to ensure the completion of all outstanding computations (like a barrier). Once all necessary computations have been completed, one can retrieve a local copy of the computed scalar (e.g., energy, property) via \texttt{exatns\_tensor\_get\_scalar}. If one needs a slice of some computed tensor instead (e.g., density matrix), \texttt{exatns\_tensor\_get\_slice} will return a local copy of the requested tensor slice. All created tensors need to be explicitly destroyed once no longer needed via \texttt{exatns\_tensor\_destroy}.

All aspects discussed above are taken care of by the implementation and cannot be changed by the user of ExaCorr. Job-specific tuning and optimization of the parallelization is, however, possible by setting environment variables and/or specific keywords in the input. This provides control over the amount of memory used on a single node, whether GPUs will be used, how OpenMP threads are to be distributed, etc.

\subsection{Molecular spinors: interface to DIRAC and ReSpect}
\label{sec:Interface}
    
The ExaCorr module was designed with modularity in mind, so it would be easy to interface with other quantum-chemical packages. For convenience we currently use the build infrastructure of DIRAC, but the code can also be compiled and used as a stand-alone program, since the minor dependencies on some specific modules of DIRAC can be easily removed.

\begin{figure}[htb]
\centering
\includegraphics[width=0.5\textwidth]{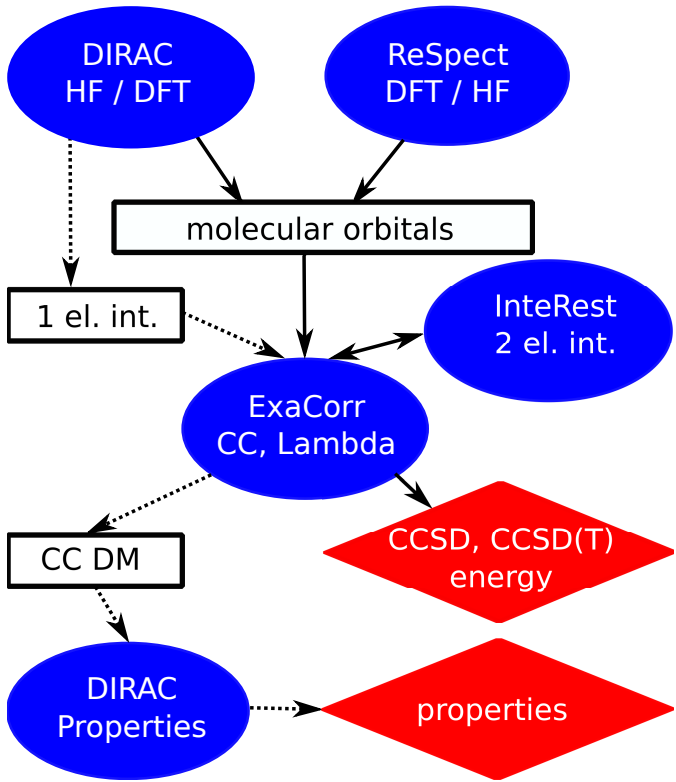}
\caption{Workflow of ExaCorr computations, details can be found in the text.}
\label{fig:program_structure}
\end{figure}

ExaCorr requires two files with information to be present: a job input file and a file containing information about the molecular spinors. A complete diagram of the interface is depicted in Figure~\ref{fig:program_structure}.\\
The input file (exacc.inp) contains the options controlling the coupled cluster computations and should at least include the definition of the active occupied and active virtual spinor spaces. Spinors outside this active space are considered as belonging to the frozen core (for the occupied spinors) or as deleted (for the virtual spinors).  In the following we will consider the occupied and virtual spaces as pertaining to these (potentially reduced) subspaces of the full spinor spaces defined in the molecular spinor file. Examples for additional options are convergence thresholds, choice of coupled cluster wave functions (CCD / CCSD / CC2 / CCSD(T)), a switch to enable the computation of the density matrix and several more technical keywords. A complete list can be found in the supporting information. These options can also be set in the DIRAC input (dirac.inp) if the ExaCorr module is called directly from DIRAC. 

The second file can either be DIRAC's molecular spinor file, DFCOEF, or the RSD\_MOS file from ReSpect\cite{Repisky2020}.
These interface files contain three different sets of data defining the canonical molecular spinors: (i) information about the basis set, (ii) the coefficients of the molecular spinors, and, (iii) the spinor energies thereof for the Fock matrix expression used in the generating SCF procedure. An optional input file (MRCONEE) containing one-electron integrals  can be generated by the MOLTRA module in DIRAC. This additional data can be used to recompute the Fock operator for open-shell cases, for which the DIRAC definition,\cite{Thyssen2001} used to define the spinor energies, differs from the simple KU formalism assumed in ExaCorr.
Results of the CC calculations are provided in the form of a text output file and an effective density matrix, in case the lambda equations are solved as well. This density matrix can be used by DIRAC to
compute a wide range of molecular properties. As DIRAC assumes a KR formalism, the latter type of calculation is currently limited to Kramers symmetric (closed shell) systems.

\subsection{Index transformation algorithms}
\label{sec:AOtoMO}
    
For relativistic calculations in which the size of the AO-space is usually an order of magnitude larger than the MO-space, the  transformation of the two-electron integrals from the atomic to the molecular basis can amount to a significant fraction of the overall computational expense. There are different approaches to implement these transformations differing in memory requirements and operation count. In ExaCorr, the current default is the standard Yoshimine\cite{Yoshimine1973} scheme with $n^5$ scaling which reads for the Coulomb interaction in the X2C models as:
\begin{align}
\left( p \lambda_{\sigma_1} | \mu \nu \right) & = 
\sum_{ \kappa }^{n_{AO}}
C^*_{\kappa_{\sigma_1} p}
\left( \kappa_{\sigma_1} \lambda_{\sigma_1} | \mu \nu \right),\\
%\quad & 
\left( p q | \mu \nu \right) &= 
\sum_{ \lambda }^{n_{AO}} \sum_{\sigma_1}
C_{\lambda_{\sigma_1} q}
\left( p \lambda_{\sigma_1} | \mu \nu \right), \\
%\nonumber \\
\left( p q | r \nu_{\sigma_2} \right) &= 
\sum_{ \mu }^{n_{AO}}
C^*_{\mu_{\sigma_2} r}
\left( p q | \mu_{\sigma_2} \nu_{\sigma_2} \right),\\
%\quad &
\left( p q | r s \right) &= 
\sum_{ \nu }^{n_{AO}} \sum_{\sigma_2}
C_{\nu_{\sigma_2} s}
\left( p q | r \nu_{\sigma_2} \right) ,
\label{eq:ao2mo_scheme1}
\end{align}
where $p, q, r, s$ are the molecular spinors, $\kappa, \lambda, \mu, \nu$ are the spatial atomic orbitals, and $\sigma_1$ and $\sigma_2$ denotes the spin for electron $1$ and $2$, respectively. In this procedure we make use of the fact that the AO spinors are defined as simple products of spatial and spin functions, so that the spin
integration reduces to additional summations in the second and fourth step of the transformation. The antisymmetrization and reordering,  $ \left\langle p r || q s \right\rangle = \left( p q | r s \right)-\left( p s | r q \right)$, is done after the index transformation is completed. By making use of permutational symmetry, only 6 unique classes of molecular integrals are used, which can be generated on basis of three classes of half-transformed integrals (vv, vo and oo, where v and o stands for virtual and occupied spaces, respectively). 
The number of atomic orbitals ($n_{AO}$) can become quite large, which makes it impractical to work with the complete atomic 2-electron integral tensor $\left( \kappa \lambda | \mu \nu \right)$. The default in the current code is to slice the last index and work with a subspace thereof. This does not increase the operation count of the algorithm and is in practice sufficient to reduce the memory footprint due to the handling of the AO integral tensor. This choice has the benefit of keeping large spaces for the other indices making the
tensor contractions optimally efficient.
    
\subsection{Coupled cluster implementation}
\label{sec:CC_implem}

The ExaTENSOR library requires the definition of the spaces that span the dimension of the tensors as outlined in Figure~\ref{fig:exatensor_outline}.
Only two spaces are needed for a standard coupled cluster algorithm, the occupied and the virtual spaces. Prior to the starting of the library these spaces are created according to the lists of active occupied and virtual spinors specified in the input. Since the transformation from atomic orbitals to spinors is also performed by ExaCorr a third space spanning the atomic orbitals is required which is defined in a way that avoids splitting shells of basis functions (see supporting information). 
In addition, the ExaCorr specific methods are registered in the ExaTENSOR library interface. Apart from the already mentioned ERI generation by InteRest this e.g. comprises methods to initialize a tensor with MO coefficients, initialize a tensor with 1-electron integrals, scale a tensor with denominators (equations \ref{eq:4order-term} and \ref{eq:5order-term}), or project on a subspace. 

After reading and processing the input data, basis set information, spinor energies, MO coefficients and, optionally, the one-electron integrals are stored as global variables and broadcasted to all nodes using MPI. 
After this preliminary step the ExaTENSOR library is started and MO integrals are computed by transforming the ERIs to the MO basis. The MP2 amplitudes are subsequently computed in order to obtain an initial guess for the CC amplitudes. 
These CC  amplitudes are then refined in an iterative procedure, the working equations for which can be found in Appendix \ref{ap:CC} or Ref.~\citenum{Papadopoulos2019}. 
As convergence of the non-linear coupled cluster iterations can be slow, we have implemented the DIIS scheme\cite{Pulay1980} and are also assessing the less memory-demanding CROP algorithm\cite{Ziolkowski2008}.
In the current implementation all necessary tensors are created before the iterative procedure starts, allowing for \textit{a priori} assessment of the maximum memory footprint of the run.

Triple excitations require tensors of the size $n_{occ}^3n_{vir}^3$. For the full triples these amplitudes need to be determined iteratively, which requires a significant amount of memory and number of operations. For the perturbative treatment of the triple excitations considered here, memory requirements can be reduced by splitting the occupied space and using three nested loops over these subspaces (of dimension $n_{red}$) to evaluate all contributions. This results in a memory requirement of n$_{red}^3$n$_{vir}^3$ in addition to the memory required for the coupled cluster amplitudes, the 2-electron integrals and the Fock matrix. Permutational symmetry is used to speed up the computation by only computing unique blocks.  
Following this evaluation of the expressions in equations \ref{eq:W-term}, \ref{eq:Y-term},~and \ref{eq:Z-term} the triples corrections are obtained by symmetrization and denomination as expressed in the equations in appendix~\ref{ap:triples}.  

For the calculation of molecular properties, the equations for the Lagrange multipliers $\mathbf{\lambda}$ have to be solved
which can be done in much the same way as described above for the CC equations, including use of DIIS to reduce the number of iterations. The $\mathbf{\lambda}$ and amplitudes tensors are then combined according to the equations in \ref{ap:density} (see also Ref.~\citenum{shee2016analytic}) to obtain the one-particle density matrix. In the case of the TAL-SH implementation the tensor elements can be accessed directly and written to file. For ExaTENSOR a local copy is first created in the form of a TAL-SH tensor which is then written. The properties module in DIRAC can read these data and compute the properties.

\section{Computational details}

All of our coupled cluster calculations have  been performed with development versions of DIRAC
and ExaTENSOR, using either the X2Cmmf~\cite{Sikkema2009} or the X2C-1e~\cite{Ilias2007,Konecny2016} Hamiltonians. For the latter, spin-orbit interactions were included via atomic mean-field integrals calculated with the AMFI~\cite{prog:amfi} code (X2C-AMFI). Details on the particular DIRAC revisions used in the calculations described below are available in the respective output files provided as part of the supplemental information.
The geometries of the systems are also included in the supplemental information~\cite{pototschnig:2021:dataset}.

The reference orbitals and single determinant reference wave functions have mostly been obtained by the SCF implementation in DIRAC, which is a Kramers restricted implementation. 
In order to enforce Kramers symmetry for systems that have an odd number of electrons or have near-degeneracies at the Fermi level, an average-of-configuration approach (AOC) is used in DIRAC \cite{DIRAC19, Saue2020, Thyssen2001}.
In contrast, the ReSpect code performs Kramers unrestricted (KU) calculations in which the Kramers symmetry is not imposed~\cite{Repisky2020}. 
For the utilization of spinors generated by the AOC procedure in DIRAC, some additional features are needed for the interface, as the definition of the Fock matrix in DIRAC differs from the KU Fock matrix, the definition assumed in  ReSpect and ExaCorr. 
For closed-shell molecules, the difference between the AOC and KU Fock matrix expressions disappears and spinor energies can be read in from the DIRAC program and are sufficient to define the reference Hamiltonian. 
For open-shell molecules one may either employ the ReSpect code or another code that has a compatible KU Fock matrix definition or recompute the Fock matrix during the CC stage of the calculation. Both cases result in the use of a KU Hartree-Fock expression for a given reference determinant that is chosen by the user of the program. This is important for perturbation treatments, because in those orbital energy differences are used which depend on the definition of the Fock matrix.

Unless otherwise noted, we employed uncontracted Dyall basis sets of double (dyall.v2z), triple (dyall.v3z) or quadruple zeta (dyall.v4z) quality~\cite{Dyall2006,Dyall2007,Gomes2010}. The set of spinors included in the correlated calculations generally consists a subset of the total set of spinors. By default, these are selected by energy thresholds corresponding to relatively high-lying occupied and low-lying virtuals, with energies between -10~Hartree and 20~Hartree.

In the case of lanthanide monofluorides (LnF) and the uranium hexafluorides (UF$_6$) dimer, geometries were optimized at DFT level with the ADF~\cite{ADF2019} code, using the scalar relativistic ZORA Hamiltonian\cite{Lenthe1999}, the PBE functional\cite{Perdew1996} and triple zeta basis sets with one polarization function (TZP).

Further, molecule-specific, computational details are listed below.

\subsection{Lanthanide monofluorides}

For LaF and YbF the AOC-SCF approach was applied using DIRAC either employing the the X2C-AMFI or the 1-component non-relativistic (NR) Hamiltonian.
In the case of EuF KU calculations were performed with ReSpect~\cite{Repisky2020} using 1-component non-relativistic or X2C-1e Hamiltonian. Several thresholds for the occupied and virtual spinors were considered for the double zeta basis set and the same values were employed for larger basis sets.

\subsection{Argon binding to gold}

For argon atoms bound to gold clusters we employed the X2Cmmf Hamiltonian~\cite{Sikkema2009}, with the structures being taken from ref.~\citenum{Nazanin2021}. The default energy thresholds in the coupled cluster step have been employed. The number of correlated electrons for the systems considered are: 46 (AuAr$^+$), 60 (AuAr$_2^+$), 74 (AuAr$_3^+$), 88 (AuAr$_4^+$) and 102 (AuAr$_5^+$).

\subsection{Uranium hexafluoride dimer}

In the case of UF$_6$ and (UF$_6$)$_2$ calculations the X2Cmmf Hamiltonian was applied\cite{Sikkema2009}, except for some smaller scaling investigations for which we used X2C-AMFI.
In these smaller computations the cc-pVDZ and dyall.v2z basis sets were selected for F and U, respectively. The larger, more accurate computations employed the corresponding triple zeta basis sets.
The energy threshold for the included spinors were -35 and 80~Hartree for the smaller computations, -10 and 8~Hartree for the triple zeta ones. 

In the computations the distance between uranium and fluorine in a monomer were fixed to the experimental value of 1.996~\AA{}\cite{Kimura1968}.
A restricted optimization using these monomers were performed for (UF$_6$)$_2$ and the structures were applied in the coupled cluster computations. 
In this case we added the dispersion correction by Grimme to the PBE functional. 
Additionally, an optimization without restrictions was performed for the UF$_6$ dimer at the DFT level in order to estimate the U-F bond distances for this level of theory. 

\subsection{Uranyl tris-nitrate complex}

Calculations of the $zz$ component of electric field gradient (EFG) at the U nucleus (q$_{zz}$) for the uranyl tris-nitrate ([UO$_2$(NO$_3$)$_3$]$^-$) complex have been performed at the X-ray structure for the RbUO$_2$(NO$_3$)$_3$ crystyal~\cite{Barclay:a04757}, employing the X2C-AMFI and taking into account the picture change of the EFG operator. In addition to CC, we have performed DFT calculations with the B3LYP, PBE0 and CAM-B3LYP density functionals.

For the property CC calculations, we considered occupied spinors with energies higher than or equal to (a) -6~Hartree (106 electrons, in which the U 5d is correlated as done for other uranyl complexes~\cite{Gomes2013}); (b) -22~Hartree (156 electrons, in which the U 4f and all electrons for the light atoms are correlated); and (c) -4500~Hartree, which amounted to correlating all 202 electrons. These three occupied spaces are combined with virtual spinors with energies up to and including (a) 5~Hartree for both double and triple bases (543 and 649 virtuals, respectively); for only double zeta bases, (b) 20
~Hartree (680 virtuals); (c) 50~Hartree (818 virtuals); (d) 150~Hartree (896 virtuals); and for triple zeta bases (e) 7~Hartree (944 virtuals). The total number of virtual spinors are 1286 and 2076 for double and triple zeta bases. We have not performed CCSD calculations with quadruple zeta bases.

\section{Results and discussion}

Our first goal was to verify the correctness of the  new implementation.
In order to do so, we compared the results of the new implementation using TAL-SH or ExaTENSOR to the results obtained by the RELCCSD implementation in DIRAC\cite{visscher1996formulation, DIRAC19, Saue2020}. Comparisons of the energies for H$_2$O, LiO and CuAr$_n^+$ are in the supporting information. 
In order to check the property implementation we compared the dipole moment, EFG, and the nuclear quadrupole coupling constant (NQCC) of CHFClBr and UF$_6$ for different implementations which can also be found in the supporting information. 

A few system were selected to show the capabilities of the new implementation in the investigation of heavy element systems.
At first, we consider the ionization energies of three lanthanide monofluorides (LnF) species, LaF, YbF and EuF since, from a methodological perspective, these calculations allow us to demonstrate the usage of our implementation for both closed and open-shell configurations.
Secondly, the binding of argon atoms to gold cations have been studied including triples corrections, which are necessary to achieve chemical accuracy.  
In the subsequent section results for uranium hexafluoride and its dimer are presented as well as some information about the scaling of the new code.  
Finally, the electric field gradient of the uranyl tris-nitrate complex was computed as an example for evaluation of electronic properties in a larger molecule.

\subsection{Ionization energies of lanthanide monofluorides}
\label{sec:LnF_results}

Lanthanides are often treated using density functional theory, but results are shown to have a strong dependence on the exchange-correlational functional that is selected\cite{Aebersold2017}. Coupled cluster theory can provide more accurate and precise results and has been applied in conjunction with more approximate methods to account for relativistic effects, like the one-component Douglas-Kroll Hamiltonian\cite{Sekiya2012} and effective core potentials.\cite{Solomonik2017}. The currrent implementation, and in particular the interface for the ReSpect code, provides a way to investigate the (generally open-shell) ground states for such systems with full inclusion of relativistic and core correlation effects in coupled cluster theory.

Before proceeding to the discussion of our results for the ionization energies themselves, we shall discuss the requirements, in terms of the number of occupied and virtual spinor necessary for obtaining reliable results. For this, we have decided to consider two sets of equilibrium structures, one for the neutral and the other for the ionized species. Our structures, obtained at the DFT level using the PBE functional, are shown in table~\ref{tab:LnF_struct}, together with experimental values and prior theoretical values. For these systems DFT produces the experimentally observed trend,  with EuF having the largest bond distance and YbF the smallest. As expected there are some deviations as well, with the DFT bond distances being smaller than the experimental values for EuF and YbF, but slightly larger for LaF.

\begin{table*}[htbp]
\caption{
Experimental reference values and structures used in the computations for Lanthanide monofluorides and DFT bond distances applied in the computations. 
}
\label{tab:LnF_struct}
\begin{tabular}{c c c c c}
 & r$_e$ (exp.) & r$_e$ (PBE) & r$_e$ (Cation, PBE) & r$_e$ (ECP,CCSD(T))\cite{Solomonik2017} \\
 \hline
LaF	& 2.0234 \cite{Bernard2000a}  &  2.0293 & 2.0150 & 2.0215
\\
EuF & 2.083\cite{Dmitriev1987} & 2.0676 & 1.9992 & 2.0750
\\
YbF	& 2.016516\cite{Dickinson2001} &  1.9868 & 1.9345 & 2.0204
\\
\hline
\end{tabular}
\end{table*}

Now concerning the coupled cluster calculations themselves, we first investigated the convergence of the energies with the number of active occupied and virtual spinors. 
The reason for such an investigation is that employing the complete set of virtual spinors is typically not needed in relativistic calculations of heavy elements. 
This is due to the use of uncontracted basis sets, which leads to a significant number of virtual spinors being mostly located in the chemically inactive core region.
These types of spinors can be deleted without affecting the results much. In the current work we identify such spinors by a simple energy criterium, relying on the observation that the large kinetic energy of these solutions puts them in the upper range of energies obtained by Fock matrix diagonalization. More advanced schemes, such as use of approximate natural orbitals are also possible and under development. 
Regarding the choice of occupued spinors to be included, one needs to take into account that for lanthanides, the closeness (in radial extent) of the open-shell 4f and other electrons which would otherwise be considered as core (4s-4d), may require that they are correlated alongside the (5s, 5p) valence. 

We present in table~\ref{tab:YbF_spin} the results of such an investigation for the YbF, which had previously been investigated by some of us~\cite{Gomes2010} and that was found to be particularly sensitive to the electron correlation treatment. We provide equivalent tables for LaF and EuF as supplemental material, due to the fact that these exhibit the same trends as discussed below. % in tables~\ref{tab:LaF_spin}, ~\ref{tab:EuF_spin}.

From table~\ref{tab:YbF_spin}, we can identify two main trends: (a) employing a too small virtual space (comprising around 21\% of the total number of virtuals), even with a fairly large number of occupied, yields a (strong) underestimation of the ionization energy at CCSD level (-1.41 eV). A modest increase in the number of virtuals (including around 30\% of the virtuals) greatly reduces this underestimation and brings values closer to the experimental value. Further increases in the number of virtuals past 60\%, yields no significant difference in the CCSD ionization energies; (b) employing a converged virtual space ($>$30\%) but not enough occupied spinors overestimates the ionization energies, though not by much (around +0.05 eV). Possible choices for the occupied space are to correlate only the F(2s$^2$2p$^6$) and Yb(5s$^2$5p$^6$4f$^{14}$) (comprising use 40\% of the occupied space) or to include the  Yb(4s$^2$4p$^6$4d$^{10}$) shells as well (63\% of the occupied space).

For reliable results we find that the following electrons need to be correlated: La(4s$^2$4p$^6$5s$^2$4d$^{10}$5p$^6$),\\ Eu(3s$^2$3p$^6$4s$^2$3d$^{10}$4p$^6$4d$^{10}$5s$^2$5p$^6$6s$^1$4f$^{7}$), and Yb(4s$^2$3d$^{10}$4p$^6$5s$^2$4d$^{10}$5p$^6$6s$^1$4f$^{14}$), which can be achieved by employing energy thresholds of and, -20, -200, -60 Hartree for LaF, EuF and YbF, respectively. The 2s$^2$ and 2p$^6$ of F are always included, the 1s$^2$ is omitted for LaF. 
In the case of Yb, the neutral molecule is an open-shell system, while the cation only has closed shells. The opposite is true for La. EuF was considered as an example with a high spin state, the neutral molecule as well as the cation have several open shells.

\begin{table*}[hbtp]
\caption{
Ionization potential in eV of YbF for different numbers of correlated spinors employing the dyall.v2z basis set. 
The number of occupied and virtual spinors refers to the neutral molecule, for the cation one of these occupied spinors becomes a virtual. 
The $\Delta$SCF ionization potential computed using the reference determinant wave functions was 5.48 eV. 
The spinor thresholds are listed in atomic units. 
}
\label{tab:YbF_spin}
\begin{tabular}{c c c c c c c c}
threshold$_{low}$ & threshold$_{high}$ & nocc & nvir & \% occ & \% vir & CCSD 
\\
\hline
%
% increasing number of occupied in correlation, increses IE
% for given occupied, increasing virtuals increases but stabilises around a value
-20 &   2.3 & 49 &  89 &  63 & 21 & 4.49 \\ % 4.489 \\
-20 &     6 & 49 & 137 &  63 & 32 & 5.89 \\ % 5.893 \\
-20 &   150 & 49 & 267 &  63 & 63 & 5.89 \\ % 5.892 \\
-20 & 10000 & 49 & 367 &  63 & 86 & 5.89 \\ % 5.893 \\
%
%
% when adding more occupied, and then increasing virtuals, decreases IE
%
-60  &  10 & 61 & 155 &  78 & 36 & 5.90 \\ % 5.903 \\
-60  &  20 & 61 & 195 &  78 & 50 & 5.90 \\ % 5.903 \\
-60  & 150 & 61 & 267 &  78 & 63 & 5.90 \\ % 5.900 \\
%
% keeping same virtual, increasing occupied, decrease IE
%
-3   & 40 & 31 & 213 &  40 & 50 & 5.95 \\ % 5.948 \\
-20  & 40 & 49 & 213 &  63 & 50 & 5.90 \\ % 5.895 \\
-40  & 40 & 51 & 213 &  65 & 50 & 5.90 \\ % 5.896 \\
-60  & 40 & 61 & 213 &  78 & 50 & 5.90 \\ % 5.900 \\
-400 & 40 & 77 & 213 &  99 & 50 & 5.90 \\ % 5.899 \\
exp  &    &    &     &   &    & 5.91$\pm 0.05$\cite{Kaledin1999} \\
\hline
\end{tabular}
\end{table*}

Following our analysis of what are the minimum requirements in terms of occupied and virtual spinors for obtaining converged ionization energies, we investigate the adiabatic and vertical ionization potential for the three molecules,  computed for basis set of increased quality, for two classes of Hamiltonians (non-relativistic and X2C). The values are listed in the table~\ref{tab:LnF_basis}.

\begin{table*}[hbtp]
\caption{
Vertical (a) and adiabatic (b) ionization energies (in eV) for the lanthanide monofluorides.
The active ranges for LaF, EuF and YbF are -20 to 40, -200 to 200, and -60 to 40 Hartree, respecitvely. 
The X2C Hamiltonians used were the X2C-AMFI for LaF and YbF (spinors obtained with the DIRAC code) and X2C-1e for EuF (spinors obtained with the ReSpect code).
The complete basis set limit values ($\infty$z) have been obtained with a 2-point extrapolation formula, based on the 3z and 4z values.
}
\label{tab:LnF_basis}
\begin{tabular}{ccc cc cc c cc cc cc}
\hline
\hline
&&& \multicolumn{2}{c}{$\Delta$SCF} & \multicolumn{3}{c}{CCSD} & \multicolumn{2}{c}{CCSD+T} & \multicolumn{2}{c}{CCSD(T)} & \multicolumn{2}{c}{CCSD-T} \\
\cline{4-5}
\cline{6-8}
\cline{9-10}
\cline{11-12}
\cline{13-14}
&& basis & (a) & (b) & (a) & (b) & T1 & (a) & (b) & (a) & (b) & (a) & (b) \\
\hline
LaF & NR  &  2z & 3.41 & 3.44 & 5.03 & 5.06 & 0.01 & 5.15 & 5.18 & 5.11 & 5.14 & 5.10 & 5.13 \\
    &     &  3z & 3.43 & 3.44 & 4.71 & 4.71 & 0.02 & 4.90 & 4.91 & 4.84 & 4.84 & 4.82 & 4.74 \\
    &     &  4z & 3.43 & 3.43 & 4.61 & 4.60 & 0.02 & 4.81 & 4.81 & 4.74 & 4.74 & 4.72 & 4.72 \\
    && $\infty$z& 3.42 & 3.43 & 4.53 & 4.53 &      & 4.74 & 4.74 & 4.66 & 4.66 & 4.65 & 4.64 \\
&&&&&&&&&&&&&\\
    & X2C &  2z & 4.93 & 4.96 & 5.91 & 5.93 & 0.01 & 5.96 & 5.99 & 5.96 & 5.98 & 5.96 & 5.98  \\   
    &     &  3z & 4.87 & 4.87 & 5.87 & 5.87 & 0.01 & 5.98 & 5.98 & 5.96 & 5.96 & 5.96 & 5.95  \\
    &     &  4z & 4.86 & 4.86 & 5.87 & 5.86 & 0.01 & 5.99 & 5.99 & 5.97 & 5.97 & 5.97 & 5.96  \\
    && $\infty$z& 4.85 & 4.84 & 5.87 & 5.86 &      & 6.00 & 5.99 & 5.98 & 5.97 & 5.97 & 5.96 \\
&&&&&&&&&&&&&\\
    & \multicolumn{2}{c}{exp.}&  \multicolumn{11}{c}{6.3$\pm 0.3$\cite{ZMBOV1968a}}\\
&&&&&&&&&&&&&\\
EuF & NR  &  2z & 4.73 & 4.75 & 5.08 & 5.12 & 0.01 & 5.10 & 5.15 & 5.10 & 5.15 & 5.11 & 5.15 \\
    &     &  3z & 4.72 & 4.74 & 5.10 & 5.13 & 0.01 & 5.13 & 5.17 & 5.13 & 5.17 & 5.13 & 5.17 \\
%   &     &  4z &  &  &  &  &  &  &  &  &  &  &  \\
%   && $\infty$z&  &  &  &  &  &  &  &  &  &  &  \\
&&&&&&&&&&&&&\\
    & X2C &  2z & 5.04 & 5.07 & 5.46 & 5.51 & 0.01 & 5.51 & 5.57 & 5.51 & 5.56 & 5.51 & 5.56 \\
    &     &  3z & 5.02 & 5.05 & 5.48 & 5.52 & 0.01 & 5.54 & 5.58 & 5.53 & 5.57 & 5.53 & 5.57 \\
%   &     &  4z &  &  &  &  &  &  &  &  &  &  &  \\
%   && $\infty$z&  &  &  &  &  &  &  &  &  &  &  \\
&&&&&&&&&&&&&\\
    & \multicolumn{2}{c}{exp.}&  \multicolumn{11}{c}{5.9$\pm 0.3$\cite{ZMBOV1967}}\\
&&&&&&&&&&&&&\\
YbF & NR  &  2z & 5.04 & 5.09 & 5.39 & 5.39 & 0.01 & 5.43 & 5.43 & 5.42 & 5.43 & 5.42 & 5.43 \\
    &     &  3z & 4.93 & 4.98 & 5.40 & 5.43 & 0.01 & 5.47 & 5.49 & 5.45 & 5.47 & 5.46 & 5.48 \\
    &     &  4z & 4.93 & 4.98 & 5.42 & 5.44 & 0.01 & 5.59 & 5.60 & 5.56 & 5.57 & 5.57 & 5.58 \\
    && $\infty$z& 4.93 & 4.98 & 5.44 & 5.45 &      & 5.67 & 5.68 & 5.64 & 5.65 & 5.65 & 5.66 \\
&&&&&&&&&&&&&\\
    & X2C &  2z & 5.48 & 5.49 & 5.90 & 5.87 & 0.05 & 6.51 & 6.44 & 5.57 & 5.56 & 5.67 & 5.65 \\
    &     &  3z & 5.44 & 5.46 & 6.00 & 5.98 & 0.09 & 7.76 & 7.73 & 5.30 & 5.29 & 5.76 & 5.75 \\
    &     &  4z & 5.44 & 5.46 & 6.00 & 5.97 & 0.05 & 6.82 & 6.78 & 5.57 & 5.56 & 5.75 & 5.74 \\
    && $\infty$z& 5.44 & 5.46 & 6.00 & 5.97 &      & 6.13 & 6.08 & 5.77 & 5.75 & 5.75 & 5.73 \\
 &&&&&&&&&&&&&\\
   & \multicolumn{2}{c}{exp.}&  \multicolumn{11}{c}{5.91$\pm 0.05$\cite{Kaledin1999}}\\
\hline
\hline
\end{tabular}
\end{table*}

The largest change of the ionization potential is due to the inclusion of relativistic contributions. 
Regarding LaF, the ionization potential is larger by about 1.3 eV (CCSD) or 1.4 eV (SCF) for the X2C-1e Hamiltonian than for the non-relativistic one. 
For EuF/YbF the changes are somewhat smaller, increases of about 0.4/0.5 eV and 0.3/0.4 eV were obtained for coupled cluster and $\Delta$SCF, respectively. 

The inclusion of electron correlation by CCSD results in an increase of the ionization potential by about 1, 0.4, and 0.5~eV for LaF, EuF, and YbF, respectively. These are the values for the X2C-1e Hamiltonian, the changes in the non-relativistic case are similar. 
The perturbative triples corrections are the smallest for EuF, probably due to the 
relatively simple high spin ground states of both the neutral and the cation, that can be well-described by the Kramers unrestricted reference wave functions obtained with ReSpect. They increase the ionization potential by less then 0.06~eV. The triples add about 0.1 eV in the case of LaF for the X2C-1e and slightly more for the non-relativistic Hamiltonian. 
While the triples in the non-relativistic case are similar (between 0.1 and 0.2 eV) for YbF, the values for X2C are much larger.
The fourth order correction increases the IP by about 1.7 eV, the fifth order correction results in values about 0.7 eV below the CCSD ones. A similar observation has been reported in Ref.~\citenum{Gomes2010}. These large values are an indication that a perturbative inclusion of triples is insufficient in agreement with the large $T_1$ values, see table~\ref{tab:LnF_basis}. 
This is probably caused by a mixing of excited states with closed and open f-shells which was observed to cause a large change in ground state polarizabilty of the Yb atom\cite{Dzuba2010} and a change in the nuclear quadrupole coupling constant in Ref.~\citenum{Pasteka2016}.

For an increase of the basis set size the ionization potential of the reference determinant becomes smaller except for the vertical transition of LaF using the non-relativistic Hamiltonian. 
Going from double to triple zeta basis sets the coupled cluster ionization potential increases for YbF and EuF, although the changes are below 0.03 eV in the latter case. Regarding LaF, a decrease of the ionization potential is observed on the coupled cluster level. 

The adiabatic IP should be smaller than the vertical one, if the equilibrium distances are correct. For the HF reference this is never the case as the electron correlation is missing in contrast to the DFT that was used to determine the bond distances.  
In the case of coupled cluster the correct order of the vertical and adiabatic IP is obtained for the X2C-1e Hamiltonian and triple or quadrupole zeta basis set, indicating that the DFT bond distances are close. EuF always shows the wrong order, probably because the bond distances are not accurate enough, which is probably also reflected in the larger differences between the theoretical and experimental values in table~\ref{tab:LnF_struct}. 

 The best estimates from table~\ref{tab:LnF_basis} are 5.97, 5.57, and 5.97~eV for LaF, EuF and YbF, respectively, while values of 6.3, 5.9, and 5.91~eV were determined in experiments (table~\ref{tab:LnF_struct}).  One of the reasons for this discrepancies are the large experimental uncertainty (especially for LaF and EuF); while also the neglect of zero  point energies in our values will play a role. Considering these sources of errors, the energies show an acceptable agreement. 

Systems with open shells, like treated above, can be difficult to describe using coupled cluster, since CC is based on a single reference determinant. The t1-amplitudes recover a portion of the static correlation, which makes a treatment possible if there is one dominant determinant, but in cases with several important configurations multireference methods are necessary\cite{Lyakh2012}. A related difficulty appearing in a 2-component treatment is that the spinors are no longer eigenfunctions of $\hat{S_Z}$. On the SCF level this can be handled by using an average-of-configuration approach\cite{Thyssen2001} which occupies all relevant configurations, resulting in spinors with varying spin up and spin down contributions. Making a proper selection
of such spinors to form a single determinant reference wave is, however, difficult
in the general case. The exception is cases with only a single unpaired electron in which either spinor of the singly occupied Kramers pair can be taken to construct
the reference determinant.
In the current work molecules were selected that are still rather easy to treat. 
YbF and LaF have only one unpaired electron and thus belong to this important special case of simple open shell molecules. For neutral and positively charged EuF, one determinant can qualitatively  correctly describe the ground states if we allow for an KU spinor optimization that is able to converge to a "high-spin" state. This is possible with Respect.

While the use of Kramers restriction in an averaged SCF is feasible for simple open shell systems, it does lead to an inconsistency in the definition of the reference Fock operator and orbital energies between the KR SCF program and the KU CC implementation. This is formally not a problem as our working equations do not require use of a diagonal Fock operator, but makes working with denominators consisting of spinor energy differences between occupied and virtual spinors more complicated. With unmodified orbital energies the energy difference between the highest occupied spinor (one of the two open shell spinors) and the lowest unoccupied spinor (its Kramers partner) would be zero. There are several ways to deal with this complication. One is to recompute the spinor energies according to the KU Fock matrix expression. This will induce an energy gap and make it possible to apply denominators. This simple approach was applied for the results for LaF and YbF presented in table~\ref{tab:LnF_basis}.

\subsection{ Binding energies of argon atoms to a gold atom }

Gold is one of the most nonreactive metals in the periodic table and noble gases are also exceptionally inert.
Nevertheless, the AuNe$^+$ dimer was reported in 1977\cite{Kapur1977} and early computations suggested a covalent bond between gold cations and noble gas atoms\cite{Pyykkoe1995}, which is supported by recent experimental results\cite{Shayeghi2015}.
A theoretical study observed a strengthening of the binding in small gold clusters if noble gas atoms are attached\cite{Ghiringhelli2015}.
This covalency is in part attributed to the relativistic nature of the heavy Au, this makes it necessary to include these contributions in theoretical studies.
Recently, a significant influence of argon atoms on the IR spectra and bonding of small gold clusters was observed\cite{Ferrari2020}. 
Here, we want to compute the interaction of a single gold atom with argon using the reliable coupled cluster method in combination with the X2C-mmf Hamiltonian. 
A summary of the current state of research on noble gas-noble metal compounds can be found in a recent review\cite{Pan2019}.

Firstly, the energy of the AuAr dimer was computed for different internuclear distances.
The equilibrium bond distances for the AuAr dimer were determined by fitting a Morse potential to about 5 points. For the dyall.v4z basis set an equilibrium bond distances of 2.50, and 2.47~\AA{} were obtained by CCSD and CCSD(T), respectively. 
The MP2 value is about 0.1~\AA{} smaller than the CCSD one, and the HF value about 0.3~\AA{} larger. The triples correction reduces the equilibrium distance by about 0.03~\AA{}, a detailed table is in the supplementary information. 
This general trend is observed for all basis sets, while the bond distance is about 0.05 smaller for 3z than for 2z. 
The structures of the larger systems were obtained by density functional theory.\cite{Nazanin2021}
Coupled cluster binding energies are listed in table~\ref{tab:AuAr_n_234z}. 
\begin{table*}[hbtp]
\caption{
Total binding energy ($\Delta E$, in eV) of AuAr$_n$ systems with dyall basis sets of different cardinal number. In all cases, spinors with energies between -10 and 20 Hartree have been included in the correlation treatment. The number of virtual spinors in each case (V) is shown, see computational details for the number of correlated electrons for each species. The complete basis set limit values ($\infty$z) have been obtained with a 2-point extrapolation formula, based on the 3z and 4z values.}
\label{tab:AuAr_n_234z}
\begin{tabular}{c  rr  rrrrrr}
\hline
\hline
   &&\multicolumn{6}{c}{$\Delta E$ }  \\
\cline{4-9}
 system & basis	 & 	V	&  HF  &  MP2   &   CCSD & CCSD+T & CCSD(T) & CCSD-T    \\
\hline
AuAr$^+$       & 2z  & 136 & -0.1341 & -0.5260 & -0.4620 & -0.5267 & -0.5165 & -0.5156 \\
               & 3z  & 230 & -0.1401 & -0.5993 & -0.4699 & -0.5519 & -0.5408 & -0.5400 \\
               & 4z  & 400 & -0.1456 & -0.6408 & -0.4817 & -0.5689 & -0.5586 & -0.5581 \\
         &$\infty$z  &     & -0.1496 & -0.6711 & -0.4904 & -0.5814 & -0.5716 & -0.5713 \\
&&&&&&\\
AuAr$_2^+$     & 2z  & 168 &  0.0845 & -1.2292 & -1.0356 & -1.1981 & -1.1723 & -1.1706 \\
               & 3z  & 288 & -0.0219 & -1.4124 & -1.0890 & -1.2813 & -1.2563 & -1.2550 \\
               & 4z  & 508 & -0.0399 & -1.4938 & -1.1161 &         &         &         \\
         &$\infty$z  &     & -0.0531 & -1.5532 & -1.1360 &         &         &         \\
&&&&&&\\
AuAr$_3^+$     & 2z  & 200 & -0.0140 & -1.3416 & -1.1491 & -1.3100 & -1.2875 & -1.2860 \\
               & 3z  & 346 & -0.0816 & -1.5202 & -1.1899 & -1.3902 & -1.3669 & -1.3652 \\
               & 4z  & 616 & -0.1019 & -1.6124 & -1.2268 &         &         &         \\
         &$\infty$z  &     & -0.1167 & -1.6797 & -1.2538 &         &         &         \\
&&&&&&\\
AuAr$_4^+$(3D) & 2z  & 232 & -0.1061 & -1.4673 & -1.2681 & -1.4267 & -1.4090 & -1.4078 \\
               & 3z  & 404 & -0.1367 & -1.6474 & -1.3009 & -1.5092 & -1.4890 & -1.4872 \\
               & 4z  & 724 & -0.1590 & -1.7527 & -1.3491 &         &         &         \\
         &$\infty$z  &     & -0.1753 & -1.8295 & -1.3843 &         &         &         \\
   &&&&&&\\
AuAr$_4^+$(2D) & 2z  & 232 & -0.1605 & -1.4326 & -1.2530 & -1.4024 & -1.3863 & -1.3852 \\
               & 3z  & 404 & -0.1783 & -1.5967 & -1.2769 & -1.4750 & -1.4563 & -1.4545 \\
               & 4z  & 724 & -0.2001 & -1.6978 & -1.3243 &         &         &         \\
         &$\infty$z  &     & -0.2160 & -1.7715 & -1.3588 &         &         &         \\
&&&&&&\\
AuAr$_5^+$     & 2z  & 264 & -0.2217 & -1.5945 & -1.3898 & -1.5398 & -1.5301 & -1.5750 \\
               & 3z  & 462 & -0.2059 & -1.7730 & -1.4126 & -1.6238 & -1.6100 & -1.6081 \\
\hline
\hline
\end{tabular}
\end{table*}
There is strong dependence on the method, Hartree-Fock underestimates the CCSD binding energies by 0.4 to 1.5~eV, MP2 overestimates them by 0.04 to 0.4 eV. 
The triples correction are also significant, they increases the binding energy by 0.16/0.20~eV for the fourth-order +T and 0.14/0.18~ for the fifth-order (T)/-T considering the dyall.v2z/dyall.v3z basis set, excluding the AuAr dimer with smaller triples contributions.
The dimer constitutes a special case as the structures were optimized at the coupled cluster level. For this reason significant HF binding energies were obtained as they are computed for larger bond distances as the CC ones.  
If the basis set is increased or extrapolated the CCSD energy increases by about 0.04 eV, except for the dimer with smaller changes. 
The growth of the CCSD+T/(T)/-T energy is about 0.08~eV  in going from the double to triple zeta basis set, excluding the AuAr dimer.
The energy per argon atom reaches its maximum for AuAr$_2^+$ with about 0.78~eV a the CBS CCSD level of theory. 
For AuAr$_4^+$ two structures have been computed to assess the relative stability of a  planar and a 3D arrangement. Independent of the basis set and method the 3 dimensional structures are found to be lower in energy. 

These preliminary findings will be incorporated in a larger investigation of Ar bound to a gold clusters in conjunction with infrared multiphoton dissociation experiments.\cite{Jamshidi.Maghari.20128qs, 10.1021/acs.jpca.0c07771}. For such investigations it is essential to be able to have reliable benchmarks of DFT calculations which will become possible with this new implementation,

\subsection{Binding energy of uranium hexafluoride dimers}

Uranium hexafluoride is used in gaseous form in enrichment methods for nuclear fuels. To simulate the behavior of this gas under different conditions, an accurate description of the intermolecular interaction potential is important. Early attempts to describe the interaction of molecules were based on potentials derived from thermophysical data and spectroscopy\cite{Beu1997a, Beu1997}. 
In quantum chemistry, the properties\cite{Malli1996, Kovacs2004, Batista2004, Manzhos2015} and reaction pathways\cite{Peluzo2018} of the monomer have been mainly studied employing relativistic DFT.
In order to describe the interaction of two such units it is important to account for relativistic\cite{Malli1996} as well as dispersion effects accurately. As the electronic structure of the dimer is not problematic and well-described by a single reference determinant,  coupled cluster theory can be used to provide accurate reference data. 
Since the computations are rather expensive due to the number of electrons that needs to be correlated, this particular system is well-suited for testing our implementation.

First, we performed computations for the UF$_6$ monomer on different numbers of nodes. Table~\ref{tab:scal_uf6_ccsd} and~\ref{tab:scal_uf6_v3z} display the obtained timings for the double and triple zeta basis set, respectively.
\begin{table*}[hbtp]
\caption{Time in seconds for integral transformation (t$_I$) and solving the coupled cluster (t$_{CCD}$, t$_{CCSD}$) and  $\Lambda$  equations (t$_\Lambda$) for UF$_6$ using the dyall.v2z basis set. The CCD, CCSD and $\Lambda$  equations, took resp. 10, 20 and 21 iterations to solve.  For the selected thresholds of -35 to 80 Hartree  110 occupied and 474 virtual spinors are included. n is the number of Summit nodes, the number in brackets in the header gives the ExaTensor blocksize.}
\label{tab:scal_uf6_ccsd}
\begin{tabular}{ r | r | r | r | r | r | r | r | r | r | r | r}
n& t$_I$ (75) & t$_I$ (50) &
t$_{CCD}$ (75) & t$_{CCD}$ (50) & t$_{CCSD}$  (75) &t$_\Lambda$ (75) 
\\
\hline
16  & 841 & 1179 &  843 & 1314 & 2046 &-
\\
24  & 654 &  902 &  711 & 1045 &  1869 & 1865 
\\
32  & 505 &  764 &  617 &  925 &  1645 & 1649
\\
48  & 413 &  686 &  594 &  861 &  1512 & 1493 
\\
64  & 375 &  634 &  600 &  810 &  1545 & 1412
\\
\end{tabular}
\end{table*}
\begin{table}[hbtp]
\caption{Time in seconds for integral transformation (t$_I$) and coupled cluster (t$_{CCSD}$) for UF$_6$ applying the dyall.v3z basis set. The CCSD equations took 21 iterations to solve. For thresholds of -10 to 8 Hartree  70 occupied and 554 virtual spinors are active. n is the number of Summit nodes, the number in brackets in the header is the ExaTensor blocksize.}
\label{tab:scal_uf6_v3z}
\begin{tabular}{  c | c | c | c | c | c | c | c }
 n &t$_I$ (75) & t$_{CCSD}$ (75) \\
 \hline
       32 &     1685 &       1279  \\
       64 &     1191 &       1205  \\
       96 &      817 &       1081  \\
      128 &      687 &       973  \\
\end{tabular}
\end{table}
As evident from the tables, our code scales up to 48 nodes for such small systems before the communication overhead and load imbalance prevent further improvement in time to solution. Load balancing is in principle better with a finer granularity of tensor blocks (which can be achieved by decreasing the dimension segment size from 75 to 50), but the increased inter- and intra-node communication overheads then lead to an overall increase of the computational time as can be seen in Table \ref{tab:scal_uf6_ccsd}. The better performance and scaling of large tensor contractions enhances the difference between the CCD and CCSD formalisms. While the additional tensor contractions related to the inclusion of single excitations are at most of order $n^5$, CCSD iterations take noticeably more time than the CCD ones as these additional contractions are computationally less efficient. The lambda equations iterations are slightly faster than the CCSD ones, but otherwise behave similarly in terms of scaling. As the size of the AO basis is much larger than that of the MO basis, in particular the first stages of the integral transformation can make up a large portion of the computational time. This is more important for larger AO sets. In Table \ref{tab:scal_uf6_v3z}, one may notice that for the smallest node count, this step even dominates the calculation.
Therefore, index transformations require special attention and will be the first target for improvements using techniques like Cholesky decomposition that allow for reduction of operation counts without impacting the accuracy.

\begin{figure*}[htb]
\centering
\begin{tabular}{ >{\centering\arraybackslash}m{1cm}  >{\centering\arraybackslash}m{0.45\textwidth}   }
D2d & \includegraphics[width=0.4\textwidth]{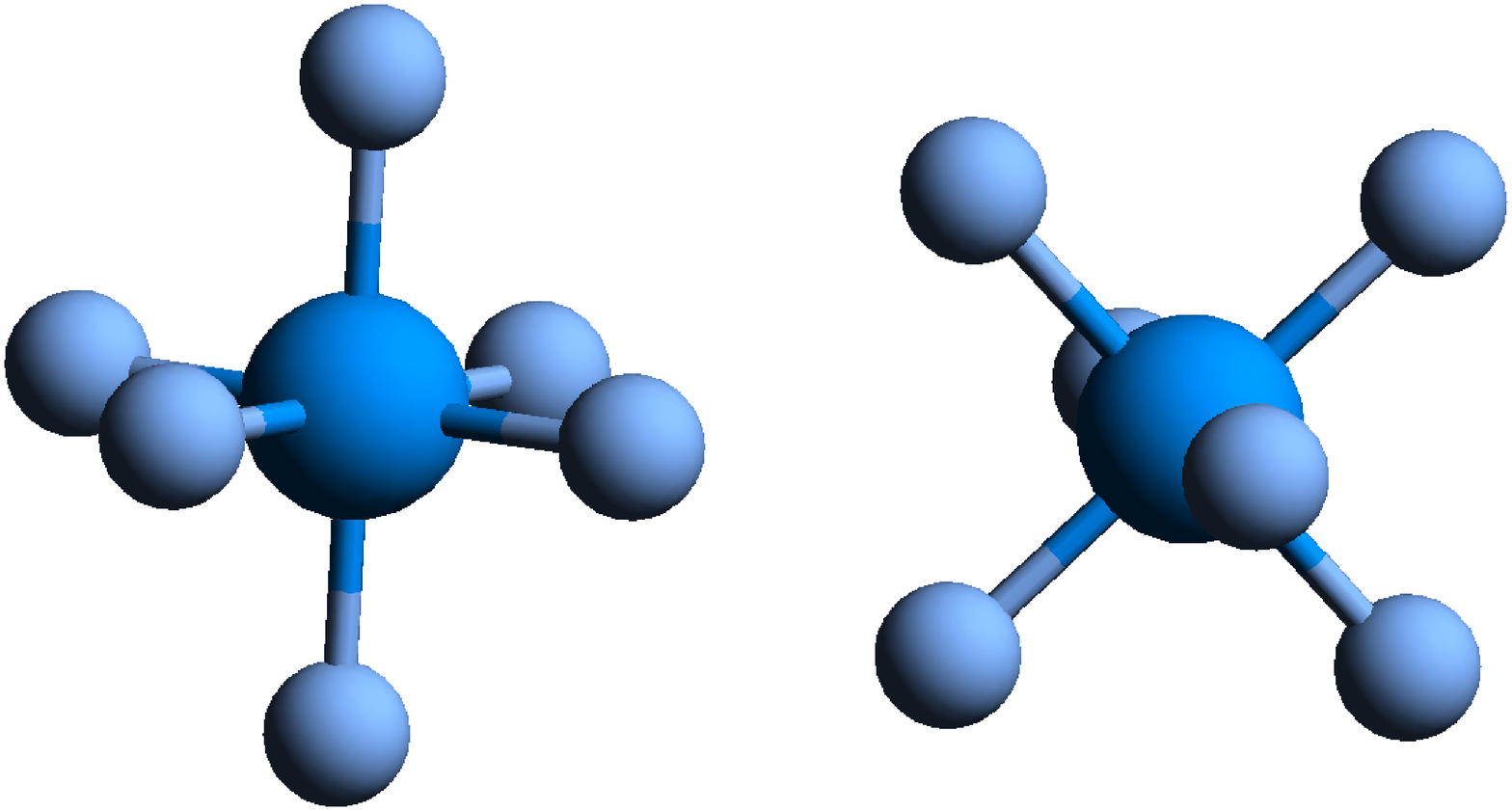}
\\
D3d & \includegraphics[width=0.4\textwidth]{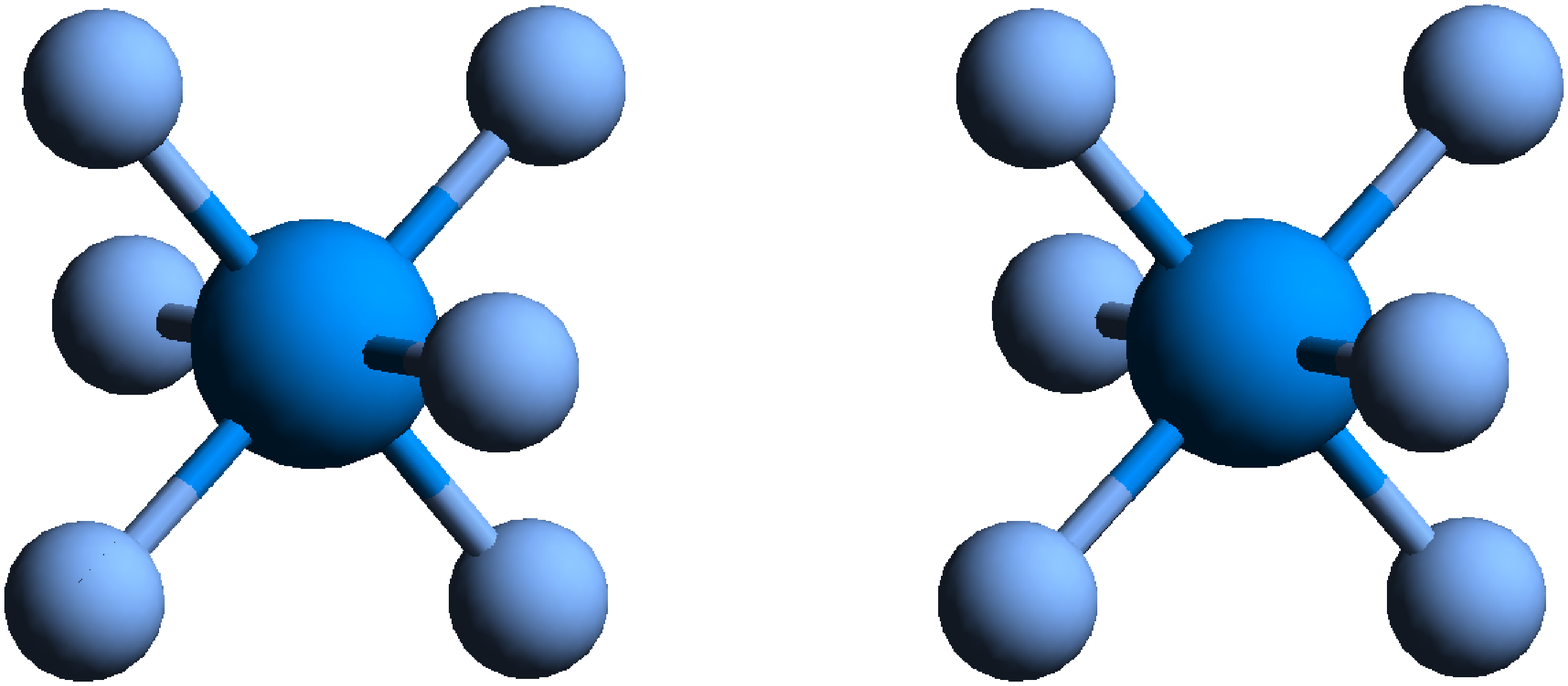}
\\
C2h & \includegraphics[width=0.4\textwidth]{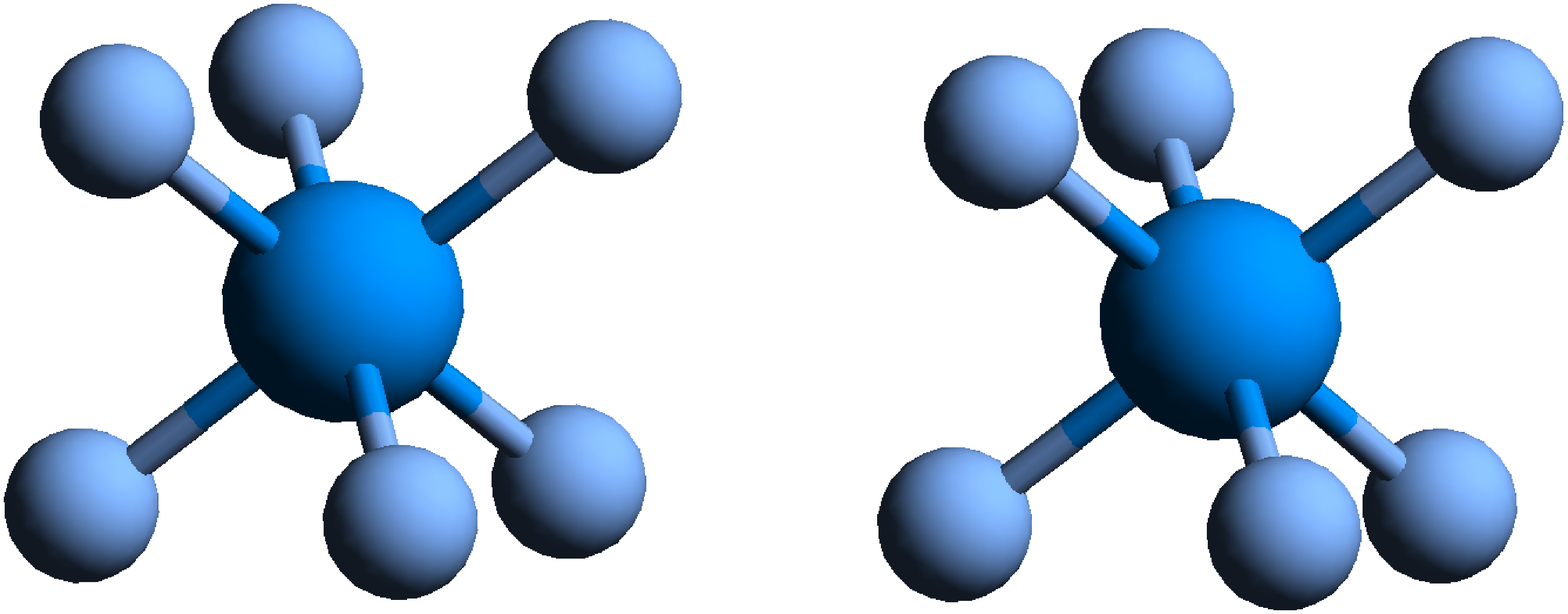}
\end{tabular}
\caption{Orientation of the UF$_6$ dimers}
\label{fig:uf6_dimer}
\end{figure*}

Currently, the scalability of our GPU-accelerated implementation is hindered by (a) the large granularity of tensor block storage, (b) the necessity of performing remote MPI\_Accumulate operations for the tensor contractions which have a relatively small output tensor. The former is caused by the necessity of processing large tensor blocks on GPU in order to amortize the cost of the Host-to-Device memory transfers. The latter is because the work distribution is trying to exploit the locality with respect to the output tensor whereas the dynamic load balancer is trying to spread the work over all MPI processes, which requires execution of a remote update via MPI\_Accumulate. Unfortunately, the implementation of the MPI\_Accumulate operation in existing MPI libraries is not efficient due to excessive synchronization, single-threaded accumulation, and lack of hardware support. 
The granularity of tensor storage and work is controlled by the segment size used for splitting tensor dimensions. In most of the presented computations the segment size of 75 was used, resulting in tensor blocks with $75^4\approx 32 \cdot 10^{6} $ elements for the large 4\textsuperscript{th} order tensors. This block size presents a reasonable granularity for processing tensor contractions on the NVIDIA Volta GPU. However, such a relatively large block size limits the scalability with respect to the number of nodes used, that might not be appropriate for CPU-only machines where a smaller segment size may result in better efficiency due to larger task pool, better load balancing and faster remote uploads.

To further assess performance on larger node counts, a larger system with a higher operation count is necessary. We therefore also investigated the (UF$_6$)$_2$ dimer to have a case with approximately 64 times more floating point operations to process. 
This system has not yet been treated at the coupled cluster level of theory, but a dimer interaction potential was computed with DFT\cite{Gagliardi1998}, including relativistic effects via a relativistic effective core potential.
Concerning the relative position and alignment of the two UF$_6$ monomers, there are three minima which are depicted in figure \ref{fig:uf6_dimer}. 
They are designated by the symmetry of the complex. The energies and U-U bond distances are listed in table~\ref{tab:uf6_dimer_struct}.
\begin{table*}[hbtp]
\small
\caption{UF$_6$ dimer bonding energies(dE, eV) for a system with 140 occupied and 1108 virtual spinors for different computational methods. The U-F distances have been fixed at 1.996~\AA{}\cite{Kimura1968}, the U-U distances were optimized with DFT and are listed in the first column in \AA{}. n is the number of Summit nodes.}
\label{tab:uf6_dimer_struct}
\begin{tabular}{ c | c | c | c | c | c | c | c | c }
 sym & U-U & \multicolumn{4}{c |}{ dE } & n & t$_I$ & t$_{CC}$/it.
\\
\hline
 &  & DFT & HF & MP2 & CCSD &  & & 
 \\ 
 \hline
D3d &  5.144  & -0.136 & 0.037 & -0.154 & -0.131 & 385 & 9244 & 687
\\
D2d &  5.139 &  -0.160 & -0.003 & -0.189 & -0.178 & 513 & 8641 & 758
\\
C2h & 5.290  & -0.150 & -0.001 & -0.163 & -0.151 & 1025 & 8356 & 665
\\
\end{tabular}
\end{table*}
For the D2d complex, the smallest U-U distance and the highest binding energy were obtained. 
The C2h complex has the largest separation of the uranium atoms in the equilibrium and for D3d the smallest binding energies were obtained. 
The trend of the energies is the same for the different levels of theory, but the absolute values vary strongly. 
In the case of Hartree-Fock, the complexes are barely bound, while MP2 overestimates the binding energy and the CCSD and DFT values are rather close ranging from 0.13 to 0.19~eV.

In the v3z computation of the dimer, 140 active occupied and 1108 active virtual spinors are taken into account in the coupled cluster computation. With a chosen segment size of 70 for both, the t$_2$ tensor consists of 1024 blocks. This means that we reach the scaling limit at 1024 MPI processes or, equivalently, 512 Summit nodes. If one compares the timings for runs with 385 and 513 nodes in table~\ref{tab:uf6_dimer_struct}, one can see a speed up for the integral transformation, but not for the coupled cluster iterations. The integral tensors participating in the time dominating tensor contractions in the integral transformation include atomic and virtual spaces, thus containing more tensor blocks to parallelize over, whereas in the coupled cluster iterations most tensors are smaller than t$_2$. This is the reason why the index transformation still shows some speed-up with increased node counts whereas this is not the case for the CCSD step. Due to the necessity of avoiding remote accumulates and maintaining large granularity of tensor blocks for GPU processing, the coupled-cluster workload simply does not have enough work items to efficiently parallelize over more than 385 nodes (for this particular system). On the other hand, due to high memory demands, we could not use less nodes for this particular calculation. Such a situation will be characteristic for molecules with a large virtual-to-occupied spinor ratio, like our current UF$_6$ dimer system with a significantly reduced occupied space (letting the occupied space include all electrons would restore the scaling to higher node counts). Currently we are working on improving our original algorithms to address this issue.

\subsection{EFG of uranium in the uranyl tris-nitrate complex}
    
As a final example of possible applications, we now turn our attention to the calculation of the $q_{zz}$ component of the EFG tensor on the uranium atom for the [UO$_2$(NO$_3$)$_3$]$^-$ complex, for which there are experimental values~\cite{Monard1974} in the solid state. 
Initial theoretical investigations of EFGs for actinyl species focused on the bare uranyl ion (UO$_2^{2+}$)~\cite{Jong1998, Infante2004}, where it was found that qualitative agreement with experiment was only achieved if the effect of the equatorial ligands was taken into account (even if through point-charge  embedding~\cite{Jong1998}). These studies nevertheless revealed that the U $q_{zz}$ value had a dominant contribution from the so-called U(6p) core-hole, arising from the depletion of charge arising from the overlap between the O(2p) and the high-lying antibonding U(6p$_\sigma$) + O(2s) spinors.

\begin{figure*}[hbtp]
\centering
\includegraphics[width=0.4\textwidth]{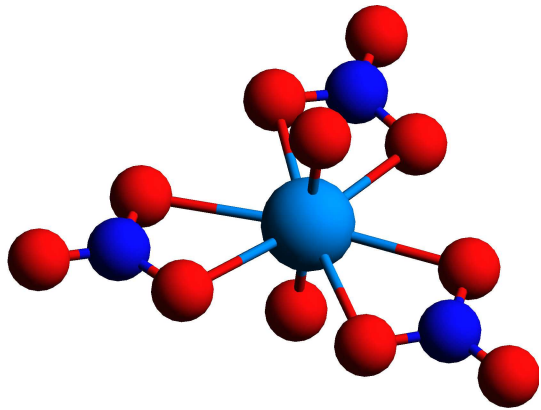}
\caption{Structure of the uranyl complex derived from the X-ray structure for the RbUO$_2$(NO$_3$)$_3$ crystyal\cite{Barclay:a04757}. }
\label{fig:uranyl}
\end{figure*}

An explicit inclusion of the contributions from the equatorial ligands to the U $q_{zz}$ value, and the associated analysis of orbital contributions to the EFG was, to the best of our knowledge, first performed by Belanzoni and coworkers~\cite{Belanzoni2005}, employing the BP86 GGA functional, the ZORA-4 Hamiltonian and QZ4P bases. They have, first, identified that the U(6p) core-hole yielded a positive contribution to the EFG, though in the bare uranyl these were offset by negative contributions due to the non-spherical electron distribution in the valence 5f shell caused by the U-O bonding. Moreover, positive contributions due to the ligands arise from the tails of the U(6p) spinor, which extends significantly to the region of the nitrate ligands, as well as from the electron donation by the nitrate groups into the U(5$f_\phi$, 6$d_\delta$) which have lobes in the equatorial plane. These calculations were found to underestimate the U $q_{zz}$ experimental value by around 4 a.u.

More recently, Autschbach and coworkers~\cite{Autschbach2012} have employed the X2C-1e Hamiltonian, triple-zeta quality basis sets (U: ANO(-h), light atoms: TZVPP), and density functional approximations (DFAs) including Hartree-Fock exchange such as B3LYP and CAMB3LYP, to revisit the U $q_{zz}$  on the [UO$_2$(HCO$_3$)$_3$]$^-$ complex that resembles fairly well the structural motif in [UO$_2$(NO$_3$)$_3$]$^-$ though for which, unfortunately, we are not aware of any experimental values. These results have demonstrated the importance of accounting for picture-change effects in the representation of the EFG operator (increases in U $q_{zz}$ of around 8 a.u., fairly consistent among the different methods), as well as the importance of including Hartree-Fock exchange in the DFAs from going from GGAs~\cite{Belanzoni2005} to hybrids~\cite{Autschbach2012} for obtaining larger values for U $q_{zz}$. They have also confirmed the large effect of the ligands on the U $q_{zz}$ found by Belanzoni and coworkers, as the U $q_{zz}$ goes from a negative value (between -8 and -7 a.u.\, depending on the DFA) in the bare uranyl to a positive ones (see table~\ref{tab:urany-trinitrite-comparison}).

Though the results by Autschbach and coworkers suggest that DFAs would work rather well in this case, it is well-known in the literature~\cite{doi:10.1098/rsta.2012.0489} that it can be difficult for these to correctly represent EFGs for transition metals~\cite{10.1002/cmr.a.20155,doi:10.1021/jz201685r} or lanthanides~\cite{Pasteka2016}, and as such it would be highly desirable to perform EFG calculations at the coupled cluster level more routinely, if not to provide benchmark values for systems larger than diatomics or triatomics. In this respect our calculations are, to the best of our knowledge, the first effort to obtain EFGs at CCSD level for uranyl species while explicitly including the equatorial ligands--in the pioneering calculation by de Jong and coworkers~\cite{Jong1998}, the structure of uranyl was investigated with CC but the EFG was only calculated at the Hartree-Fock level. As such, our calculations are also the first to  investigate the effects of increasing the number of correlated occupied and virtual spinors on the EFG values of uranyl complexes. 

Our results are shown in tables~\ref{tab:uranyl-trinitrite-ccsd-calcs} and~\ref{tab:urany-trinitrite-comparison}. We note that we have restricted ourselves to the X2C-AMFI Hamiltonian to remain close to the setup of prior calculations~\cite{Autschbach2012}. Furthermore, due to constraints on the available memory and other resources, we were only able to carry out calculations with significantly extended virtual spaces for double-zeta calculations. From table~\ref{tab:uranyl-trinitrite-ccsd-calcs} a first interesting result is the behavior of the T$_1$ diagnostic. For the smallest calculation (1), we have a value slightly over than 0.012, which is higher that what is usually observed for molecules containing light elements, but in line with our experience~\cite{C0CP02534H, Gomes2010, Gomes2013,Kervazo2019} of CCSD calculations on heavy elements in which we did not correlate electrons below the U 5d shell. When we increase the number of correlated electrons the  T$_1$ value progressively drops, and reaches around 0.009 for the calculations in which all electrons are correlated (4--7). This would suggest, therefore, that in prior calculations the relatively high T$_1$ values (with respect to the prescriptions originally put forth for light-element molecules) arise mostly due to an incomplete occupied correlating space, instead of being the sign of a growing multireference character for the heavy elements.

\begin{table*}[hbtp]
\caption{Comparison of times to solution (TTS, in hours, equal to the total wall time for each calculation and the part spent in the T and $\Lambda$ equations), T$_1$ diagnostic and $q_{zz}$ component of the EFG tensor (in atomic units), for CCSD expectation value calculations for the uranyl tris-nitrate complex ([UO$_2$(NO$_3$)$_3$]$^-$) with the X2C-AMFI Hamiltonian. O: number of occupied spinors correlated; V: number of virtual spinors correlated; M: number of ranks in the parallel calculation; C: cost estimate for the calculation (scaled by 1.0E+14); NormC: cost divided by the number of ranks, and normalized with respect to the value for the smallest calculation. See text for details.}
\label{tab:uranyl-trinitrite-ccsd-calcs}
\begin{tabular}{cccccrrccrcr}
\hline
\hline
            &&&&&&&&\multicolumn{2}{c}{TTS} &\\
\cline{9-10}
calculation & O	 & $x_o$ &	V	& $x_v$ &   M	& C &	NormC	&	total	& T+$\Lambda$&	T1	&	$q_{zz}$	 \\
\hline
\multicolumn{11}{c}{dyall.v2z} \\ 
\hline
 1 & 106	& 0.53 &	534	& 0.42 &	160	&   9 &	1.00	&	3h34	&  2h55 &	0.0126	& 10.02 \\
 2 & 156	& 0.77 &	534	& 0.42 & 	160	&  20 &	2.17	&	5h33	&  5h06 &	0.0103	&  9.70 \\
 3 & 156	& 0.77 & 	818	& 0.64 &	360	& 109 &	5.30	&	11h21	& 10h43 &	0.0102	&  8.48 \\
 4 & 202	& 1.00 & 	534	& 0.42 &	400	&  33 &	1.45	&	8h02	&  7h29	&   0.0091	&  9.73 \\
 5 & 202	& 1.00 &	680	& 0.53 &   2050 &  87 &	0.75	&	12h28	& 11h58 & 	0.0090	&  8.89 \\
 6 & 202	& 1.00 &	818	& 0.64 &   2050	& 183 &	1.56	&	18h23	& 17h49 & 	0.0089	&  8.59 \\
 7 & 202	& 1.00 &	896	& 0.68 &   2050	& 263 &	2.25	&	23h04	& 22h21 &	0.0089	&  8.54 \\
\hline
\multicolumn{11}{c}{dyall.v3z} \\ 
\hline
8  & 106    & 0.53 &    694 & 0.33 &    480 &  26 & 0.95    &    6h     &  4h08 &   0.0139  &  10.29 \\ % & 10.91\\
9  & 156    & 0.72 &    694 & 0.33 &    480 &  57 & 2.06    &   10h24   &  8h02 &   0.0114  &  10.04 \\ % & 10.57\\
10 & 156    & 0.72 &    944 & 0.46 &   2050 & 193 & 1.65    &   17h48   & 15h44 &   0.0104  &   9.99 \\ % & 9.34 \\
\hline
\hline
\end{tabular}
\end{table*}

\begin{table*}[hbtp]
\caption{Comparison between electronic structure methods and experimental results for the $q_{zz}$ component of the EFG tensor for the uranyl tris-nitrate complex ([UO$_2$(NO$_3$)$_3$]$^-$), for basis sets of different cardinal number. The coupled cluster calculation employs 202 occupied and 896 virtual spinors. All values in atomic units. The X2C-AMFI Hamiltonian is employed in all calculations.}
\label{tab:urany-trinitrite-comparison}
\begin{tabular}{l rrr} 
\hline
\hline
Method  & 2z & 3z & 4z \\
\hline
HF	        &  14.67	& 14.85 &  14.97 \\
B3LYP	    &	6.64	&  6.79 &   6.89 \\
CAMB3LYP    &	7.24	&  7.40 &   7.50 \\
PBE0	    &	7.58	&  7.75 &   7.85 \\
CCSD      	&	8.54	&       &        \\
&&&\\
ref.~\citenum{Belanzoni2005}, BP86, ZORA-4 & & & 4.11 \\ 
ref.~\citenum{Autschbach2012}$^a$, HF, X2C-1e & & 15.17 & \\ 
ref.~\citenum{Autschbach2012}$^a$, B3LYP, X2C-1e & & 6.81 & \\
ref.~\citenum{Autschbach2012}$^a$, CAMB3LYP, X2C-1e & & 8.33 & \\
&&&\\
exp\cite{Monard1974}	        &	\multicolumn{3}{c}{8.38$\pm$0.13} \\
\hline
\hline
$^a$ calculations on [UO$_2$(HCO$_3$)$_3$]$^-$
\end{tabular}
\end{table*}

Concerning the value of $q_{zz}$, we have fairly large variations going from the calculations in which  the correlation space is rather small (1) to the largest calculation performed (7). Comparing calculations in which
the number of occupied spinors is increased but the number of virtual spinors is kept constant (1 and 2, and 1 and 4), to those in which the reverse is true (2 and 3), we have that the size of the virtual space is the most significant variable to control. From our calculations, we observe that it is only after including roughly 60\% of the virtual space in the correlating space (more than 680 virtual spinors for double-zeta bases) that the $q_{zz}$ values start to converge to a value around 8.5 a.u., and though we have a much more limited set of data for triple-zeta bases, we seem to observe a similar pattern, with rather large variations in the $q_{zz}$ values with the change in the number of correlated virtual spinors. Unfortunately, with the largest calculations we carried out we are not able to include more than roughly 45\% of the virtual space in the correlating space, which appears not to be sufficient for obtaining a converged $q_{zz}$ value.

Strictly speaking, the calculations presented in table~\ref{tab:uranyl-trinitrite-ccsd-calcs} cannot be used to characterize the strong scaling of the code. This is because the memory required to store the 2-electron integral tensor grows sharply as we increase the number of correlated occupied and virtuals, requiring that the number of Summit nodes used was changed for each run in order to fit this tensor in. It is nevertheless possible to extract some information on the code's weak scaling. To do so, we define the following two metrics: (a) the cost of calculation $n$, taken to be proportional to the number of operations of the costlier contractions for both amplitude and $\Lambda$ equations ($C[n] = O^2V^4[n]$, with O: number of correlated occupied spinors; and V: number of correlated virtual spinors); and (b) cost per rank $M$ employed in the parallel calculation, normalized by the cost of calculation 1, the smallest considered here (NormC[n] = C[n]/(M*C[1]). We see that the significant cost increase resulting from the augmentation of the occupied and virtual spaces (a 28-fold increase from calculation 1 to calculation 7) can be offset by the increase in the number of ranks in the parallel calculation (a 13-fold increase comparing the same calculations) such that the time to solution remains within reasonable bounds (less than 24h). At the same time, NormC for the different calculations remains, with a few exceptions, between 1 and 2, and that including both the double and triple-zeta calculations. This indicates the code can handle high workloads in roughly the same manner as it does for much smaller workloads.

Finally, in table~\ref{tab:urany-trinitrite-comparison} we compare our best CCSD results to uncorrelated (Hartree-Fock), DFT calculations performed with DIRAC for hybrid DFAs, to the results from Belanzoni and coworkers~\cite{Belanzoni2005}, Autschbach and coworkers~\cite{Autschbach2012}, and to experiment~\cite{Monard1974}. We observe that the Hartree-Fock calculations provide largely overestimated values with respect to experiment, with differences of over 6 a.u.\ for the double-, triple- and quadruple-zeta bases, whereas all DFT calculation provide somewhat underestimated values--among the functionals compared, B3LYP fares the worst (differences from -1.74 to -1.49  a.u.), while PBE0 (differences from -0.80 and -0.53 a.u.) fares slightly better than CAMB3LYP (differences from -1.14 and -0.88 a.u.). Taken together, these results indicate that for the mean field approaches, an increase in basis set quality translates into an increase in $q_{zz}$ of slightly under 0.2 a.u.\ going from double to triple-zeta, with an additional increase of 0.1 a.u.\ when going from triple to quadruple zeta, and that all DFT results appear to move in the direction of experiment. We observe that our Hartree-Fock and DFT values are slightly smaller than those obtained by Autschbach and coworkers~\cite{Autschbach2012}, though a somewhat larger discrepancy is seen for CAMB3LYP than for B3LYP or Hartree-Fock, which suggests something other than purely the difference in molecular system could be at play here, and that the very good agreement to experiment with CAMB3LYP in the literature seems fortuitous.

Our best double-zeta CCSD result (calculation 7) shows only a slight overestimation (0.16 a.u.) with respect to experiment, which is a significantly better result than any of the mean-field ones. That said, since  we were not able to perform accurate calculations with more flexible basis sets, and from the DFT trends, it would not be surprising that these results, compared to future calculations, would be found to underestimate the $q_{zz}$ value at least a few tenths of an a.u., pointing to an overall overestimation of the experimental results at CCSD level. Once we implement approaches which allow us to efficiently truncate our (artificially, due to the use of uncontracted basis sets) large virtual spaces without loss of accuracy, we intend to revisit this question.

\section{Conclusion}

A reimplementation of the Kramers-unrestricted coupled cluster method was presented and shown to be able to exploit hundreds of GPU-accelerated nodes, while correctly reproducing the results of the prior implementation for a range of test cases. With this implementation, which currently does not exploit any rank reduction techniques or other approximations, we were already able to investigate systems for which about two hundred electrons and around a thousand virtual molecular spinors were taken into account in the correlated calculations.

Current functionalities include CC2, CCD and CCSD wavefuctions, and perturbative triples corrections to the ground-state. Ground-state expectation values are also available for CCD and CCSD, through the computation of one-body density matrices. The code is now interfaced to the DIRAC and ReSpect packages, but interfaces to other software packages employing Gaussian-type atomic orbitals can be implemented in a straightforward manner as they only require functionality to read in the molecular spinors.

Unlike the original implementation, this reimplementation does not exploit double group point symmetry. For the aforementioned functionality, this is a disadvantage for small, highly symmetric systems (i.e.\ around 10 atoms or 100 electrons), though we consider that in practice this shortcoming is offset by the large-scale parallelization that allows us to treat such systems upon distortions (bent/twisted configurations) with triple or quadruple zeta basis sets without hitting the wall of transitioning from real to complex algebra (as is the case in the original code when symmetry is reduced), inefficient communication over hundreds of nodes, and the extensive use of disk storage of intermediate quantities. For larger systems, which often have little to no symmetry, the current implementation makes correlated calculations feasible.

We have employed the code to investigate the properties of different heavy element complexes: ionization energies for lanthanide monofluorides (LaF, EuF and YbF), energies of formation of gold-argon clusters of different sizes, and the electric field gradient ($q^U_{zz}$) at the uranium nucleus in a uranyl tris-nitrate complex.

For the lanthanide monofluorides, we have shown that we need to correlate at least the Ln (4s4p4d4f) electron shell to obtain converged ionization energies. We observed that LaF and EuF, the perturbative triples-corrected ionization energies for the triple-zeta quality basis sets are already within the experimental error bounds, with adiabatic values showing a slightly better agreement to experiment than vertical ones. For YbF, we have encountered the same issue previously described in the literature, in that spin-orbit coupled calculations show a surprisingly large $T_1$ diagnostic value around the equilibrium distance, which makes perturbative triples results unreliable, but we see a smooth convergence of the CCSD values towards the experimental values.

For the gold-argon clusters, substantial binding energy is obtained at the coupled cluster level for closed shell noble metal Au$^+$ and noble gas Ar, with negligible values at the Hartree-Fock level (below 0.6 eV) and overestimation of the binding energy up to 0.5~eV for MP2. Therefore, a selection of a reliable method is important. For AuAr$_4^+$ two different structures which are close in energy have been computed and the 3D structure is about 0.2/0.3~eV lower than the 2D structure at the CCSD/CCSD(T) level.

For the uranyl tris-nitrate complex, we have found that CCSD wavefunctions are capable of providing $q^U_{zz}$ values in very good agreement to experiment, with non-negligible improvements over DFT calculations. This improvement, however, appears to comes at the cost of including high-lying virtual spinors in the CCSD calculation. We have been limited to performing CCSD calculations with double-zeta basis due to the number of virtual spinors that appears to be required to converge the $q^U_{zz}$ value. While the DFT results suggest non-negligible basis set effects, due to the use of uncontracted basis sets, for reasons of computational cost it was not possible to include enough high-lying virtual spinors in triple-zeta calculations and therefore the calculations we were able to performed are still far from converged. We are currently working on implementing approaches to compress the virtual space while retaining precision in the correlated treatment, in order to enable efficient calculations with larger basis sets. We expect these will allow us not only to revisit this system in the near future but also enable calculations on significantly larger systems.

%%%%%%%%%%%%%%%%%%%%%%%%%%%%%%%%%%%%%%%%%%%%%%%%%%%%%%%%%%%%%%%%%%%%%
%% The "Acknowledgement" section can be given in all manuscript
%% classes.  This should be given within the "acknowledgement"
%% environment, which will make the correct section or running title.
%%%%%%%%%%%%%%%%%%%%%%%%%%%%%%%%%%%%%%%%%%%%%%%%%%%%%%%%%%%%%%%%%%%%%
\begin{acknowledgement}
This research used resources of the Oak Ridge Leadership Computing Facility, which is a DOE Office of Science User Facility supported under Contract DE-AC05-00OR22725. Some computer codes used in this research (ExaTENSOR, TAL-SH, partly ExaCorr) were developed during the OLCF-4 Center for Accelerated Application Readiness (CAAR) program funded by the US Department of Energy at the Oak Ridge National Laboratory. ASPG and LH acknowledge support from PIA ANR project CaPPA (ANR-11-LABX-0005-01), the Franco-German project CompRIXS (Agence nationale de la recherche ANR-19-CE29-0019, Deutsche Forschungsgemeinschaft JA 2329/6-1), I-SITE ULNE projects OVERSEE, the French Ministry of Higher Education and Research, region Hauts de France council and European Regional Development Fund (ERDF) project CPER CLIMIBIO, and the French national supercomputing facilities (grants DARI A0070801859 and Joliot Curie grands challenges 2019 gch0417). ASPG, LH, JVP and LV acknowledge support from MESONM International Associated Laboratory (LAI) (ANR-16-IDEX-0004). JVP acknowledges funding by the Austrian Science Fund(FWF):J 4177-N36. MR acknowledges the funding support from the Research Council of Norway through a Center of Excellence Grant (Grant No. 262695).
\end{acknowledgement}

\section{Appendix}

In this Appendix we list the equations implemented in the current ExaCorr code.

\subsection{Working equations for CCSD amplitude equations}
\label{ap:CC}
The working equations for coupled cluster presented in ref.~\citenum{visscher1996formulation} were optimized to reduce the operation count. In the current implementation all the tensors are kept in memory and, therefore, it is desirable to avoid large intermediates, even if the number of operations is slightly increased. The B intermediate in ref.~\citenum{visscher1996formulation} has a size of $n_{vir}^4$ and can be avoided by using the following equations:
\begin{strip}
\begin{align}
P_{pq}^{-} f(p,q) & = f(p,q) - f(q,p) \\
\tau^{ab}_{ij} & = t^{ab}_{ij}+P^{-}_{ij}\left(t^{a}_{i}t^{b}_{j}\right)
\\
H^{a}_{c} & = F^{a}_{c} + \frac{1}{2} \, \sum_{kld} V^{kl}_{cd}\tau^{da}_{kl}
\\
G^{a}_{c} & = H^{a}_{c}+ \sum_{kd} V^{ak}_{cd}t^{d}_{k} - \sum_{k} F^{k}_{c}t^{a}_{k}
\end{align}
\end{strip}
\begin{strip}
\begin{align}
H^{k}_{i} & =-F^{k}_{i} + \frac{1}{2} \, \sum_{lcd} V^{kl}_{cd}\tau^{cd}_{li}
\\
G^{k}_{i} & =H^{k}_{i}- \sum_{lc} V^{kl}_{ic}t^{c}_{l}- \sum_{c} F^{k}_{c}t^{c}_{i}
\\
H^{k}_{c} & =F^{k}_{c}+ \sum_{ld} V^{kl}_{cd}t^{d}_{l}
\\
H^{ak}_{ci} & = V^{ak}_{ci} 
+ \sum_{d} V^{ak}_{cd}t^{d}_{i} - \sum_{l} V^{kl}_{ic}t^{a}_{l}
+ \sum_{ld} V^{kl}_{cd} \left( \frac{1}{2} \, t^{ad}_{li} + t^{a}_{l}t^{d}_{i} \right)
\\
C^{kb}_{ij} & =V^{ \, ij\dagger}_{kb}
+ \sum_{c} P^{-}_{ij}\left( V^{bk}_{ci}t^{c}_{j} \right)
- \frac{1}{2} \, \sum_{cd} V^{bk}_{cd}\tau^{cd}_{ij}
\\
A^{kl}_{ij} & =V^{kl}_{ij} + \sum_{c} P^{-}_{ij}\left( V^{kl}_{ic}t^{c}_{j} \right) + \frac{1}{2}\,
\sum_{cd} V^{kl}_{cd}\tau^{cd}_{ij}
\\
S^{ab}_{ij} & =  
  \sum_{kc} P_{ij}^{-}P_{ab}^{-} \left(H^{ak}_{ci}t^{cb}_{jk}\right)
\nonumber \\
 &  + \sum_{k} P_{ji}^{-} G^{k}_{j}t^{ab}_{ik}
+ \sum_{c} P^{-}_{ij}\left( V^{cj\dagger}_{ab}t^{c}_{i} \right)
+ \sum_{c} P_{ab}^{-} G^{a}_{c}t^{cb}_{ij}
 + \sum_{k} P^{-}_{ab}\left(C^{ka}_{ij}t^{b}_{k}\right)
 \nonumber \\
& + \frac{1}{2} \sum_{kl} A^{kl}_{ij}\tau^{ab}_{kl} + \frac{1}{2}\sum_{cd} V^{ab}_{cd}\tau^{cd}_{ij}
+ V^{ij\dagger}_{ab}
\\
S^{a}_{i} & =F^{i\dagger}_{a} 
+ \sum_{c} H^{a}_{c}t^{c}_{i} + \sum_{k} H^{k}_{i}t^{a}_{k} + \sum_{kc} H^{k}_{c}t^{ac}_{ik}
\nonumber \\
& + \sum_{k} \left(\sum_{c} \left(H^{k}_{c}-2 \, F^{k}_{c} \right) t^{c}_{i} \right)t^{a}_{k}
+ \frac{1}{2}\sum_{kcd} V^{ak}_{cd}\tau^{cd}_{ik}
+ \frac{1}{2} \sum_{klc} V^{kl}_{ic}\tau^{ca}_{kl}
- \sum_{kc} V^{ak}_{ci}t^{c}_{k}
\label{eq:CC_general}
\end{align}

\subsection{Working equations for CC2 amplitude equations}
\label{ap:cc2}

In the CC2 method~\cite{cc2-christiansen-1995} the T$_1$ amplitude equations in~\ref{ap:CC} are unchanged whereas the T$_2$ amplitude equations are given by
\begin{align}
	S^{ab}_{ij} = &  V^{ab}_{ij} 
	+\sum_e P^{-}_{ij}  t^{e}_{i} W^{ab}_{ej} {}_{CC2} -\sum_m
	P^{-}_{ab}  t^{a}_{m}
	W^{mb}_{ij} {}_{CC2}
	\nonumber \\
	&
	+\sum_{m<n} P^{-}_{ab}t^{a}_{m} t^{b}_{n} W^{mn}_{ij} {}_{CC2}
	+
	\sum_{e<f}P^{-}_{ij} t^{e}_{i} t^{f}_{j} W^{ab}_{ef} {}_{CC2}
	\nonumber
	\\
	& + \sum_{e} P^{-}_{ab} f^{b}_{e} t^{ae}_{ij} -\sum_{m} P^{-}_{ij}
	f^{m}_{j} t^{ab}_{im}
\end{align}
\noindent with :
\begin{align}
	W^{mn}_{ij} {} _{CC2}= & V^{mn}_{ij} + \sum_{e} P^{-}_{ij}
	t^{e}_{j} V^{mn}_{ie}  + \frac{1}{4}
	\sum_{ef}	P^{-}_{ij} t^{e}_{i}  t^{f}_{j} V^{mn}_{ef} \\
	W^{ab}_{ef} {}_{CC2}=&  V^{ab}_{ef} -\sum_{m} P^{-}_{ab} t^{b}_{m}
	V^{am}_{ef}
	+\frac{1}{4} \sum_{mn}  P^{-}_{ab}t^{a}_{m} t^{b}_{n}
	V^{mn}_{ef}  \\
	W^{ab}_{ej}  {}_{CC2}=&V^{ab}_{ej} -\frac{1}{2} \sum_{m}P^{-}_{ab}
	t^{a}_{m} V^{mb}_{ej} \\
	W^{mb}_{ij}  {}_{CC2}=&V^{mb}_{ij} +\frac{1}{2}\sum_{e}P^{-}_{ij}
	t^{e}_{i} V^{mb}_{ej}
\end{align}

\subsection{Working equations for perturbative Triples}
\label{ap:triples}

In contrast to ref.~\citenum{visscher1996formulation} the full term is used, not only the upper triangle. The following expression needs to be evaluated
for the older fourth order term 
\begin{align}
& E^{(+T)} =- \sum_{i,j,k}^{n_{occ}} \sum_{a,b,c}^{n_{vir}}   \frac{
                \left( W^{abc}_{ijk}+W^{bca}_{ijk}+W^{cab}_{ijk} \right)
                \left( W^{abc}_{ijk}+W^{bca}_{ijk}+W^{cab}_{ijk} \right)
                }{36 \, D^{abc}_{ijk}}  ,
\label{eq:4order-term}
\end{align}
where $D^{abc}_{ijk}$ is the energy denominator computed from the spinor energies and the intermediate is defined in following manner
\begin{align}
W^{abc}_{ijk}=& \sum_e \left\langle i e || a b \right\rangle t_{jk}^{ec}
                        +  \sum_e \left\langle j e || a b \right\rangle t_{ki}^{ec}
                        + \sum_e \left\langle k e || a b \right\rangle t_{ij}^{ec} 
                       \nonumber \\
                        & +    \sum_m \left\langle i j || a m \right\rangle t_{km}^{bc}
                        +   \sum_m \left\langle j k || a m \right\rangle t_{im}^{bc}
                        +   \sum_m \left\langle k i || a m \right\rangle t_{jm}^{bc} 
.
\label{eq:W-term}
\end{align}
The fifth order correction is given by
\begin{align}
 & E^{((T))} = E^{(+T)} +  \sum_{i,j,k}^{n_{occ}} \sum_{a,b,c}^{n_{vir}}  \frac{
                \left( W^{abc}_{ijk}+W^{bca}_{ijk}+W^{cab}_{ijk} \right)
                \left(  Z^{abc}_{ijk} + Z^{cab}_{ijk} + Z^{bca}_{ijk}\right)
                }{36 \, D^{abc}_{ijk}} 
,
\label{eq:5order-term}
\end{align}
using the intermediate
\begin{align}
Z^{abc}_{ijk}=&
           \left\langle i j || a b \right\rangle^*  t_k^c +  t_{ij}^{ab \, *} f_k^c
      +  \left\langle j k || a b \right\rangle^*  t_i^c +  t_{jk}^{ab \, *} f_i^c
      +  \left\langle k i || a b \right\rangle^*  t_j^c +  t_{ki}^{ab \, *} f_j^c
.
\label{eq:Z-term}
\end{align} 
An alternative fifth order expression\cite{Deegan1994} is defined by
\begin{align}
& E^{(-T)} = E^{(+T)} +  \sum_{i,j,k}^{n_{occ}} \sum_{a,b,c}^{n_{vir}}  \frac{ 
                \left( W^{abc}_{ijk}+W^{bca}_{ijk}+W^{cab}_{ijk} \right)
                \left( Y^{abc}_{ijk}+Y^{cab}_{ijk}+Y^{bca}_{ijk}\right)
                }{36} 
,
\end{align}
applying the expression
\begin{align}
Y^{abc}_{ijk}=& \frac{1}{3} \left(
         t_i^a t_j^b t_k^c - t_j^a t_i^b t_k^c 
      + t_j^a t_k^b t_i^c - t_k^a t_j^b t_i^c
      + t_k^a t_i^b t_j^c - t_i^a t_k^b t_j^c
      \right)
      + t_{ij}^{ab} t_k^c + t_{jk}^{ab} t_i^c + t_{ki}^{ab} t_j^c .
\label{eq:Y-term}
\end{align}

All three expressions are evaluated in our implementation so that a user of the program may easily assess the resulting differences in energies.

\subsection{Working equations for CCSD lambda equations}
\label{ap:lambda}

This equations are close to the equations presented in ref.~\citenum{shee2016analytic}, there was a slight error in the equations in the reference, which is corrected here. 
These are the corrected equations:
 \begin{align}
       W^{ij}_{mn} =& V^{ij}_{mn} + P^{-}_{mn} \sum_e V^{ij}_{en} t^e_m + \sum_{e<f} V^{ij}_{ef} \tau^{ef}_{mn}
       \\
        W^{mb}_{ej} = & V^{mb}_{ej} + \sum_f V^{mb}_{ef} t^f_j - \sum_n V^{mn}_{ej} t^b_n \\
        & -  \sum_{nf} V^{mn}_{ef} (t^{fb}_{jn} + t^f_j t^b_n)
        \\
        W^{ef}_{am} = & V^{ef}_{am} + P^{-}_{ef} \sum_{ng} V^{en}_{ag} t^{fg}_{mn} + \sum_{g} W^{ef}_{ag} t^g_m \\
        & + \sum_n \overline{F}^n_a t^{ef}_{mn} + \sum_{n>o} V^{no}_{am} \tau^{ef}_{no} - P^{-}_{ef} \sum_n \overline{W}^{nf}_{am} t^e_n
        \\
    \mathcal{G}_a^e=& - \sum_{f,m<n}\lambda^{mn}_{af} t^{ef}_{mn}
        \\
    \mathcal{G}_m^i=& \sum_{n,e<f} \lambda^{in}_{ef} t^{ef}_{mn} 
        \\
	    0 = & V^{ij}_{ab} + P^{-}_{ab} \sum_e \lambda^{ij}_{ae} \overline{F}^e_b - P^{-}_{ij} \sum_m \lambda^{im}_{ab} \overline{F}^j_m + \sum_{m>n} \lambda^{mn}_{ab} W^{ij}_{mn} + P^{-}_{ij}P_{ab} \sum_{me} \lambda^{im}_{ae} W^{je}_{bm} \\ 
	    & + P^{-}_{ab} \sum_e V^{ij}_{ae} \mathcal{G}^e_b - P^{-}_{ab} \sum_m \lambda^m_a W^{ij}_{mb} - P^{-}_{ij} \sum_m V^{im}_{ab} \mathcal{G}^j_m + P^{-}_{ij} \sum_e \lambda^i_e W^{ej}_{ab} \\
	    & + P^{-}_{ij}P^{-}_{ab} \lambda^i_a \overline{F}^j_b + \sum_{e>f} \lambda^{ij}_{ef} W^{ef}_{ab}\\
	    0 = & \overline{F}_i^a + \sum_{e} \lambda_e^i \overline{F}_a^e - \sum_m \lambda_a^m \overline{F}_m^i - \sum_{mn} \mathcal{G}^n_m W^{mi}_{na} - \sum_{ef} \mathcal{G}^f_e W^{ei}_{fa} \\
	    & + \sum_{me} \lambda^m_e W^{ie}_{am} - \sum_{m>n,e} \lambda^{mn}_{ae} W^{ie}_{mn} + \sum_{m,e<f} \lambda^{im}_{ef} W^{ef}_{am} \\
	    \overline{F}^e_a =& f^e_a - \sum_m f^m_a t^e_m + \sum_{mf} V^{me}_{fa} t^f_m - \sum_{m>n,f} V^{mn}_{af} \tau^{ef}_{mn} 
\end{align}
The full list including intermediates can be found in ref.~\citenum{shee2016analytic}

\subsection{Working equations for Lagrange 1-body density matrix}
\label{ap:density}

\begin{align}
    \gamma_{ij} & = \frac{1}{2} \sum_{mef} t_{im}^{ef} \lambda_{ef}^{mj} - \sum_{e} t_{i}^{e} \lambda_{e}^{j} \\
    \gamma_{ia} & = \lambda_{a}^{i} \\
    \gamma_{ai} & = t_{i}^{a} + \sum_{me} \lambda_{e}^{m} (t_{im}^{ae} - t_{i}^{e}t_{m}^{a}) - \frac{1}{2} \sum_{mnef} \lambda_{ef}^{mn} (t_{in}^{ef} t_{m}^{a} + t_i^e t_{mn}^{af}) \\
    \gamma_{ab} & = \frac{1}{2} \sum_{mne} t_{mn}^{ae} \lambda_{be}^{mn} + \sum_{m} t_m^a \lambda_b^m
\end{align}

The density matrix is symmetrized before use in property calculations.
\end{strip}
%%%%%%%%%%%%%%%%%%%%%%%%%%%%%%%%%%%%%%%%%%%%%%%%%%%%%%%%%%%%%%%%%%%%%
%% The appropriate \bibliography command should be placed here.
%% Notice that the class file automatically sets \bibliographystyle
%% and also names the section correctly.
%%%%%%%%%%%%%%%%%%%%%%%%%%%%%%%%%%%%%%%%%%%%%%%%%%%%%%%%%%%%%%%%%%%%%
\bibliography{exacorr-paper}

\providecommand{\latin}[1]{#1}
\makeatletter
\providecommand{\doi}
  {\begingroup\let\do\@makeother\dospecials
  \catcode`\{=1 \catcode`\}=2 \doi@aux}
\providecommand{\doi@aux}[1]{\endgroup\texttt{#1}}
\makeatother
\providecommand*\mcitethebibliography{\thebibliography}
\csname @ifundefined\endcsname{endmcitethebibliography}
  {\let\endmcitethebibliography\endthebibliography}{}
\begin{mcitethebibliography}{103}
\providecommand*\natexlab[1]{#1}
\providecommand*\mciteSetBstSublistMode[1]{}
\providecommand*\mciteSetBstMaxWidthForm[2]{}
\providecommand*\mciteBstWouldAddEndPuncttrue
  {\def\EndOfBibitem{\unskip.}}
\providecommand*\mciteBstWouldAddEndPunctfalse
  {\let\EndOfBibitem\relax}
\providecommand*\mciteSetBstMidEndSepPunct[3]{}
\providecommand*\mciteSetBstSublistLabelBeginEnd[3]{}
\providecommand*\EndOfBibitem{}
\mciteSetBstSublistMode{f}
\mciteSetBstMaxWidthForm{subitem}{(\alph{mcitesubitemcount})}
\mciteSetBstSublistLabelBeginEnd
  {\mcitemaxwidthsubitemform\space}
  {\relax}
  {\relax}

\bibitem[Thiel and Hummer(2013)Thiel, and Hummer]{Thiel2013}
Thiel,~W.; Hummer,~G. Methods for computational chemistry. \emph{Nature}
  \textbf{2013}, \emph{504}, 96--97\relax
\mciteBstWouldAddEndPuncttrue
\mciteSetBstMidEndSepPunct{\mcitedefaultmidpunct}
{\mcitedefaultendpunct}{\mcitedefaultseppunct}\relax
\EndOfBibitem
\bibitem[Bak \latin{et~al.}(2001)Bak, Gauss, Jørgensen, Olsen, Helgaker, and
  Stanton]{Bak2001}
Bak,~K.~L.; Gauss,~J.; Jørgensen,~P.; Olsen,~J.; Helgaker,~T.; Stanton,~J.~F.
  The accurate determination of molecular equilibrium structures. \emph{J.
  Chem. Phys.} \textbf{2001}, \emph{114}, 6548--6556\relax
\mciteBstWouldAddEndPuncttrue
\mciteSetBstMidEndSepPunct{\mcitedefaultmidpunct}
{\mcitedefaultendpunct}{\mcitedefaultseppunct}\relax
\EndOfBibitem
\bibitem[Coriani \latin{et~al.}(2005)Coriani, Marchesan, Gauss, Hättig,
  Helgaker, and Jørgensen]{Coriani2005}
Coriani,~S.; Marchesan,~D.; Gauss,~J.; Hättig,~C.; Helgaker,~T.;
  Jørgensen,~P. The accuracy of ab initio molecular geometries for systems
  containing second-row atoms. \emph{J. Chem. Phys.} \textbf{2005}, \emph{123},
  184107\relax
\mciteBstWouldAddEndPuncttrue
\mciteSetBstMidEndSepPunct{\mcitedefaultmidpunct}
{\mcitedefaultendpunct}{\mcitedefaultseppunct}\relax
\EndOfBibitem
\bibitem[Bartlett(2012)]{Bartlett2012}
Bartlett,~R.~J. Coupled-cluster theory and its equation-of-motion extensions.
  \emph{WIREs Comput Mol Sci} \textbf{2012}, \emph{2}, 126--138\relax
\mciteBstWouldAddEndPuncttrue
\mciteSetBstMidEndSepPunct{\mcitedefaultmidpunct}
{\mcitedefaultendpunct}{\mcitedefaultseppunct}\relax
\EndOfBibitem
\bibitem[Shee \latin{et~al.}(2018)Shee, Saue, Visscher, and Severo
  Pereira~Gomes]{shee2018equation}
Shee,~A.; Saue,~T.; Visscher,~L.; Severo Pereira~Gomes,~A. Equation-of-motion
  coupled-cluster theory based on the 4-component Dirac--Coulomb (--Gaunt)
  Hamiltonian. Energies for single electron detachment, attachment, and
  electronically excited states. \emph{J. Chem. Phys.} \textbf{2018},
  \emph{149}, 174113\relax
\mciteBstWouldAddEndPuncttrue
\mciteSetBstMidEndSepPunct{\mcitedefaultmidpunct}
{\mcitedefaultendpunct}{\mcitedefaultseppunct}\relax
\EndOfBibitem
\bibitem[Haase \latin{et~al.}(2020)Haase, Eliav, Iliaš, and
  Borschevsky]{Haase2020}
Haase,~P. A.~B.; Eliav,~E.; Iliaš,~M.; Borschevsky,~A. Hyperfine Structure
  Constants on the Relativistic Coupled Cluster Level with Associated
  Uncertainties. \emph{J. Phys. Chem. A} \textbf{2020}, \emph{124},
  3157--3169\relax
\mciteBstWouldAddEndPuncttrue
\mciteSetBstMidEndSepPunct{\mcitedefaultmidpunct}
{\mcitedefaultendpunct}{\mcitedefaultseppunct}\relax
\EndOfBibitem
\bibitem[Kervazo \latin{et~al.}(2019)Kervazo, Réal, Virot, Severo
  Pereira~Gomes, and Vallet]{Kervazo2019}
Kervazo,~S.; Réal,~F.; Virot,~F.; Severo Pereira~Gomes,~A.; Vallet,~V.
  Accurate Predictions of Volatile Plutonium Thermodynamic Properties.
  \emph{Inorg. Chem.} \textbf{2019}, \emph{58}, 14507--14521\relax
\mciteBstWouldAddEndPuncttrue
\mciteSetBstMidEndSepPunct{\mcitedefaultmidpunct}
{\mcitedefaultendpunct}{\mcitedefaultseppunct}\relax
\EndOfBibitem
\bibitem[Denis \latin{et~al.}(2019)Denis, Haase, Timmermans, Eliav, Hutzler,
  and Borschevsky]{Denis2019}
Denis,~M.; Haase,~P. A.~B.; Timmermans,~R. G.~E.; Eliav,~E.; Hutzler,~N.~R.;
  Borschevsky,~A. Enhancement factor for the electric dipole moment of the
  electron in the BaOH and YbOH molecules. \emph{Phys. Rev. A} \textbf{2019},
  \emph{99}, 042512\relax
\mciteBstWouldAddEndPuncttrue
\mciteSetBstMidEndSepPunct{\mcitedefaultmidpunct}
{\mcitedefaultendpunct}{\mcitedefaultseppunct}\relax
\EndOfBibitem
\bibitem[Zelovich \latin{et~al.}(2017)Zelovich, Borschevsky, Eliav, and
  Kaldor]{Zelovich2017}
Zelovich,~T.; Borschevsky,~A.; Eliav,~E.; Kaldor,~U. Relativistic coupled
  cluster calculation of Mössbauer isomer shifts of iodine compounds.
  \emph{Molecular Physics} \textbf{2017}, \emph{115}, 138--143\relax
\mciteBstWouldAddEndPuncttrue
\mciteSetBstMidEndSepPunct{\mcitedefaultmidpunct}
{\mcitedefaultendpunct}{\mcitedefaultseppunct}\relax
\EndOfBibitem
\bibitem[van Stralen and Visscher(2002)van Stralen, and Visscher]{Stralen2002}
van Stralen,~J. N.~P.; Visscher,~L. The nuclear quadrupole moment of
  {$^{115}$In} from molecular data. \emph{J. Chem. Phys.} \textbf{2002},
  \emph{117}, 3103--3108\relax
\mciteBstWouldAddEndPuncttrue
\mciteSetBstMidEndSepPunct{\mcitedefaultmidpunct}
{\mcitedefaultendpunct}{\mcitedefaultseppunct}\relax
\EndOfBibitem
\bibitem[Whitten(1973)]{Whitten1973}
Whitten,~J.~L. Coulombic potential energy integrals and approximations.
  \emph{J. Chem. Phys.} \textbf{1973}, \emph{58}, 4496--4501\relax
\mciteBstWouldAddEndPuncttrue
\mciteSetBstMidEndSepPunct{\mcitedefaultmidpunct}
{\mcitedefaultendpunct}{\mcitedefaultseppunct}\relax
\EndOfBibitem
\bibitem[Vahtras \latin{et~al.}(1993)Vahtras, {Alml\"of}, and
  Feyereisen]{Vahtras1993}
Vahtras,~O.; {Alml\"of},~J.; Feyereisen,~M.~W. Integral approximations for
  LCAO-SCF calculations. \emph{Chemical Physics Letters} \textbf{1993},
  \emph{213}, 514--518\relax
\mciteBstWouldAddEndPuncttrue
\mciteSetBstMidEndSepPunct{\mcitedefaultmidpunct}
{\mcitedefaultendpunct}{\mcitedefaultseppunct}\relax
\EndOfBibitem
\bibitem[{F\"orster} and Visscher(2020){F\"orster}, and Visscher]{Foerster2020}
{F\"orster},~A.; Visscher,~L. Low-Order Scaling G0W0 by Pair Atomic Density
  Fitting. \emph{J. Chem. Theory Comput.} \textbf{2020}, \emph{16},
  7381--7399\relax
\mciteBstWouldAddEndPuncttrue
\mciteSetBstMidEndSepPunct{\mcitedefaultmidpunct}
{\mcitedefaultendpunct}{\mcitedefaultseppunct}\relax
\EndOfBibitem
\bibitem[Beebe and Linderberg(1977)Beebe, and Linderberg]{Beebe1977}
Beebe,~N. H.~F.; Linderberg,~J. Simplifications in the generation and
  transformation of two-electron integrals in molecular calculations.
  \emph{Int. J. Quantum Chem.} \textbf{1977}, \emph{12}, 683--705\relax
\mciteBstWouldAddEndPuncttrue
\mciteSetBstMidEndSepPunct{\mcitedefaultmidpunct}
{\mcitedefaultendpunct}{\mcitedefaultseppunct}\relax
\EndOfBibitem
\bibitem[Pedersen \latin{et~al.}(2009)Pedersen, Aquilante, and
  Lindh]{Pedersen2009}
Pedersen,~T.~B.; Aquilante,~F.; Lindh,~R. Density fitting with auxiliary basis
  sets from Cholesky decompositions. \emph{Theoretical Chemistry Accounts}
  \textbf{2009}, \emph{124}, 1--10\relax
\mciteBstWouldAddEndPuncttrue
\mciteSetBstMidEndSepPunct{\mcitedefaultmidpunct}
{\mcitedefaultendpunct}{\mcitedefaultseppunct}\relax
\EndOfBibitem
\bibitem[Peng and Kowalski(2017)Peng, and Kowalski]{Peng2017}
Peng,~B.; Kowalski,~K. Highly Efficient and Scalable Compound Decomposition of
  Two-Electron Integral Tensor and Its Application in Coupled Cluster
  Calculations. \emph{Journal of Chemical Theory and Computation}
  \textbf{2017}, \emph{13}, 4179--4192\relax
\mciteBstWouldAddEndPuncttrue
\mciteSetBstMidEndSepPunct{\mcitedefaultmidpunct}
{\mcitedefaultendpunct}{\mcitedefaultseppunct}\relax
\EndOfBibitem
\bibitem[Helmich-Paris \latin{et~al.}(2019)Helmich-Paris, Repisky, and
  Visscher]{Helmich-Paris2019}
Helmich-Paris,~B.; Repisky,~M.; Visscher,~L. Relativistic Cholesky-decomposed
  density matrix MP2. \emph{Chemical Physics} \textbf{2019}, \emph{518}, 38 --
  46\relax
\mciteBstWouldAddEndPuncttrue
\mciteSetBstMidEndSepPunct{\mcitedefaultmidpunct}
{\mcitedefaultendpunct}{\mcitedefaultseppunct}\relax
\EndOfBibitem
\bibitem[Sikkema \latin{et~al.}(2009)Sikkema, Visscher, Saue, and
  Iliaš]{Sikkema2009}
Sikkema,~J.; Visscher,~L.; Saue,~T.; Iliaš,~M. The molecular mean-field
  approach for correlated relativistic calculations. \emph{J. Chem. Phys.}
  \textbf{2009}, \emph{131}, 124116\relax
\mciteBstWouldAddEndPuncttrue
\mciteSetBstMidEndSepPunct{\mcitedefaultmidpunct}
{\mcitedefaultendpunct}{\mcitedefaultseppunct}\relax
\EndOfBibitem
\bibitem[de~Jong \latin{et~al.}(1998)de~Jong, Visscher, and
  Nieuwpoort]{Jong1998}
de~Jong,~W.~A.; Visscher,~L.; Nieuwpoort,~W.~C. On the bonding and the electric
  field gradient of the uranyl ion. \emph{Journal of Molecular Structure:
  THEOCHEM} \textbf{1998}, \emph{458}, 41--52\relax
\mciteBstWouldAddEndPuncttrue
\mciteSetBstMidEndSepPunct{\mcitedefaultmidpunct}
{\mcitedefaultendpunct}{\mcitedefaultseppunct}\relax
\EndOfBibitem
\bibitem[Infante and Visscher(2004)Infante, and Visscher]{Infante2004}
Infante,~I.; Visscher,~L. QM/MM study of aqueous solvation of the uranyl
  fluoride {[UO$_{2}$F]} complex. \emph{J. Comput. Chem.} \textbf{2004},
  \emph{25}, 386--392\relax
\mciteBstWouldAddEndPuncttrue
\mciteSetBstMidEndSepPunct{\mcitedefaultmidpunct}
{\mcitedefaultendpunct}{\mcitedefaultseppunct}\relax
\EndOfBibitem
\bibitem[Visscher \latin{et~al.}(1996)Visscher, Lee, and
  Dyall]{visscher1996formulation}
Visscher,~L.; Lee,~T.~J.; Dyall,~K.~G. Formulation and implementation of a
  relativistic unrestricted coupled-cluster method including noniterative
  connected triples. \emph{J. Chem. Phys.} \textbf{1996}, \emph{105},
  8769--8776\relax
\mciteBstWouldAddEndPuncttrue
\mciteSetBstMidEndSepPunct{\mcitedefaultmidpunct}
{\mcitedefaultendpunct}{\mcitedefaultseppunct}\relax
\EndOfBibitem
\bibitem[Pernpointner and Visscher(2003)Pernpointner, and
  Visscher]{Pernpointner2003}
Pernpointner,~M.; Visscher,~L. Parallelization of four-component calculations.
  II. Symmetry-driven parallelization of the 4-Spinor CCSD algorithm. \emph{J.
  Comput. Chem.} \textbf{2003}, \emph{24}, 754--759\relax
\mciteBstWouldAddEndPuncttrue
\mciteSetBstMidEndSepPunct{\mcitedefaultmidpunct}
{\mcitedefaultendpunct}{\mcitedefaultseppunct}\relax
\EndOfBibitem
\bibitem[Lyakh(2019)]{lyakh2019domain}
Lyakh,~D.~I. Domain-specific virtual processors as a portable programming and
  execution model for parallel computational workloads on modern heterogeneous
  high-performance computing architectures. \emph{Int. J. Quant. Chem.}
  \textbf{2019}, \emph{119}, e25926\relax
\mciteBstWouldAddEndPuncttrue
\mciteSetBstMidEndSepPunct{\mcitedefaultmidpunct}
{\mcitedefaultendpunct}{\mcitedefaultseppunct}\relax
\EndOfBibitem
\bibitem[Papadopoulos(2019)]{Papadopoulos2019}
Papadopoulos,~S. Working towards a Massively Parallel Equation-of-Motion
  Coupled-Cluster Implementation in DIRAC. M.Sc.\ thesis, Vrije Universiteit
  Amsterdam, 2019\relax
\mciteBstWouldAddEndPuncttrue
\mciteSetBstMidEndSepPunct{\mcitedefaultmidpunct}
{\mcitedefaultendpunct}{\mcitedefaultseppunct}\relax
\EndOfBibitem
\bibitem[Parrish \latin{et~al.}(2013)Parrish, Hohenstein, Schunck, Sherrill,
  and Martinez]{Parrish.Martinez.2013pxo}
Parrish,~R.~M.; Hohenstein,~E.~G.; Schunck,~N.~F.; Sherrill,~C.~D.;
  Martinez,~T.~J. {Exact Tensor Hypercontraction: A Universal Technique for the
  Resolution of Matrix Elements of Local Finite-Range N-Body Potentials in
  Many-Body Quantum Problems}. \emph{Physical Review Letters} \textbf{2013},
  \emph{111}, 132505\relax
\mciteBstWouldAddEndPuncttrue
\mciteSetBstMidEndSepPunct{\mcitedefaultmidpunct}
{\mcitedefaultendpunct}{\mcitedefaultseppunct}\relax
\EndOfBibitem
\bibitem[Peng and Kowalski(2017)Peng, and Kowalski]{Peng2017a}
Peng,~B.; Kowalski,~K. Low-rank factorization of electron integral tensors and
  its application in electronic structure theory. \emph{Chemical Physics
  Letters} \textbf{2017}, \emph{672}, 47--53\relax
\mciteBstWouldAddEndPuncttrue
\mciteSetBstMidEndSepPunct{\mcitedefaultmidpunct}
{\mcitedefaultendpunct}{\mcitedefaultseppunct}\relax
\EndOfBibitem
\bibitem[Almlöf(1991)]{Almlof1991}
Almlöf,~J. {Elimination of energy denominators in Møller—Plesset
  perturbation theory by a Laplace transform approach}. \emph{Chemical Physics
  Letters} \textbf{1991}, \emph{181}, 319 -- 320\relax
\mciteBstWouldAddEndPuncttrue
\mciteSetBstMidEndSepPunct{\mcitedefaultmidpunct}
{\mcitedefaultendpunct}{\mcitedefaultseppunct}\relax
\EndOfBibitem
\bibitem[Visscher and Dyall(1997)Visscher, and Dyall]{Visscher1997a}
Visscher,~L.; Dyall,~K.~G. Dirac-Fock atomic electronic structure calculations
  using different nuclear charge distributions. \emph{Atomic Data and Nuclear
  Data Tables} \textbf{1997}, \emph{67}, 207--224\relax
\mciteBstWouldAddEndPuncttrue
\mciteSetBstMidEndSepPunct{\mcitedefaultmidpunct}
{\mcitedefaultendpunct}{\mcitedefaultseppunct}\relax
\EndOfBibitem
\bibitem[Iliaš and Saue(2007)Iliaš, and Saue]{Ilias2007}
Iliaš,~M.; Saue,~T. An infinite-order two-component relativistic Hamiltonian
  by a simple one-step transformation. \emph{J. Chem. Phys.} \textbf{2007},
  \emph{126}, 064102\relax
\mciteBstWouldAddEndPuncttrue
\mciteSetBstMidEndSepPunct{\mcitedefaultmidpunct}
{\mcitedefaultendpunct}{\mcitedefaultseppunct}\relax
\EndOfBibitem
\bibitem[Konecny \latin{et~al.}(2016)Konecny, Kadek, Komorovsky, Malkina, Ruud,
  and Repisky]{Konecny2016}
Konecny,~L.; Kadek,~M.; Komorovsky,~S.; Malkina,~O.~L.; Ruud,~K.; Repisky,~M.
  Acceleration of Relativistic Electron Dynamics by Means of X2C
  Transformation: Application to the Calculation of Nonlinear Optical
  Properties. \emph{Journal of Chemical Theory and Computation} \textbf{2016},
  \emph{12}, 5823--5833\relax
\mciteBstWouldAddEndPuncttrue
\mciteSetBstMidEndSepPunct{\mcitedefaultmidpunct}
{\mcitedefaultendpunct}{\mcitedefaultseppunct}\relax
\EndOfBibitem
\bibitem[Schimmelpfennig()]{prog:amfi}
Schimmelpfennig,~B. {\textit{AMFI, an atomic mean-field spin-orbit integral
  program}}. {University of Stockholm, Stockholm, Sweden; 1999}\relax
\mciteBstWouldAddEndPuncttrue
\mciteSetBstMidEndSepPunct{\mcitedefaultmidpunct}
{\mcitedefaultendpunct}{\mcitedefaultseppunct}\relax
\EndOfBibitem
\bibitem[Kramers(1930)]{Kramers1930}
Kramers,~H.~A. Th{\'e}orie g{\'e}n{\'e}rale de la rotation paramagn{\'e}tique
  dans les cristaux. \emph{Proceedings of the Royal Netherlands Academy of Arts
  and Sciences} \textbf{1930}, \emph{33}, 959--972\relax
\mciteBstWouldAddEndPuncttrue
\mciteSetBstMidEndSepPunct{\mcitedefaultmidpunct}
{\mcitedefaultendpunct}{\mcitedefaultseppunct}\relax
\EndOfBibitem
\bibitem[Saue and Jensen(1999)Saue, and Jensen]{Saue1999}
Saue,~T.; Jensen,~H. J.~{\relax Aa}. Quaternion symmetry in relativistic
  molecular calculations: The Dirac-Hartree-Fock method. \emph{Journal of
  Chemical Physics} \textbf{1999}, \emph{111}, 6211--6222\relax
\mciteBstWouldAddEndPuncttrue
\mciteSetBstMidEndSepPunct{\mcitedefaultmidpunct}
{\mcitedefaultendpunct}{\mcitedefaultseppunct}\relax
\EndOfBibitem
\bibitem[Repisky \latin{et~al.}(2020)Repisky, Komorovsky, Kadek, Konecny,
  Ekstr{\"o}m, Malkin, Kaupp, Ruud, Malkina, and Malkin]{Repisky2020}
Repisky,~M.; Komorovsky,~S.; Kadek,~M.; Konecny,~L.; Ekstr{\"o}m,~U.;
  Malkin,~E.; Kaupp,~M.; Ruud,~K.; Malkina,~O.~L.; Malkin,~V.~G. ReSpect:
  Relativistic spectroscopy DFT program package. \emph{J. Chem. Phys.}
  \textbf{2020}, \emph{152}, 184101\relax
\mciteBstWouldAddEndPuncttrue
\mciteSetBstMidEndSepPunct{\mcitedefaultmidpunct}
{\mcitedefaultendpunct}{\mcitedefaultseppunct}\relax
\EndOfBibitem
\bibitem[Visscher \latin{et~al.}(1995)Visscher, Dyall, and Lee]{Visscher1995}
Visscher,~L.; Dyall,~K.~G.; Lee,~T.~J. Kramers-restricted closed-shell CCSD
  theory. \emph{Int. J. Quantum Chem.} \textbf{1995}, \emph{56}, 411--419\relax
\mciteBstWouldAddEndPuncttrue
\mciteSetBstMidEndSepPunct{\mcitedefaultmidpunct}
{\mcitedefaultendpunct}{\mcitedefaultseppunct}\relax
\EndOfBibitem
\bibitem[Pulay(1980)]{Pulay1980}
Pulay,~P. Convergence acceleration of iterative sequences. the case of scf
  iteration. \emph{Chemical Physics Letters} \textbf{1980}, \emph{73},
  393--398\relax
\mciteBstWouldAddEndPuncttrue
\mciteSetBstMidEndSepPunct{\mcitedefaultmidpunct}
{\mcitedefaultendpunct}{\mcitedefaultseppunct}\relax
\EndOfBibitem
\bibitem[Raghavachari \latin{et~al.}(1989)Raghavachari, Trucks, Pople, and
  Head-Gordon]{Raghavachari1989}
Raghavachari,~K.; Trucks,~G.~W.; Pople,~J.~A.; Head-Gordon,~M. A fifth-order
  perturbation comparison of electron correlation theories. \emph{Chemical
  Physics Letters} \textbf{1989}, \emph{157}, 479--483\relax
\mciteBstWouldAddEndPuncttrue
\mciteSetBstMidEndSepPunct{\mcitedefaultmidpunct}
{\mcitedefaultendpunct}{\mcitedefaultseppunct}\relax
\EndOfBibitem
\bibitem[Deegan and Knowles(1994)Deegan, and Knowles]{Deegan1994}
Deegan,~M. J.~O.; Knowles,~P.~J. Perturbative corrections to account for triple
  excitations in closed and open shell coupled cluster theories. \emph{Chemical
  Physics Letters} \textbf{1994}, \emph{227}, 321--326\relax
\mciteBstWouldAddEndPuncttrue
\mciteSetBstMidEndSepPunct{\mcitedefaultmidpunct}
{\mcitedefaultendpunct}{\mcitedefaultseppunct}\relax
\EndOfBibitem
\bibitem[Shee \latin{et~al.}(2016)Shee, Visscher, and Saue]{shee2016analytic}
Shee,~A.; Visscher,~L.; Saue,~T. Analytic one-electron properties at the
  4-component relativistic coupled cluster level with inclusion of spin-orbit
  coupling. \emph{J. Chem. Phys.} \textbf{2016}, \emph{145}, 184107\relax
\mciteBstWouldAddEndPuncttrue
\mciteSetBstMidEndSepPunct{\mcitedefaultmidpunct}
{\mcitedefaultendpunct}{\mcitedefaultseppunct}\relax
\EndOfBibitem
\bibitem[Wigner(1959)]{Wigner}
Wigner,~E.~P. \emph{Group Theory and its Application to the Quantum Mechanics
  of Atomic Spectra}; Academic, New York, 1959\relax
\mciteBstWouldAddEndPuncttrue
\mciteSetBstMidEndSepPunct{\mcitedefaultmidpunct}
{\mcitedefaultendpunct}{\mcitedefaultseppunct}\relax
\EndOfBibitem
\bibitem[Saue(1996. May be retrieved from http://diracprogram.org)]{Saue1996}
Saue,~T. Principles and Applications of Relativistic Molecular Calculations.
  Ph.D.\ thesis, University of Oslo, 1996. May be retrieved from
  http://diracprogram.org\relax
\mciteBstWouldAddEndPuncttrue
\mciteSetBstMidEndSepPunct{\mcitedefaultmidpunct}
{\mcitedefaultendpunct}{\mcitedefaultseppunct}\relax
\EndOfBibitem
\bibitem[dal()]{daltonprogram}
DALTON, a molecular electronic structure program, see
  http://daltonprogram.org.\relax
\mciteBstWouldAddEndPunctfalse
\mciteSetBstMidEndSepPunct{\mcitedefaultmidpunct}
{}{\mcitedefaultseppunct}\relax
\EndOfBibitem
\bibitem[Shiozaki(2017)]{Shiozaki2017}
Shiozaki,~T. An efficient solver for large structured eigenvalue problems in
  relativistic quantum chemistry. \emph{Molecular Physics} \textbf{2017},
  \emph{115}, 5--12\relax
\mciteBstWouldAddEndPuncttrue
\mciteSetBstMidEndSepPunct{\mcitedefaultmidpunct}
{\mcitedefaultendpunct}{\mcitedefaultseppunct}\relax
\EndOfBibitem
\bibitem[Repisky(2018)]{Repisky2018}
Repisky,~M. InteRest, An integral library for relativistic quantum chemistry.
  2018\relax
\mciteBstWouldAddEndPuncttrue
\mciteSetBstMidEndSepPunct{\mcitedefaultmidpunct}
{\mcitedefaultendpunct}{\mcitedefaultseppunct}\relax
\EndOfBibitem
\bibitem[Stanton and Havriliak(1984)Stanton, and Havriliak]{Stanton1984}
Stanton,~R.~E.; Havriliak,~S. Kinetic balance: A partial solution to the
  problem of variational safety in Dirac calculations. \emph{J. Chem. Phys.}
  \textbf{1984}, \emph{81}, 1910--1918\relax
\mciteBstWouldAddEndPuncttrue
\mciteSetBstMidEndSepPunct{\mcitedefaultmidpunct}
{\mcitedefaultendpunct}{\mcitedefaultseppunct}\relax
\EndOfBibitem
\bibitem[Obara and Saika(1986)Obara, and Saika]{Obara1986}
Obara,~S.; Saika,~A. Efficient recursive computation of molecular integrals
  over Cartesian Gaussian functions. \emph{J. Chem. Phys.} \textbf{1986},
  \emph{84}, 3963--3974\relax
\mciteBstWouldAddEndPuncttrue
\mciteSetBstMidEndSepPunct{\mcitedefaultmidpunct}
{\mcitedefaultendpunct}{\mcitedefaultseppunct}\relax
\EndOfBibitem
\bibitem[Kadek \latin{et~al.}(2019)Kadek, Repisky, and Ruud]{Kadek2019}
Kadek,~M.; Repisky,~M.; Ruud,~K. All-electron fully relativistic Kohn-Sham
  theory for solids based on the Dirac-Coulomb Hamiltonian and Gaussian-type
  functions. \emph{Phys. Rev. B} \textbf{2019}, \emph{99}, 205103\relax
\mciteBstWouldAddEndPuncttrue
\mciteSetBstMidEndSepPunct{\mcitedefaultmidpunct}
{\mcitedefaultendpunct}{\mcitedefaultseppunct}\relax
\EndOfBibitem
\bibitem[{Komorovsk\'{y}} \latin{et~al.}(2008){Komorovsk\'{y}}, {Repisk\'{y}},
  Malkina, Malkin, Malkin~{Ond\'{i}}k, and Kaupp]{Komorovsky2008}
{Komorovsk\'{y}},~S.; {Repisk\'{y}},~M.; Malkina,~O.~L.; Malkin,~V.~G.;
  Malkin~{Ond\'{i}}k,~I.; Kaupp,~M. A fully relativistic method for calculation
  of nuclear magnetic shielding tensors with a restricted magnetically balanced
  basis in the framework of the matrix Dirac-Kohn-Sham equation. \emph{J. Chem.
  Phys.} \textbf{2008}, \emph{128}, 104101\relax
\mciteBstWouldAddEndPuncttrue
\mciteSetBstMidEndSepPunct{\mcitedefaultmidpunct}
{\mcitedefaultendpunct}{\mcitedefaultseppunct}\relax
\EndOfBibitem
\bibitem[{Komorovsk\'{y}} \latin{et~al.}(2010){Komorovsk\'{y}}, {Repisk\'{y}},
  Malkina, and Malkin]{Komorovsky2010}
{Komorovsk\'{y}},~S.; {Repisk\'{y}},~M.; Malkina,~O.~L.; Malkin,~V.~G. Fully
  relativistic calculations of NMR shielding tensors using restricted
  magnetically balanced basis and gauge including atomic orbitals. \emph{J.
  Chem. Phys.} \textbf{2010}, \emph{132}, 154101\relax
\mciteBstWouldAddEndPuncttrue
\mciteSetBstMidEndSepPunct{\mcitedefaultmidpunct}
{\mcitedefaultendpunct}{\mcitedefaultseppunct}\relax
\EndOfBibitem
\bibitem[Lyakh()]{TALSH}
Lyakh,~D.~I. TAL-SH: Numerical tensor algebra library for shared-memory
  heterogeneous nodes equipped with multicore CPU and NVIDIA GPU.
  \url{https://github.com/DmitryLyakh/TAL_SH}\relax
\mciteBstWouldAddEndPuncttrue
\mciteSetBstMidEndSepPunct{\mcitedefaultmidpunct}
{\mcitedefaultendpunct}{\mcitedefaultseppunct}\relax
\EndOfBibitem
\bibitem[Thyssen(2001)]{Thyssen2001}
Thyssen,~J. Development and Applications of Methods for Correlated Relativistic
  Calculations of Molecular Properties. Ph.D.\ thesis, University of Southern
  Denmark, 2001\relax
\mciteBstWouldAddEndPuncttrue
\mciteSetBstMidEndSepPunct{\mcitedefaultmidpunct}
{\mcitedefaultendpunct}{\mcitedefaultseppunct}\relax
\EndOfBibitem
\bibitem[Yoshimine(1973)]{Yoshimine1973}
Yoshimine,~M. Construction of the hamiltonian matrix in large configuration
  interaction calculations. \emph{Journal of Computational Physics}
  \textbf{1973}, \emph{11}, 449--454\relax
\mciteBstWouldAddEndPuncttrue
\mciteSetBstMidEndSepPunct{\mcitedefaultmidpunct}
{\mcitedefaultendpunct}{\mcitedefaultseppunct}\relax
\EndOfBibitem
\bibitem[{Zi\'o\l{}owski} \latin{et~al.}(2008){Zi\'o\l{}owski}, Weijo,
  {J\o{}rgensen}, and Olsen]{Ziolkowski2008}
{Zi\'o\l{}owski},~M.; Weijo,~V.; {J\o{}rgensen},~P.; Olsen,~J. An efficient
  algorithm for solving nonlinear equations with a minimal number of trial
  vectors: Applications to atomic-orbital based coupled-cluster theory.
  \emph{J. Chem. Phys.} \textbf{2008}, \emph{128}, 204105\relax
\mciteBstWouldAddEndPuncttrue
\mciteSetBstMidEndSepPunct{\mcitedefaultmidpunct}
{\mcitedefaultendpunct}{\mcitedefaultseppunct}\relax
\EndOfBibitem
\bibitem[Pototschnig \latin{et~al.}(2021)Pototschnig, Papadopoulos, Lyakh,
  Repisky, Halbert, Gomes, Jensen, and Visscher]{pototschnig:2021:dataset}
Pototschnig,~J.~V.; Papadopoulos,~A.; Lyakh,~D.~I.; Repisky,~M.; Halbert,~L.;
  Gomes,~A. S.~P.; Jensen,~H. J.~A.; Visscher,~L. Dataset: Implementation of
  relativistic coupled cluster theory for massively parallel GPU-accelerated
  computing architectures. 2021;
  \url{https://doi.org/10.5281/zenodo.4589358}\relax
\mciteBstWouldAddEndPuncttrue
\mciteSetBstMidEndSepPunct{\mcitedefaultmidpunct}
{\mcitedefaultendpunct}{\mcitedefaultseppunct}\relax
\EndOfBibitem
\bibitem[DIR()]{DIRAC19}
{DIRAC}, a relativistic ab initio electronic structure program, Release
  {DIRAC19} (2019), written by A.~S.~P.~Gomes, T.~Saue, L.~Visscher,
  H.~J.~{\relax Aa}.~Jensen, and R.~Bast, with contributions from I.~A.~Aucar,
  V.~Bakken, K.~G.~Dyall, S.~Dubillard, U.~Ekstr{\"o}m, E.~Eliav,
  T.~Enevoldsen, E.~Fa{\ss}hauer, T.~Fleig, O.~Fossgaard, L.~Halbert,
  E.~D.~Hedeg{\aa}rd, B.~Heimlich--Paris, T.~Helgaker, J.~Henriksson,
  M.~Ilia{\v{s}}, Ch.~R.~Jacob, S.~Knecht, S.~Komorovsk{\'y}, O.~Kullie,
  J.~K.~L{\ae}rdahl, C.~V.~Larsen, Y.~S.~Lee, H.~S.~Nataraj, M.~K.~Nayak,
  P.~Norman, G.~Olejniczak, J.~Olsen, J.~M.~H.~Olsen, Y.~C.~Park,
  J.~K.~Pedersen, M.~Pernpointner, R.~di~Remigio, K.~Ruud, P.~Sa{\l}ek,
  B.~Schimmelpfennig, B.~Senjean, A.~Shee, J.~Sikkema, A.~J.~Thorvaldsen,
  J.~Thyssen, J.~van~Stralen, M.~L.~Vidal, S.~Villaume, O.~Visser, T.~Winther,
  and S.~Yamamoto (available at \url{http://dx.doi.org/10.5281/zenodo.3572669},
  see also \url{http://www.diracprogram.org})\relax
\mciteBstWouldAddEndPuncttrue
\mciteSetBstMidEndSepPunct{\mcitedefaultmidpunct}
{\mcitedefaultendpunct}{\mcitedefaultseppunct}\relax
\EndOfBibitem
\bibitem[Saue \latin{et~al.}(2020)Saue, Bast, Gomes, Jensen, Visscher, Aucar,
  Di~Remigio, Dyall, Eliav, Fasshauer, Fleig, Halbert, Hedeg{\aa}rd,
  Helmich-Paris, Ilia{\v{s}}, Jacob, Knecht, Laerdahl, Vidal, Nayak,
  Olejniczak, Olsen, Pernpointner, Senjean, Shee, Sunaga, and van
  Stralen]{Saue2020}
Saue,~T.; Bast,~R.; Gomes,~A. S.~P.; Jensen,~H. J.~{\relax Aa}.; Visscher,~L.;
  Aucar,~I.~A.; Di~Remigio,~R.; Dyall,~K.~G.; Eliav,~E.; Fasshauer,~E.;
  Fleig,~T.; Halbert,~L.; Hedeg{\aa}rd,~E.~D.; Helmich-Paris,~B.;
  Ilia{\v{s}},~M.; Jacob,~C.~R.; Knecht,~S.; Laerdahl,~J.~K.; Vidal,~M.~L.;
  Nayak,~M.~K.; Olejniczak,~M.; Olsen,~J. M.~H.; Pernpointner,~M.; Senjean,~B.;
  Shee,~A.; Sunaga,~A.; van Stralen,~J. N.~P. The DIRAC code for relativistic
  molecular calculations. \emph{J. Chem. Phys.} \textbf{2020}, \emph{152},
  204104\relax
\mciteBstWouldAddEndPuncttrue
\mciteSetBstMidEndSepPunct{\mcitedefaultmidpunct}
{\mcitedefaultendpunct}{\mcitedefaultseppunct}\relax
\EndOfBibitem
\bibitem[Dyall(2006)]{Dyall2006}
Dyall,~K.~G. Relativistic Quadruple-Zeta and Revised Triple-Zeta and
  Double-Zeta Basis Sets for the 4p, 5p, and 6p Elements. \emph{Theoretical
  Chemistry Accounts} \textbf{2006}, \emph{115}, 441--447\relax
\mciteBstWouldAddEndPuncttrue
\mciteSetBstMidEndSepPunct{\mcitedefaultmidpunct}
{\mcitedefaultendpunct}{\mcitedefaultseppunct}\relax
\EndOfBibitem
\bibitem[Dyall(2007)]{Dyall2007}
Dyall,~K.~G. Relativistic double-zeta, triple-zeta, and quadruple-zeta basis
  sets for the actinides Ac-Lr. \emph{Theoretical Chemistry Accounts}
  \textbf{2007}, \emph{117}, 491--500\relax
\mciteBstWouldAddEndPuncttrue
\mciteSetBstMidEndSepPunct{\mcitedefaultmidpunct}
{\mcitedefaultendpunct}{\mcitedefaultseppunct}\relax
\EndOfBibitem
\bibitem[Gomes \latin{et~al.}(2010)Gomes, Dyall, and Visscher]{Gomes2010}
Gomes,~A. S.~P.; Dyall,~K.~G.; Visscher,~L. Relativistic double-zeta,
  triple-zeta, and quadruple-zeta basis sets for the lanthanides La-Lu.
  \emph{Theoretical Chemistry Accounts} \textbf{2010}, \emph{127},
  369--381\relax
\mciteBstWouldAddEndPuncttrue
\mciteSetBstMidEndSepPunct{\mcitedefaultmidpunct}
{\mcitedefaultendpunct}{\mcitedefaultseppunct}\relax
\EndOfBibitem
\bibitem[ADF()]{ADF2019}
{ADF2019, SCM, Theoretical Chemistry, Vrije Universiteit, Amsterdam, The
  Netherlands, https://www.scm.com}\relax
\mciteBstWouldAddEndPuncttrue
\mciteSetBstMidEndSepPunct{\mcitedefaultmidpunct}
{\mcitedefaultendpunct}{\mcitedefaultseppunct}\relax
\EndOfBibitem
\bibitem[van Lenthe \latin{et~al.}(1999)van Lenthe, Ehlers, and
  Baerends]{Lenthe1999}
van Lenthe,~E.; Ehlers,~A.; Baerends,~E.-J. Geometry optimizations in the zero
  order regular approximation for relativistic effects. \emph{J. Chem. Phys.}
  \textbf{1999}, \emph{110}, 8943--8953\relax
\mciteBstWouldAddEndPuncttrue
\mciteSetBstMidEndSepPunct{\mcitedefaultmidpunct}
{\mcitedefaultendpunct}{\mcitedefaultseppunct}\relax
\EndOfBibitem
\bibitem[Perdew \latin{et~al.}(1996)Perdew, Burke, and Ernzerhof]{Perdew1996}
Perdew,~J.~P.; Burke,~K.; Ernzerhof,~M. Generalized Gradient Approximation Made
  Simple. \emph{Phys. Rev. Lett.} \textbf{1996}, \emph{77}, 3865--3868\relax
\mciteBstWouldAddEndPuncttrue
\mciteSetBstMidEndSepPunct{\mcitedefaultmidpunct}
{\mcitedefaultendpunct}{\mcitedefaultseppunct}\relax
\EndOfBibitem
\bibitem[Pototschnig \latin{et~al.}(2021)Pototschnig, Visscher, and
  Jamshidi]{Nazanin2021}
Pototschnig,~J.~V.; Visscher,~L.; Jamshidi,~Z. Bonding of argon to gold
  cluster. \textbf{2021}, in preparation\relax
\mciteBstWouldAddEndPuncttrue
\mciteSetBstMidEndSepPunct{\mcitedefaultmidpunct}
{\mcitedefaultendpunct}{\mcitedefaultseppunct}\relax
\EndOfBibitem
\bibitem[Kimura \latin{et~al.}(1968)Kimura, Schomaker, Smith, and
  Weinstock]{Kimura1968}
Kimura,~M.; Schomaker,~V.; Smith,~D.~W.; Weinstock,~B. Electron‐Diffraction
  Investigation of the Hexafluorides of Tungsten, Osmium, Iridium, Uranium,
  Neptunium, and Plutonium. \emph{J. Chem. Phys.} \textbf{1968}, \emph{48},
  4001--4012\relax
\mciteBstWouldAddEndPuncttrue
\mciteSetBstMidEndSepPunct{\mcitedefaultmidpunct}
{\mcitedefaultendpunct}{\mcitedefaultseppunct}\relax
\EndOfBibitem
\bibitem[Barclay \latin{et~al.}(1965)Barclay, Sabine, and
  Taylor]{Barclay:a04757}
Barclay,~G.~A.; Sabine,~T.~M.; Taylor,~J.~C. {The crystal structure of rubidium
  uranyl nitrate: A neutron-diffraction study}. \emph{Acta Crystallographica}
  \textbf{1965}, \emph{19}, 205--209\relax
\mciteBstWouldAddEndPuncttrue
\mciteSetBstMidEndSepPunct{\mcitedefaultmidpunct}
{\mcitedefaultendpunct}{\mcitedefaultseppunct}\relax
\EndOfBibitem
\bibitem[Gomes \latin{et~al.}(2013)Gomes, Jacob, Réal, Visscher, and
  Vallet]{Gomes2013}
Gomes,~A. S.~P.; Jacob,~C.~R.; Réal,~F.; Visscher,~L.; Vallet,~V. Towards
  systematically improvable models for actinides in condensed phase: the
  electronic spectrum of uranyl in {Cs$_2$UO$_2$Cl$_4$} as a test case.
  \emph{Phys. Chem. Chem. Phys.} \textbf{2013}, \emph{15}, 15153--15162\relax
\mciteBstWouldAddEndPuncttrue
\mciteSetBstMidEndSepPunct{\mcitedefaultmidpunct}
{\mcitedefaultendpunct}{\mcitedefaultseppunct}\relax
\EndOfBibitem
\bibitem[Aebersold \latin{et~al.}(2017)Aebersold, Yuwono, Schoendorff, and
  Wilson]{Aebersold2017}
Aebersold,~L.~E.; Yuwono,~S.~H.; Schoendorff,~G.; Wilson,~A.~K. Efficacy of
  Density Functionals and Relativistic Effective Core Potentials for
  Lanthanide-Containing Species: The Ln54 Molecule Set. \emph{Journal of
  Chemical Theory and Computation} \textbf{2017}, \emph{13}, 2831--2839\relax
\mciteBstWouldAddEndPuncttrue
\mciteSetBstMidEndSepPunct{\mcitedefaultmidpunct}
{\mcitedefaultendpunct}{\mcitedefaultseppunct}\relax
\EndOfBibitem
\bibitem[Sekiya \latin{et~al.}(2012)Sekiya, Noro, Koga, and
  Shimazaki]{Sekiya2012}
Sekiya,~M.; Noro,~T.; Koga,~T.; Shimazaki,~T. Relativistic segmented
  contraction basis sets with core-valence correlation effects for atoms La-57
  through Lu-71: Sapporo-DK-nZP sets (n = D, T, Q). \emph{Theoretical Chemistry
  Accounts} \textbf{2012}, \emph{131}, 1247\relax
\mciteBstWouldAddEndPuncttrue
\mciteSetBstMidEndSepPunct{\mcitedefaultmidpunct}
{\mcitedefaultendpunct}{\mcitedefaultseppunct}\relax
\EndOfBibitem
\bibitem[Solomonik and Smirnov(2017)Solomonik, and Smirnov]{Solomonik2017}
Solomonik,~V.~G.; Smirnov,~A.~N. Toward Chemical Accuracy in ab Initio
  Thermochemistry and Spectroscopy of Lanthanide Compounds: Assessing
  Core-Valence Correlation, Second-Order Spin-Orbit Coupling, and Higher Order
  Effects in Lanthanide Diatomics. \emph{J. Chem. Theory Comput.}
  \textbf{2017}, \emph{13}, 5240--5254\relax
\mciteBstWouldAddEndPuncttrue
\mciteSetBstMidEndSepPunct{\mcitedefaultmidpunct}
{\mcitedefaultendpunct}{\mcitedefaultseppunct}\relax
\EndOfBibitem
\bibitem[Bernard \latin{et~al.}(2000)Bernard, Effantin, d'Incan, and
  Verges]{Bernard2000a}
Bernard,~A.; Effantin,~C.; d'Incan,~J.; Verges,~J. The {1 $^1\Pi$, 2
  $^1\Sigma^+$ $->$ X $^1\Sigma^+$ } transitions of LaF. \emph{Journal of
  Molecular Spectroscopy} \textbf{2000}, \emph{202}, 163--165\relax
\mciteBstWouldAddEndPuncttrue
\mciteSetBstMidEndSepPunct{\mcitedefaultmidpunct}
{\mcitedefaultendpunct}{\mcitedefaultseppunct}\relax
\EndOfBibitem
\bibitem[Dmitriev \latin{et~al.}(1987)Dmitriev, Kaledin, Kobyliansky, Kulikov,
  Shenyavskaya, and Gurvich]{Dmitriev1987}
Dmitriev,~Y.~N.; Kaledin,~L.~A.; Kobyliansky,~A.~I.; Kulikov,~A.~N.;
  Shenyavskaya,~E.~A.; Gurvich,~L.~V. Electronic spectra of diatomic molecules
  containing f-elements: GdO, EuF and UO. \emph{Acta Physica Hungarica}
  \textbf{1987}, \emph{61}, 51--54\relax
\mciteBstWouldAddEndPuncttrue
\mciteSetBstMidEndSepPunct{\mcitedefaultmidpunct}
{\mcitedefaultendpunct}{\mcitedefaultseppunct}\relax
\EndOfBibitem
\bibitem[Dickinson \latin{et~al.}(2001)Dickinson, Coxon, Walker, and
  Gerry]{Dickinson2001}
Dickinson,~C.~S.; Coxon,~J.~A.; Walker,~N.~R.; Gerry,~M. C.~L. Fourier
  transform microwave spectroscopy of the {$^2\Sigma^+$} ground states of YbX
  (X=F, Cl, Br): Characterization of hyperfine effects and determination of the
  molecular geometries. \emph{Journal of Chemical Physics} \textbf{2001},
  \emph{115}, 6979--6989\relax
\mciteBstWouldAddEndPuncttrue
\mciteSetBstMidEndSepPunct{\mcitedefaultmidpunct}
{\mcitedefaultendpunct}{\mcitedefaultseppunct}\relax
\EndOfBibitem
\bibitem[Kaledin \latin{et~al.}(1999)Kaledin, Heaven, and Field]{Kaledin1999}
Kaledin,~L.~A.; Heaven,~M.~C.; Field,~R.~W. Thermochemical properties
  (D{$^0_O$} and IP) of the lanthanide monohalides. \emph{Journal of Molecular
  Spectroscopy} \textbf{1999}, \emph{193}, 285--292\relax
\mciteBstWouldAddEndPuncttrue
\mciteSetBstMidEndSepPunct{\mcitedefaultmidpunct}
{\mcitedefaultendpunct}{\mcitedefaultseppunct}\relax
\EndOfBibitem
\bibitem[Zmbov and Margrave(1968)Zmbov, and Margrave]{ZMBOV1968a}
Zmbov,~K.~F.; Margrave,~J.~L. Mass Spectrometric Studies of Scandium Yttrium
  Lanthanum and Rare-earth Fluorides. \emph{Advances in Chemistry Series}
  \textbf{1968}, 267--\&\relax
\mciteBstWouldAddEndPuncttrue
\mciteSetBstMidEndSepPunct{\mcitedefaultmidpunct}
{\mcitedefaultendpunct}{\mcitedefaultseppunct}\relax
\EndOfBibitem
\bibitem[Zmbov and Margrave(1967)Zmbov, and Margrave]{ZMBOV1967}
Zmbov,~K.~F.; Margrave,~J.~L. Mass Spectrometric Studies at High Temperatures
  .13. Stabilities of Samarium Europium and Gadolinium Mono- and Difluorides.
  \emph{Journal of Inorganic \& Nuclear Chemistry} \textbf{1967}, \emph{29},
  59\relax
\mciteBstWouldAddEndPuncttrue
\mciteSetBstMidEndSepPunct{\mcitedefaultmidpunct}
{\mcitedefaultendpunct}{\mcitedefaultseppunct}\relax
\EndOfBibitem
\bibitem[Dzuba and Derevianko(2010)Dzuba, and Derevianko]{Dzuba2010}
Dzuba,~V.~A.; Derevianko,~A. Dynamic polarizabilities and related properties of
  clock states of the ytterbium atom. \emph{Journal of Physics B: Atomic,
  Molecular and Optical Physics} \textbf{2010}, \emph{43}, 074011\relax
\mciteBstWouldAddEndPuncttrue
\mciteSetBstMidEndSepPunct{\mcitedefaultmidpunct}
{\mcitedefaultendpunct}{\mcitedefaultseppunct}\relax
\EndOfBibitem
\bibitem[Pasteka \latin{et~al.}(2016)Pasteka, Mawhorter, and
  Schwerdtfeger]{Pasteka2016}
Pasteka,~L.~F.; Mawhorter,~R.~J.; Schwerdtfeger,~P. Relativistic
  coupled-cluster calculations of the {$^{173}$Yb} nuclear quadrupole coupling
  constant for the YbF molecule. \emph{Molecular Physics} \textbf{2016},
  \emph{114}, 1110--1117\relax
\mciteBstWouldAddEndPuncttrue
\mciteSetBstMidEndSepPunct{\mcitedefaultmidpunct}
{\mcitedefaultendpunct}{\mcitedefaultseppunct}\relax
\EndOfBibitem
\bibitem[Lyakh \latin{et~al.}(2012)Lyakh, Musiał, Lotrich, and
  Bartlett]{Lyakh2012}
Lyakh,~D.~I.; Musiał,~M.; Lotrich,~V.~F.; Bartlett,~R.~J. Multireference
  Nature of Chemistry: The Coupled-Cluster View. \emph{Chem. Rev.}
  \textbf{2012}, \emph{112}, 182--243\relax
\mciteBstWouldAddEndPuncttrue
\mciteSetBstMidEndSepPunct{\mcitedefaultmidpunct}
{\mcitedefaultendpunct}{\mcitedefaultseppunct}\relax
\EndOfBibitem
\bibitem[Kapur and {M\"uller}(1977)Kapur, and {M\"uller}]{Kapur1977}
Kapur,~S.; {M\"uller},~E.~W. Metal-neon compound ions in slow field
  evaporation. \emph{Surface Science} \textbf{1977}, \emph{62}, 610--620\relax
\mciteBstWouldAddEndPuncttrue
\mciteSetBstMidEndSepPunct{\mcitedefaultmidpunct}
{\mcitedefaultendpunct}{\mcitedefaultseppunct}\relax
\EndOfBibitem
\bibitem[{Pyykk\"o}(1995)]{Pyykkoe1995}
{Pyykk\"o},~P. Predicted Chemical Bonds between Rare Gases and Au+. \emph{J.
  Am. Chem. Soc.} \textbf{1995}, \emph{117}, 2067--2070\relax
\mciteBstWouldAddEndPuncttrue
\mciteSetBstMidEndSepPunct{\mcitedefaultmidpunct}
{\mcitedefaultendpunct}{\mcitedefaultseppunct}\relax
\EndOfBibitem
\bibitem[Shayeghi \latin{et~al.}(2015)Shayeghi, Johnston, Rayner, Sch{\"a}fer,
  and Fielicke]{Shayeghi2015}
Shayeghi,~A.; Johnston,~R.~L.; Rayner,~D.~M.; Sch{\"a}fer,~R.; Fielicke,~A. The
  Nature of Bonding between Argon and Mixed Gold-Silver Trimers. \emph{Angew.
  Chem. Int. Ed.} \textbf{2015}, \emph{54}, 10675--10680\relax
\mciteBstWouldAddEndPuncttrue
\mciteSetBstMidEndSepPunct{\mcitedefaultmidpunct}
{\mcitedefaultendpunct}{\mcitedefaultseppunct}\relax
\EndOfBibitem
\bibitem[Ghiringhelli and Levchenko(2015)Ghiringhelli, and
  Levchenko]{Ghiringhelli2015}
Ghiringhelli,~L.~M.; Levchenko,~S.~V. Strengthening gold-gold bonds by
  complexing gold clusters with noble gases. \emph{Inorganic Chemistry
  Communications} \textbf{2015}, \emph{55}, 153--156\relax
\mciteBstWouldAddEndPuncttrue
\mciteSetBstMidEndSepPunct{\mcitedefaultmidpunct}
{\mcitedefaultendpunct}{\mcitedefaultseppunct}\relax
\EndOfBibitem
\bibitem[Ferrari \latin{et~al.}(2020)Ferrari, Hou, Lushchikova, Calvo, Bakker,
  and Janssens]{Ferrari2020}
Ferrari,~P.; Hou,~G.-L.; Lushchikova,~O.~V.; Calvo,~F.; Bakker,~J.~M.;
  Janssens,~E. The structures of cationic gold clusters probed by far-infrared
  spectroscopy. \emph{Phys. Chem. Chem. Phys.} \textbf{2020}, \emph{22},
  11572--11577\relax
\mciteBstWouldAddEndPuncttrue
\mciteSetBstMidEndSepPunct{\mcitedefaultmidpunct}
{\mcitedefaultendpunct}{\mcitedefaultseppunct}\relax
\EndOfBibitem
\bibitem[Pan \latin{et~al.}(2019)Pan, Jana, Merino, and Chattaraj]{Pan2019}
Pan,~S.; Jana,~G.; Merino,~G.; Chattaraj,~P.~K. Noble-Noble Strong Union: Gold
  at Its Best to Make a Bond with a Noble Gas Atom. \emph{ChemistryOpen}
  \textbf{2019}, \emph{8}, 173--187\relax
\mciteBstWouldAddEndPuncttrue
\mciteSetBstMidEndSepPunct{\mcitedefaultmidpunct}
{\mcitedefaultendpunct}{\mcitedefaultseppunct}\relax
\EndOfBibitem
\bibitem[Jamshidi \latin{et~al.}(2012)Jamshidi, Far, and
  Maghari]{Jamshidi.Maghari.20128qs}
Jamshidi,~Z.; Far,~M.~F.; Maghari,~A. {Binding of Noble Metal Clusters with
  Rare Gas Atoms: Theoretical Investigation}. \emph{The Journal of Physical
  Chemistry A} \textbf{2012}, \emph{116}, 12510 -- 12517\relax
\mciteBstWouldAddEndPuncttrue
\mciteSetBstMidEndSepPunct{\mcitedefaultmidpunct}
{\mcitedefaultendpunct}{\mcitedefaultseppunct}\relax
\EndOfBibitem
\bibitem[Jamshidi \latin{et~al.}(2020)Jamshidi, Lushchikova, Bakker, and
  Visscher]{10.1021/acs.jpca.0c07771}
Jamshidi,~Z.; Lushchikova,~O.~V.; Bakker,~J.~M.; Visscher,~L. {Not Completely
  Innocent: How Argon Binding Perturbs Cationic Copper Clusters}. \emph{The
  Journal of Physical Chemistry A} \textbf{2020}, \emph{124}, 9004--9010\relax
\mciteBstWouldAddEndPuncttrue
\mciteSetBstMidEndSepPunct{\mcitedefaultmidpunct}
{\mcitedefaultendpunct}{\mcitedefaultseppunct}\relax
\EndOfBibitem
\bibitem[Beu \latin{et~al.}(1997)Beu, Onoe, and Takeuchi]{Beu1997a}
Beu,~T.~A.; Onoe,~J.; Takeuchi,~K. Calculations of structure and IR-spectrum
  for small UF6 clusters. \emph{J. Chem. Phys.} \textbf{1997}, \emph{106},
  5910--5919\relax
\mciteBstWouldAddEndPuncttrue
\mciteSetBstMidEndSepPunct{\mcitedefaultmidpunct}
{\mcitedefaultendpunct}{\mcitedefaultseppunct}\relax
\EndOfBibitem
\bibitem[Beu \latin{et~al.}(1997)Beu, Onoe, and Takeuchi]{Beu1997}
Beu,~T.~A.; Onoe,~J.; Takeuchi,~K. Structure and frequency shift calculations
  for small UF6 clusters. \emph{Journal of Molecular Structure} \textbf{1997},
  \emph{410-411}, 295--298\relax
\mciteBstWouldAddEndPuncttrue
\mciteSetBstMidEndSepPunct{\mcitedefaultmidpunct}
{\mcitedefaultendpunct}{\mcitedefaultseppunct}\relax
\EndOfBibitem
\bibitem[Malli and Styszynski(1996)Malli, and Styszynski]{Malli1996}
Malli,~G.~L.; Styszynski,~J. Ab initio all‐electron Dirac-Fock-Breit
  calculations for {UF$_6$}. \emph{J. Chem. Phys.} \textbf{1996}, \emph{104},
  1012--1017\relax
\mciteBstWouldAddEndPuncttrue
\mciteSetBstMidEndSepPunct{\mcitedefaultmidpunct}
{\mcitedefaultendpunct}{\mcitedefaultseppunct}\relax
\EndOfBibitem
\bibitem[Kovacs and Konings(2004)Kovacs, and Konings]{Kovacs2004}
Kovacs,~A.; Konings,~R. J.~M. Theoretical study of {UX$_6$} and {UO$_2$X$_2$}
  (X=F, Cl, Br, I). \emph{Journal of Molecular Structure: THEOCHEM}
  \textbf{2004}, \emph{684}, 35--42\relax
\mciteBstWouldAddEndPuncttrue
\mciteSetBstMidEndSepPunct{\mcitedefaultmidpunct}
{\mcitedefaultendpunct}{\mcitedefaultseppunct}\relax
\EndOfBibitem
\bibitem[Batista \latin{et~al.}(2004)Batista, Martin, Hay, Peralta, and
  Scuseria]{Batista2004}
Batista,~E.~R.; Martin,~R.~L.; Hay,~P.~J.; Peralta,~J.~E.; Scuseria,~G.~E.
  Density functional investigations of the properties and thermochemistry of
  {UF$_6$} and {UF$_5$} using valence-electron and all-electron approaches.
  \emph{J. Chem. Phys.} \textbf{2004}, \emph{121}, 2144--2150\relax
\mciteBstWouldAddEndPuncttrue
\mciteSetBstMidEndSepPunct{\mcitedefaultmidpunct}
{\mcitedefaultendpunct}{\mcitedefaultseppunct}\relax
\EndOfBibitem
\bibitem[Manzhos \latin{et~al.}(2015)Manzhos, Carrington, Laverdure, and
  Mosey]{Manzhos2015}
Manzhos,~S.; Carrington,~T.; Laverdure,~L.; Mosey,~N. Computing the Anharmonic
  Vibrational Spectrum of {UF$_6$} in 15 Dimensions with an Optimized Basis Set
  and Rectangular Collocation. \emph{J. Phys. Chem. A} \textbf{2015},
  \emph{119}, 9557--9567\relax
\mciteBstWouldAddEndPuncttrue
\mciteSetBstMidEndSepPunct{\mcitedefaultmidpunct}
{\mcitedefaultendpunct}{\mcitedefaultseppunct}\relax
\EndOfBibitem
\bibitem[Peluzo and Galvão(2018)Peluzo, and Galvão]{Peluzo2018}
Peluzo,~B. M. T.~C.; Galvão,~B. R.~L. Theoretical study on the structure and
  reactions of uranium fluorides. \emph{Journal of Molecular Modeling}
  \textbf{2018}, \emph{24}, 197\relax
\mciteBstWouldAddEndPuncttrue
\mciteSetBstMidEndSepPunct{\mcitedefaultmidpunct}
{\mcitedefaultendpunct}{\mcitedefaultseppunct}\relax
\EndOfBibitem
\bibitem[Gagliardi \latin{et~al.}(1998)Gagliardi, Willetts, Skylaris, Handy,
  Spencer, Ioannou, and Simper]{Gagliardi1998}
Gagliardi,~L.; Willetts,~A.; Skylaris,~C.-K.; Handy,~N.~C.; Spencer,~S.;
  Ioannou,~A.~G.; Simper,~A.~M. A Relativistic Density Functional Study on the
  Uranium Hexafluoride and Plutonium Hexafluoride Monomer and Dimer Species.
  \emph{J. Am. Chem. Soc.} \textbf{1998}, \emph{120}, 11727--11731\relax
\mciteBstWouldAddEndPuncttrue
\mciteSetBstMidEndSepPunct{\mcitedefaultmidpunct}
{\mcitedefaultendpunct}{\mcitedefaultseppunct}\relax
\EndOfBibitem
\bibitem[Monard \latin{et~al.}(1974)Monard, Huray, and Thomson]{Monard1974}
Monard,~J.~A.; Huray,~P.~G.; Thomson,~J.~O. M\"ossbauer studies of electric
  hyperfine interactions in $^{234}\mathrm{U}$, $^{236}\mathrm{U}$,
  $^{238}\mathrm{U}$. \emph{PRB} \textbf{1974}, \emph{9}, 2838--2845\relax
\mciteBstWouldAddEndPuncttrue
\mciteSetBstMidEndSepPunct{\mcitedefaultmidpunct}
{\mcitedefaultendpunct}{\mcitedefaultseppunct}\relax
\EndOfBibitem
\bibitem[Belanzoni \latin{et~al.}(2005)Belanzoni, Baerends, and
  Van~Lenthe]{Belanzoni2005}
Belanzoni,~P.; Baerends,~E.~J.; Van~Lenthe,~E. The uranyl ion revisited: the
  electric field gradient at U as a probe of environmental effects.
  \emph{Molecular Physics} \textbf{2005}, \emph{103}, 775--787\relax
\mciteBstWouldAddEndPuncttrue
\mciteSetBstMidEndSepPunct{\mcitedefaultmidpunct}
{\mcitedefaultendpunct}{\mcitedefaultseppunct}\relax
\EndOfBibitem
\bibitem[Autschbach \latin{et~al.}(2012)Autschbach, Peng, and
  Reiher]{Autschbach2012}
Autschbach,~J.; Peng,~D.; Reiher,~M. Two-Component Relativistic Calculations of
  Electric-Field Gradients Using Exact Decoupling Methods: Spin-orbit and
  Picture-Change Effects. \emph{J. Chem. Theory Comput.} \textbf{2012},
  \emph{8}, 4239--4248\relax
\mciteBstWouldAddEndPuncttrue
\mciteSetBstMidEndSepPunct{\mcitedefaultmidpunct}
{\mcitedefaultendpunct}{\mcitedefaultseppunct}\relax
\EndOfBibitem
\bibitem[Autschbach(2014)]{doi:10.1098/rsta.2012.0489}
Autschbach,~J. Relativistic calculations of magnetic resonance parameters:
  background and some recent developments. \emph{Philosophical Transactions of
  the Royal Society A: Mathematical, Physical and Engineering Sciences}
  \textbf{2014}, \emph{372}, 20120489\relax
\mciteBstWouldAddEndPuncttrue
\mciteSetBstMidEndSepPunct{\mcitedefaultmidpunct}
{\mcitedefaultendpunct}{\mcitedefaultseppunct}\relax
\EndOfBibitem
\bibitem[Autschbach \latin{et~al.}(2010)Autschbach, Zheng, and
  Schurko]{10.1002/cmr.a.20155}
Autschbach,~J.; Zheng,~S.; Schurko,~R.~W. Analysis of electric field gradient
  tensors at quadrupolar nuclei in common structural motifs. \emph{Concepts in
  Magnetic Resonance Part A} \textbf{2010}, \emph{36A}, 84--126\relax
\mciteBstWouldAddEndPuncttrue
\mciteSetBstMidEndSepPunct{\mcitedefaultmidpunct}
{\mcitedefaultendpunct}{\mcitedefaultseppunct}\relax
\EndOfBibitem
\bibitem[Srebro and Autschbach(2012)Srebro, and
  Autschbach]{doi:10.1021/jz201685r}
Srebro,~M.; Autschbach,~J. Does a Molecule-Specific Density Functional Give an
  Accurate Electron Density? The Challenging Case of the CuCl Electric Field
  Gradient. \emph{The Journal of Physical Chemistry Letters} \textbf{2012},
  \emph{3}, 576--581, PMID: 26286152\relax
\mciteBstWouldAddEndPuncttrue
\mciteSetBstMidEndSepPunct{\mcitedefaultmidpunct}
{\mcitedefaultendpunct}{\mcitedefaultseppunct}\relax
\EndOfBibitem
\bibitem[Tecmer \latin{et~al.}(2011)Tecmer, Gomes, Ekström, and
  Visscher]{C0CP02534H}
Tecmer,~P.; Gomes,~A. S.~P.; Ekström,~U.; Visscher,~L. Electronic spectroscopy
  of {UO$_2^{2+}$, NUO$^+$} and NUN: an evaluation of time-dependent density
  functional theory for actinides. \emph{Phys. Chem. Chem. Phys.}
  \textbf{2011}, \emph{13}, 6249--6259\relax
\mciteBstWouldAddEndPuncttrue
\mciteSetBstMidEndSepPunct{\mcitedefaultmidpunct}
{\mcitedefaultendpunct}{\mcitedefaultseppunct}\relax
\EndOfBibitem
\bibitem[Christiansen \latin{et~al.}(1995)Christiansen, Koch, and
  Jørgensen]{cc2-christiansen-1995}
Christiansen,~O.; Koch,~H.; Jørgensen,~P. The second-order approximate coupled
  cluster singles and doubles model CC2. \emph{Chemical Physics Letters}
  \textbf{1995}, \emph{243}, 409--418\relax
\mciteBstWouldAddEndPuncttrue
\mciteSetBstMidEndSepPunct{\mcitedefaultmidpunct}
{\mcitedefaultendpunct}{\mcitedefaultseppunct}\relax
\EndOfBibitem
\end{mcitethebibliography}

%%%%%%%%%%%%%%%%%%%%%%%%%%%%%%%%%%%%%%%%%%%%%%%%%%%%%%%%%%%%%%%%%%%%%
%% The same is true for Supporting Information, which should use the
%% suppinfo environment.
%%%%%%%%%%%%%%%%%%%%%%%%%%%%%%%%%%%%%%%%%%%%%%%%%%%%%%%%%%%%%%%%%%%%%
\onecolumn

\begin{suppinfo}
This is the supporting information for the manuscript entitled: "Implementation of relativistic coupled cluster theory for massively parallel GPU-accelerated computing architectures" by J. V. Pototschnig, A. Papadopoulos, D. I. Lyakh, M. Repisky, L.Halbert, A. S. P. Gomes, H. J. Aa. Jensen, and L. Visscher. 
At first tables comparing the new implemtnation with the reference one are presented.
Subsequently, tables to determine the active occupied and virtual spinors for LaF and EuF are shown. 
Next, are some results for the structure of molecules, first for for AuAr$^+_n$, then for UF$_6$. Afterwards there is a chapter containing a list of the input options in the new code and an example for an input file. 
In the following two parts some code examples are given for ExaTENSOR and TAL-SH and their usage is discussed. 
Finally, we discuss the different integral transformation routines available in the code and present an argument for the default. 

\section{Verify Accuracy by comparing to RELCCSD}

\subsection{Energy}

At first we want to check that the correct numbers are produced by the code. 
H$_2$O and LiO were selected as small examples for an closed- and open-shell molecule, respectively. The differences between the new TAL-SH implementation and the RELCCSD results are collected in table~\ref{tab:comp_RELCC_TALSH_H2O_LiO}.
\begin{table*}[hbtp]
\caption{Differences between RELCCSD and TAL-SH for H$_2$O and LiO obtained for an convergence threshold of 1.0E-8. Values are in Hartree.}
\label{tab:comp_RELCC_TALSH_H2O_LiO}
\begin{tabular}{ | c | c | c | c | c | c | c | c | c | c | c | c |}
\hline
Molecule  & $\Delta$ CCSD & $\Delta$ CCSD+T & $\Delta$  CCSD(T) & $\Delta$ CCSD-T \\ 
 \hline
H2O(DZ)&5.10E-11&4.80E-11&4.50E-11&4.50E-11
\\
H2O(TZ)&1.10E-11&2.00E-11&2.10E-11&2.00E-11
\\
H2O(QZ)&1.20E-11&2.80E-11&2.60E-11&2.50E-11
\\
LiO(DZ)&1.00E-08&1.30E-07&2.20E-08&1.30E-07
\\
LiO(TZ)&1.00E-08&8.20E-07&2.20E-07&8.20E-07
\\
LiO(QZ)&9.90E-09&6.10E-07&1.40E-07&6.10E-07
\\
 \hline
\end{tabular}
\end{table*}
Correspondingly, table~\ref{tab:comp_RELCC_ExaCorr_H2O_LiO} contains the results for ExaTENSOR.
\begin{table*}[hbtp]
\caption{Differences between RELCCSD and ExaCorr for H$_2$O and LiO obtained for an convergence threshold of 1.0E-8. Values are in Hartree.}
\label{tab:comp_RELCC_ExaCorr_H2O_LiO}
\begin{tabular}{ | c | c | c | c | c | c | c | c | c | c | c | c |}
\hline
Molecule  & $\Delta$ CCSD & $\Delta$ CCSD+T & $\Delta$  CCSD(T) & $\Delta$ CCSD-T \\ 
 \hline
H2O(DZ)&5.10E-11&4.80E-11&4.50E-11&4.50E-11
\\
H2O(TZ)&1.10E-11&2.10E-11&2.10E-11&2.00E-11
\\
H2O(QZ)&2.20E-11&3.80E-11&3.60E-11&3.50E-11
\\
LiO(DZ)&1.00E-08&2.00E-07&1.80E-07&2.00E-07
\\
LiO(TZ)&1.00E-08&6.00E-07&1.00E-06&6.00E-07
\\
LiO(QZ)&9.90E-09&4.40E-07&8.60E-07&4.40E-07
\\
 \hline
\end{tabular}
\end{table*}
If one looks at the tables the differences between the closed- and open-shell system are noticeable. For the closed-shell molecule the errors are below the convergence threshold of 1.0E-8, but for LiO this only holds for the CCSD energy. The perturbative triples show slightly larger deviations.   

In order to test a larger system with different symmetries we selected CuAr$_n^+$ which was recently studied\cite{10.1021/acs.jpca.0c07771}.
The agreement is satisfactory, see table~\ref{tab:comp_RELCC_ExaCorr_CuAr}. 
\begin{table*}[hbtp]
\caption{Differences between RELCCSD and ExaCorr for CuAr$_n$ obtained for an convergence threshold of 1.0E-7. Values are in Hartree.}
\label{tab:comp_RELCC_ExaCorr_CuAr}
\begin{tabular}{ | c | c | c | c | c | c | c | c |}
\hline
Molecule  & $\Delta$ CCSD & $\Delta$ CCSD+T & $\Delta$  CCSD(T) & $\Delta$ CCSD-T \\ 
 \hline
CuAr$_1^+$ & 1.5E-07 & 1.4E-07 & 1.5E-07 & 1.5E-07 \\
CuAr$_2^+$ & 1.8E-07 & 1.8E-07 & 1.8E-07 & 1.8E-07 \\
CuAr$_3^+$ & 2.0E-07 & 2.0E-07 & 2.0E-07 & 2.0E-07 \\
 \hline
\end{tabular}
\end{table*}
In the recent publication also the open-shell molecules Cu$_2$Ar$_n^+$ were studied besides CuAr$_n^+$. In this case a direct comparison is not possible since in RELCCSD symmetry was used to deal with spinor degeneracy obtained the SCF level. In ExaCorr the spinor energies are recomputed, but this is not sufficient and non-convergence is observed in the CC iterations. A solution is to use a level shift for the virtual spinors, which results in convergence. Nevertheless, there is a difference in the binding energy of the two approaches of about 1.0E-3 Hartree. 

\subsection{Properties}

The properties were checked by comparing the two implementations (TAL-SH, ExaTENSOR) to the well tested DIRAC module. A non-symmetric molecule was selected and the results are listed in table~\ref{tab:comp_RELCC_ExaCorr_CHFClBr}. As can be see satisfactory agreement was obtained. 
 \begin{table*}[hbtp]
\caption{Differences between RELCCSD, ExaTENSOR and TAL-SH properties for CHFClBr obtained for an convergence threshold of 1.0E-8. Dipole moments and electric field gradients are in atomic units, the nuclear quadrupole coupling constant in MHz.}
\label{tab:comp_RELCC_ExaCorr_CHFClBr}
\begin{tabular}{ | l | l | l | l | l | l | l | l |}
\hline
Property  & xyz & RELCCSD & TAL-SH & ExaTENSOR & $\Delta$ (TAL-SH) & $\Delta$ (ExaTENSOR)\\ 
 \hline
 DM&X&-0.32275&-0.32275&-0.32275&$<$ 1E-10&$<$ 1E-10
 \\ 
&y&-0.37547&-0.37547&-0.37547&$<$ 1E-10&$<$ 1E-10
 \\ 
&z&-0.08469&-0.08469&-0.08469&$<$ 1E-10&$<$ 1E-10
 \\ 
EFG Br&qxx&-4.30878&-4.30878&-4.30878&1E-09&1E-09
 \\ 
&qyy&-4.07288&-4.07288&-4.07288&1E-09&1E-09
 \\ 
&qzz&8.38165&8.38165&8.38165&-2E-09&-2E-09
 \\ 
EFG Cl&qxx&4.04825&4.04825&4.04825&-2E-09&-2E-09
 \\ 
&qyy&-1.95116&-1.95116&-1.95116&1E-09&1E-09
 \\ 
&qzz&-2.09708&-2.09708&-2.09708&1E-09&1E-09
 \\ 
EFG F&qxx&-1.60295&-1.60295&-1.60295&$<$ 1E-10&$<$ 1E-10
 \\ 
&qyy&3.06056&3.06056&3.06056&$<$ 1E-10&$<$ 1E-10
 \\ 
&qzz&-1.45760&-1.45760&-1.45760&$<$ 1E-10&$<$ 1E-10
 \\ 
EFG C&qxx&-0.39569&-0.39569&-0.39569&-2E-10&-2E-10
 \\ 
&qyy&0.27723&0.27723&0.27723&$<$ 1E-10&$<$ 1E-10
 \\ 
&qzz&0.11845&0.11845&0.11845&2E-10&2E-10
 \\ 
EFG H&qxx&0.30145&0.30145&0.30145&$<$ 1E-10&$<$ 1E-10
 \\ 
&qyy&-0.14711&-0.14711&-0.14711&1E-10&1E-10
 \\ 
&qzz&-0.15434&-0.15434&-0.15434&$<$ 1E-10&$<$ 1E-10
 \\ 
NQCC Br&Xxx&-316.88&-316.88&-316.88&9E-08&1E-07
 \\ 
&Xyy&-299.54&-299.54&-299.54&2E-08&4E-08
 \\ 
&Xzz&616.42&616.42&616.42&-1E-07&-1E-07
 \\ 
&Eta&0.02814&0.02814&0.02814&$<$ 1E-10&$<$ 1E-10
 \\ 
NQCC Br&Xxx&-264.75&-264.75&-264.75&7E-08&9E-08
 \\ 
&Xyy&-250.25&-250.25&-250.25&1E-08&3E-08
 \\ 
&Xzz&515.00&515.00&515.00&-9E-08&-1E-07
 \\ 
&Eta&0.02814&0.02814&0.02814&$<$ 1E-10&$<$ 1E-10
 \\ 
NQCC Cl&Xxx&-77.67&-77.67&-77.67&4E-08&4E-08
 \\ 
&Xyy&37.43&37.43&37.43&-2E-08&-2E-08
 \\ 
&Xzz&40.23&40.23&40.23&-2E-08&-2E-08
 \\ 
&Eta&0.03604&0.03604&0.03604&$<$ 1E-10&$<$ 1E-10
 \\ 
NQCC Cl&Xxx&-61.21&-61.21&-61.21&3E-08&3E-08
 \\ 
&Xyy&29.50&29.50&29.50&-2E-08&-2E-08
 \\ 
&Xzz&31.71&31.71&31.71&-2E-08&-2E-08
 \\ 
&Eta&0.03604&0.03604&0.03604&$<$ 1E-10&$<$ 1E-10
 \\ 
NQCC H&Xxx&0.20&0.20&0.20&$<$ 1E-10&$<$ 1E-10
 \\ 
&Xyy&-0.10&-0.10&-0.10&$<$ 1E-10&$<$ 1E-10
 \\ 
&Xzz&-0.10&-0.10&-0.10&1E-10&1E-10
 \\ 
&Eta&0.02400&0.02400&0.02400&$<$ 1E-10&$<$ 1E-10
\\
 \hline
\end{tabular}
\end{table*}
In order to test a large and heavier molecule UF$_6$ was computed, see table~\ref{tab:comp_RELCC_ExaCorr_UF6}. 
\begin{table*}[hbtp]
\caption{Differences between RELCCSD and ExaCorr properties for UF$_6$ obtained for an convergence threshold of 1.0E-8. Dipole moments and electric field gradients are in atomic units, the nuclear quadrupole coupling constant in MHz.}
\label{tab:comp_RELCC_ExaCorr_UF6}
\begin{tabular}{ | l | l | l | l | l | l | l | l |}
\hline
Property  & xyz & RELCCSD & ExaTENSOR  \\ 
 \hline
DM&X&0.00000&0.00000
\\
&y&0.00000&0.00000
\\
&z&0.00000&0.00000
\\
EFG U&qxx&0.00000&0.00000
\\
&qyy&0.00000&0.00001
\\
&qzz&0.00000&-0.00001
\\
EFG F&qxx&-0.37729
\\
&qyy&-0.37729&-0.37729
\\
&qzz&0.75459&0.75459
\\
NQCC U&Xxx&0.00&0.00
\\
&Xyy&0.00&0.01
\\
&Xzz&0.00&-0.01
\\
&Eta&0.42310&0.55373
\\
NQCC F&Xxx&0.00&0.00
\\
&Xyy&0.00&0.00
\\
&Xzz&0.00&-0.01
\\
&Eta&0.42310&0.55373
\\
 \hline
\end{tabular}
\end{table*}

\clearpage

\section{Additional tables for the monofluorides}

\begin{table*}[hbtp]
\caption{
Ionization potential in eV of LaF for different numbers of correlated spinors. 
The number of occupied and virtual spinors refers to the neutral molecule, one spinor changes from occupied to virtual for the cation. 
The ionization potential of the reference determinant in all computations was 4.9346 eV. 
The spinor thresholds are listed in atomic units.
}
\label{tab:LaF_spin}
\begin{tabular}{c c c c c c c c}
threshold$_{low}$ & threshold$_{high}$ & nocc & nvir & \% occ & \% vir & CCSD 
\\
\hline
-3 & 40 & 18 & 136  & 27 & 44 & 5.9097\\
-20 & 40 & 36 & 136 & 55 & 44 & 5.9067\\
-60 & 40 & 56 & 136 & 85 & 44 & 5.9067\\
-300 & 40 &64 & 136 & 97 & 44 & 5.9066\\
\hline
-20 & 2 & 36 & 74   & 55 & 24 & 5.8782\\
-20 & 10 & 36 & 110 & 55 & 36 & 5.9049\\
-20 & 40 & 36 & 136 & 55 & 44 & 5.9067\\
-20 & 200 & 36 & 184& 55 & 60 & 5.9067\\
\hline
exp. &    &    &    &    &    & 6.3$\pm 0.3$\cite{ZMBOV1968a} \\
\end{tabular}
\end{table*}

\begin{table*}[hbtp]
\caption{
Ionization potential in eV of EuF for different numbers of correlated spinors. 
The number of occupied and virtual spinors refers to the neutral molecule, one spinor changes from occupied to virtual for the cation. 
The ionization potential of the reference determinant in all computations was 5.0399 eV. 
The spinor thresholds are listed in atomic units.
}
\label{tab:EuF_spin}
\begin{tabular}{c c c c c c c c}
threshold$_{low}$ & threshold$_{high}$ & nocc & nvir & \% occ & \% vir & CCSD 
\\
\hline
 -20 &   26 & 42 & 218 & 47 & 41 & 5.460 \\
 -50 &   26 & 54 & 218 & 75 & 41 & 5.462 \\
 -200 &  26 & 62 & 218 & 86 & 41 & 5.461 \\
 -270 &  26 & 66 & 218 & 92 & 41 & 5.461\\
 \hline
 -50 &  26 & 54 & 218 & 75 & 41 & 5.462 \\
 -50 &  40 & 54 & 234 & 75 & 44 & 5.462 \\
 -50 &  80 & 54 & 258 & 75 & 48 & 5.464 \\
 -50 & 200 & 54 & 303 & 75 & 55 & 5.465 \\
 -50 & 500 & 54 & 326 & 75 & 61 & 5.465 \\
 \hline
-200 & 200 & 62 & 303 & 86 & 55 & 5.464 \\
\hline
exp. &     &    &     &    &    & 5.9$\pm 0.3$\cite{ZMBOV1967} \\
\end{tabular}
\end{table*}
\clearpage

\clearpage

\section{Bond distances for the gold argon cation} 

\begin{table}[hbtp]
\small
\caption{AuAr$^+$ bond distances for different basis set sizes and levels of theory. 
They were obtained by fitting a Morse potentials to the lowest points of the potential. 
}
\label{tab:auar_dimer_distances}
\begin{tabular}{ c | c | c | c | c | c | c | c }
 basis	 & V & HF  &  MP2   &   CCSD & CCSD+T & CCSD(T) & CCSD-T    \\
\hline
2z  & 136 & 2.907 & 2.498 & 2.546 & 2.515 & 2.522 & 2.523
\\
3z  & 230 & 2.862 & 2.438 & 2.515 & 2.480 & 2.485 & 2.485
\\
4z  & 400 & 2.862 & 2.414 & 2.505 & 2.465 & 2.471 & 2.471
\\
\end{tabular}
\end{table}

\section{DFT optimization for the uranium hexafluoride dimer}

\begin{table}[hbtp]
\small
\caption{UF$_6$ dimer bonding energies(eV) and distances (\AA{}) as obtained in DFT computations. The U-F distances have been fixed for the experimental structures column, in the optimized case they range from 2.028 to 2.034~\AA{} with the mean values listed in the table. The U-U distances were optimized in all DFT computations.}
\label{tab:uf6_dimer_dft}
\begin{tabular}{ c | c | c | c | c | c | c }

 & \multicolumn{3}{c |}{DFT (opt. struct) } & \multicolumn{3}{c }{DFT (exp. struct)}
\\
\hline
 sym. & U-U & U-F & dE & U-U & U-F & dE 
 \\ 
 \hline
D2d & 5.142 & 2.030 & -0.552 & 5.139 & 1.996 & -0.160 
\\
D3d & 5.182 & 2.030 & -0.526 & 5.144 & 1.996 & -0.136 
\\
C2h & 5.298 & 2.030 & -0.541 & 5.290 & 1.996 & -0.150 
\\
\end{tabular}
\end{table}
The completely optimized structures in table~\ref{tab:uf6_dimer_dft} show a U-F bond distance which is about 0.034~\AA{} larger than the experimentally determined ones,\cite{Kimura1968} with the U-U internuclear separations also being larger in the fully optimized case. The bonding energy increases by about 0.34~eV for the fully relaxed structures.

\section{Input}

This section contains useful information for setting up an input file. 
The possible keywords for ExaCorr are listed in table~\ref{tab:keywords}. 
A definition of the occupied and virtual spinors using .OCCUPIED and .VIRTUAL is necessary. 
The keyword .EXATENSOR is required to activate the multi-node version, the default is to use the single node TAL-SH code. 

\begin{table}[htbp]
\caption{Keywords of the ExaCorr module}
\label{tab:keywords}
\begin{tabular}{ | p{3.5cm} | p{3.5cm} | p{7cm} |}
\hline
 keyword & example & description
 \\ \hline
 .EXATENSOR & &
Keyword activates the multi-node implementation using ExaTENSOR,
otherwise the single-node TAL-SH implementation is used  
 \\ \hline
.NOTRIPLES & &
keyword to deactivate triples 
 \\ \hline
.LAMBDA & &
Solve lambda equations, needs to be activated for properties
 \\ \hline
.TALSH\_BUFF & 50 &  
Maximum memory used in TAL-SH / for density matrix
\newline 50 GB default, set to 200 GB on summit
 \\ \hline
.CCDOUBLES & &
only compute CCD, default is CCSD
 \\ \hline
.CC2 & &
CC2 calculation, energies working, properties in progress
 \\ \hline
.OCCUPIED & 2..6,9,23..27 &
obligatory keyword, as list of MOs or energy range  
 \\ \hline
.VIRTUAL & energy 0.0 20.0 0.1 &
obligatory keyword, as list of MOs or energy range  
\\ \hline
.MOINT\_SCHEME & 4 &
1-4 available for ExaTENSOR,
42 uses Cholesky decomposition
\\ \hline
.OCC\_BETA & 2..6,9,23..27 &
list of beta spin MOs or energy range, alpha spin defined by .OCCUPIED, only use in open-shell computations  
\\ \hline
.VIR\_BETA & energy 0.0 2.0 0.1 &
list of beta spin MOs or energy range, alpha spin defined by .VIRTUAL, only use in open-shell computations  
\\ \hline
.EXA\_BLOCKSIZE & 75 &
Number to branch the tensors, should be $<$ nocc
\\ \hline
.PRINT & 0 &  
print level 
\\ \hline
.LSHIFT & 0.0D-0 &
Level shift 
\\ \hline
.NCYCLES & 30 &
Set the number of CC cycles
\\ \hline
.TCONVERG & 1.0D-9 &
Set convergence criterria (CC iterations, lambda equations)
\\ \hline
.CHOLESKY & 1.0D-9&
threshold for Cholesky decomposition
\\ \hline
\end{tabular}
\end{table}

\clearpage
Below is an example for an input file, where ExaCorr is called by DIRAC. 
\renewcommand{\baselinestretch}{1.0}
\begin{verbatim}
**DIRAC
.WAVE FUNCTIONS 
.PROPERTIES
**PROPERTIES
.DIPOLE
.EFG
.NQCC
**HAMILTONIAN
.X2C
**INTEGRALS
*READIN
.UNCONTRACT
**WAVE FUNCTIONS
.SCF
.EXACC
*SCF
.ERGCNV
1.0E-9
.MAXITR
50
**EXACC
.EXATENSOR
.OCCUPIED
energy -6.0 -0.4 0.001
.VIRTUAL
energy -0.4 5.0 0.001
.NOTRIPLES
.LAMBDA
.MOINT_SCHEME
1
.TCONVERG
1.0E-8
.NCYCLES
80
**MOLECULE
*SYMMETRY
.NOSYM
*BASIS
.DEFAULT
dyall.v2z
*END OF
\end{verbatim}
An additional xyz-file specifying the coordinates of the molecule is necessary.
It can then be run by:
\begin{verbatim}
PATH_TO_DIRAC/pam --mpi=#mpi --mol=molecule.xyz --inp=dirac.inp
\end{verbatim}

\section{ExaTENSOR code examples}

In this part we will give hints and examples for implementing code based on the ExaTENSOR library. 

\subsection{Spaces}

Before the library can be started the spaces and method used in the library need to be defined. In order to set up the spaces an basis needs to be define first using the subspace\_basis\_t data type. 
The basis is set up using
\begin{verbatim}
call basis%subspace_basis_ctor(basis_size,ierr)
\end{verbatim}
where the basis functions can be assigned a color (optional argument) to differentiate them:
\begin{verbatim}
call basis%set_basis_func(i_function,BASIS_ABSTRACT,ierr,symm=color).
\end{verbatim}
This assignment of colors assures that the indices are not split if they have the same color. 
The space is then registered using the basis
\begin{verbatim}
ierr=exatns_space_register('space',basis,space_id,space,branch_factor)
\end{verbatim}
where space\_id is an integer identifying the space and space is an object of the class h\_space\_t. The branching factor determines in how many pieces the tensor is split up during execution and was determined by
\begin{verbatim}
branch_factor = max(2, (int(get_num_segments(int(nvir,8),
int(exa_input%exa_blocksize,8)),4) - 1)/tavp_mng_depth + 1)
\end{verbatim}

\subsection{Methods}

By extending the tens\_method\_uni\_t type user-defined methods can be created, 
like for example a method to initialize a tensor representing a delta function
\begin{verbatim}
type, extends(tens_method_uni_t), public:: delta_t
  integer(INTL), private :: ind(4)
  contains
    procedure, public:: delta_t_init
    procedure, public:: reset=>delta_t_reset
    procedure, public:: pack=>delta_t_pack
    procedure, public:: unpack=>delta_t_unpack
    procedure, public:: apply=>delta_t_apply
end type delta_t
\end{verbatim}
Such a type can contain private information, like in this case an integer that determines where to put the unit values in the otherwise zero tensor. For the initalizer of the MO coefficients the coefficients themselves are stored in the type. 
There is one method to set this data before the method is set up, in this case delta\_t\_init. There is also the possibility to reset private data, 
but than it needs to communicated with the pack and unpack procedures. 
If the method is called the apply function is executed on each node.
In order to get the local part of the tensor the following function is used
\begin{verbatim}
tens=tensor%get_dense_adapter(ierr)
\end{verbatim}
The variable tens is of the tens\_dense\_t type and contains information about the size of the local tensor as well as an pointer to it, which can be accessed as an Fortran array
\begin{verbatim}
call c_f_pointer(tens%body_ptr,tens_fortran,tens%dims(1:tens%num_dims))
\end{verbatim}
which can be manipulated with standard Fortran operations. 
It is also possible to use external libraries as for example InteRest\cite{Repisky2018}.

The methods defined in such a way have to be registered before the library is started and if there are any private variables that need to be initialized it has to be done beforehand.  
\begin{verbatim}
call f_delta%delta_t_init(ind_tens)
ierr=exatns_method_register('Delta_F',f_delta)
\end{verbatim}

During runtime of the library the private variables can be updated via
\begin{verbatim}
call f_delta%reset(ind_tens)
\end{verbatim}

\subsection{Usage}

The library is then started using the command
\begin{verbatim}
ierr=exatns_start()
\end{verbatim}
where as an optional parameter the MPI communicator from the code using the ExaTENSOR library can be passed on. 
A tensor first needs to be created
\begin{verbatim}
ierr=exatns_tensor_create(tensor,"tensor_name",tensor_id,tensor_root,EXA_DATA_KIND_C8)
\end{verbatim},
where tensor is a variable of the type tens\_rcrsv\_t, the name of the tensor i given by tensor\_name, tensor\_id is a vector containing the space id's for all the dimensions, and tensor\_root determines the actual length for these dimensions. 
Every tensor needs to be initialized either with a user-defined method (in this case ZERO is replaced by the name of the method) or a constant (ZERO is the default value of the same type as the tensor) using the command 
\begin{verbatim}
ierr=exatns_tensor_init(tensor,ZERO)
\end{verbatim}.

The main operation of the code is a tensor contraction
\begin{verbatim}
ierr=exatns_tensor_contract("S(a,b,i,j)+=V(a,b,c,d)*T(c,d,i,j)",s2,vvvv,t2,scalar)
\end{verbatim},
where the string defines the contraction, s2/vvvv/t2 are tens\_rcrsv\_t tensors and the scalar allows to scale the result. 
The application of user-defined is also important
\begin{verbatim}
ierr=exatns_tensor_transform(tensor,method)
\end{verbatim}.
The method can either be called by its name or passed as an object, if it is updated. 

After usage the memory of a tensor is released by
\begin{verbatim}
ierr=exatns_tensor_destroy(tensor)
\end{verbatim}

There are two ways to get values back, one is to create a scalar tensor and use the function
\begin{verbatim}
ierr=exatns_tensor_get_scalar(tensor,value)
\end{verbatim}.
An alternative is to create a TAL-SH tensor on the main node and obtain a part of the tensor or the whole tensor by
\begin{verbatim}
ierr=exatns_tensor_get_slice(tensor,TALSH_tensor)
\end{verbatim}.

\section{TAL-SH code examples}

This part contains some hints regarding the implementations with TAL-SH.
This library doesn't require a set up of spaces or methods before it starts, as it is designed for single node without requiring communicating information. 
The library is started by
\begin{verbatim}
ierr=talsh_init(buf_size)
\end{verbatim}
There are two modes to initialize the library. Either the buffer size (buf\_size) is provided by the user and the library only uses the given amount of memory, or the arrays are allocated dynamical and use the memory they require. 

The tensors are created by the command
\begin{verbatim}
ierr=talsh_tensor_construct(tensor,C8,dims,init_val=ZERO)
\end{verbatim},
where tensor is a object of the type talsh\_tens\_t, C8 defines the datatype, dims defines the size of the tensor and is a one dimensional array of integers. 
These tensors can be initialized during creation with the optional parameter init\_val.

By using pointers the tensors can be accessed by following commands
\begin{verbatim}
ierr=talsh_tensor_get_body_access(tensor,body_p,C8,0,DEV_HOST)
call c_f_pointer(body_p,tens,dims)
\end{verbatim}.
The pointer is set to the position of the body of the tensor with the first command. 
The C pointer (C\_PTR) body\_p is then used to link the fortran array tens to a tensor of the dimensions dims. This enables the initialization of tensors, operations on the elements and extraction of elements. It is possible to use external libraries, e.~g. InteRest in these executions. 

The contractions are defined by a string for the input tensors t2 and vvvv, and the output tensor s2
\begin{verbatim}
ierr=talsh_tensor_contract("S(a,b,i,j)+=V(a,b,c,d)*T(c,d,i,j)",s2,vvvv,t2,scale=NUM)
\end{verbatim}.
The result can be scaled by a number NUM.

\begin{verbatim}
ierr=talsh_tensor_destruct(tensor)
\end{verbatim}
is used to delete the tensors and release the memory.

\section{Alternative integral transformations}

The integral transformation takes some time, especially in the MOLTRA module in DIRAC. 
An efficient implementation is desired and for this reason several approaches were tested and compared. 
In the ExaTENSOR part of ExaCorr four different integral transformation schemes are available. The 1\textsuperscript{st},  3\textsuperscript{rd}, and 4\textsuperscript{th} 
follow equation~\ref{eq:ao2mo_scheme1}. The latter two reuse the half transformed integrals which increases efficiency. The 4\textsuperscript{th} scheme uses batches to avoid the large AO tensor. 
An alternative is to construct the density matrix and use it for the transformation (scheme 2): 
\begin{align}
D_{\kappa p \lambda q} = C^\dagger_{\kappa p} C_{\lambda q} ,
\\
\left[ p q | \mu \nu \right] = 
\sum_{\lambda \kappa }^{n_{AO}} D_{\kappa p \lambda q}
\left[ \kappa \lambda | \mu \nu \right] ,
\\
\left[ p q | r s \right] = 
\sum_{\mu \nu}^{n_{AO}} D_{\mu r \nu s}
\left[ p q | \mu \nu \right] 
\label{eq:ao2mo_scheme2}
\end{align}
This scheme has a N$^6$ scaling and the density matrix requires additional memory as well. It is meant as a precursor for algorithms that exploit sparsity in the density matrices.
The 3\textsuperscript{rd} and 4\textsuperscript{th} scheme are currently the best performing ones and scheme 4 is taken as a default as it has the smallest memory footprint.

\begin{table}[hbtp]
\caption{Compare the time in seconds (t$_I$) of different integral transformation schemes for different systems alongside the time for the CCSD computations (t$_{CC}$). The number of occupied (n$_{occ}$) and virtual (n$_{vir}$) spinors are listed in the table. n(nodes) is the number of used summit nodes.  }

\label{tab:integral_trans}
\begin{tabular}{ | c | c | c | c | c | c | c | c | c | }
\hline
 system & n$_{occ}$ & n$_{vir}$ & n(nodes) & t$_I$ (1) & t$_I$ (2) & t$_I$ (3) & t$_I$ (4) & t$_{CC}$\\ 
\hline
SF6 (DZ) &  48 & 252   &  8 &  56       & 56       &  39        &  62  & $\approx$ 380 \\
\hline
SF6 (TZ) &  48 & 458   &  32 &  232      & 356      &  152       &  209 & $\approx$ 1050 \\ 
\hline
LaF3 (DZ) &  50 & 240  &   8 &  129      & 129      &  60        &  129 & $\approx$ 300\\
\hline
UF6 (DZ)  &  66 & 314  &  40 &  430      & 477       & 219       & 277 & $\approx$ 2000\\
 \hline
\end{tabular}
\end{table}

\end{suppinfo}

\twocolumn

\end{document}